\documentclass[a4paper,11pt]{article}
\pdfoutput=1 

\usepackage{jheppub} 

\usepackage[T1]{fontenc} 

\usepackage{tikzit}

\tikzstyle{red_ball}=[fill=red, draw=red, shape=circle, tikzit category=Ball]
\tikzstyle{blue_ball}=[fill=blue, draw=blue, shape=circle, tikzit category=Ball]
\tikzstyle{medium_box}=[fill=white, draw=black, shape=rectangle, tikzit category=boxes, minimum width=0.75cm, minimum height=1cm]
\tikzstyle{blank_edge}=[line width=0.2cm]
\tikzstyle{small_box}=[fill=white, draw=black, shape=rectangle, tikzit category=boxes, minimum height=0.4cm, minimum width=0.4cm]
\tikzstyle{big_box}=[fill=white, draw=black, shape=rectangle, tikzit category=boxes, minimum height=2.0cm, minimum width=1.5cm]
\tikzstyle{small_box_2}=[fill=white, draw=black, shape=rectangle, tikzit category=boxes, minimum height=0.6cm, minimum width=0.6cm]

\tikzstyle{white_edge}=[-, draw=white, line width=3.2pt]
\tikzstyle{blue_edge}=[-, draw=blue]
\tikzstyle{red_edge}=[-, draw=red]
\tikzstyle{red_arrow}=[->, draw=red]
\tikzstyle{blue_arrow}=[->, draw=blue]
\tikzstyle{arrow}=[->]
\tikzstyle{blue_filled_line}=[-, fill={rgb,255: red,255; green,193; blue,207}, draw=black]
\tikzstyle{pink_filled}=[-, fill={rgb,255: red,165; green,255; blue,235}]

\usepackage{mathtools} 
\usepackage{amsthm} 
\usepackage{braket} 
\usepackage{subcaption} 
\usepackage{import} 
\usepackage{comment} 
\theoremstyle{definition}
\newtheorem{definition}{Definition}[section]
\newtheorem{theorem}{Theorem}[section]

  \newcommand{\Tr}{\operatorname{Tr}}
  
  \newcommand{\Vol}{\operatorname{Vol}}
  
  \newcommand{\wes}[1]{\textcolor{blue}{#1}}

\title{\boldmath On a Matrix Ensemble for Arbitrary Complex Quantum Systems}

\author[a,b,1]{William E. Salazar,}
\author[c]{Juan Diego Urbina}
\author[a]{and Javier Madroñero}

\affiliation[a]{Centre for Bioinformatics and Photonics (CIBioFi),
Universidad del Valle, Edificio E20 No. 1069, 760032 Cali, Colombia}
\affiliation[b]{ICTP South American Institute for Fundamental Research, Instituto de Física Teórica, UNESP Univ. Estadual Paulista, Rua Dr. Bento Teobaldo Ferraz 271, 01140-070, São Paulo, SP, Brazil}
\affiliation[c]{Institut für Theoretische Physik, Universität Regensburg, 93040 Regensburg, Germany}

\emailAdd{william.esteban@unesp.br}
\emailAdd{juan-diego.urbina(at)ur.de}
\emailAdd{javier.madronero@correounivalle.edu.co}

\abstract{We present a comprehensive analytical study of a variation of the eigenvector ensemble initially proposed by Deutsch \cite{Deutsch/great_idea/1991,Deutsch/ETH_famous/1991} for the foundations of the Eigenstate Thermalization Hypothesis (ETH). This ensemble, called the $C$-ensemble, incorporates additional system-dependent information, enabling the study of complex quantum systems beyond the universal predictions of Random Matrix Theory (RMT). Specifically, we focus on how system-specific details influence late-time behavior in correlation functions, such as the spectral form factor, and how explicit Hamiltonian corrections not captured by RMT can be included. We demonstrate the consistency of this ensemble with respect to the universal (Haar) results by showing that it defines a unitary 1-design for arbitrary systems and for strongly chaotic systems it becomes an approximated 2-design. Universal expressions for two- and four-point ensemble-averaged correlation functions are derived, revealing how system-dependent information is spectrally decoupled. Furthermore, we show that for small energy windows, the correlation functions defined by this ensemble reduce to the predictions made by the ETH.}

\begin{document} 
\maketitle
\flushbottom

\section{Introduction}

To what extent can the dynamics of individual chaotic many-body quantum systems be accurately described by sampling over random and uniformly distributed unitaries? A nowadays standard answer, mainly supported by evidence from black-hole physics \cite{Page/Black_hole_entropy/1993,Preskill/Black_holes_mirrors/2008,susskind/Fast_scramblers/2008,Shenker/Black_holes_Butterfly/2014,Susskind/Complementarity/1993}, is to argue that this replacement can be made at late times when the chaotic physics of the individual system is well captured by the universal content of random matrix ensembles \cite{Atland/Holographic_quantum_chaos/2021}. Here, late times means times-scales greater than thermalization/ergodic time\footnote{For the single-particle case, this time scale can be understood as the time scale beyond which a chaotic flow uniformly
covers the classical energy shell in phase space.}, defined as the characteristic time scale taken for the connected part of two-point correlations between local operators to decay.

A simple but insightful argument to understand the emergence of the seemingly random (pseudo-random) dynamics from the underlying chaotic structure of the individual many-body Hamiltonian was provided in \cite{Yoshida/Chaos_design/2017,Preskill/Black_holes_mirrors/2008} and can be summarized as follows. In a typical chaotic many-body system, the Hamiltonian $H$ is a sum of strings of low-weight operators, \emph{e.g.} for a $k$-local $N$-qubit system \cite{Susskind/Second_law/2018}

\begin{equation}
 H = \sum_{i_{1}<i_{2}\dots<i_{k}}\, J_{i_{1},\dots,i_{k}} \sigma_{i_{1}} \sigma_{i_{2}} \sigma_{i_{3}} \dots \sigma_{i_{k}}, 
\end{equation}
with $J_{i_{1},\dots,i_{k}}$ the couplings, and $\sigma_{i_{l}}$ single Pauli operators. The unitary dynamics generated by $H$ is specified at all times by $U(t) =e^{-itH}= \mathbb{I}-it H+\dots$, where as time increases higher powers of $H$ become dominant. In terms of the Pauli strings contributing to the dynamics, after a certain time, the length of the strings will grow up to $\sim N$, and $U(t)$ will become a sum of long-Pauli strings pseudo-randomly weighted by higher powers of the couplings $J_{i_{1},\dots,i_{k}}$. At late times, the contribution from those pseudo-randomly weighted strings only can increase, therefore the replacement of $U(t)$ by a uniform distributed unitary $U$ appears to hold\footnote{The same argument also holds directly for the dynamics of an arbitrary local operator $W$, \emph{e.g.} $W(t)=e^{itH}We^{-itH}=W+it[H,W]+\dots$, and as time increases Higher commutators will dominate.}.

However, as natural as the approximation of the chaotic dynamics with the pure random one appears to be, here we desire to remark that this is indeed not the case, and to replace the chaotic dynamics of individual many-body systems with a uniform random distributed unitary is rather a crude approximation. To argue in favor of this claim we provide the following two-side argument:

\begin{itemize}
    \item First, the effect of replacing the late-time dynamics with a randomly distributed unitary is that, at the level of observables, such as correlation functions, the late-time behavior will be simply obtained by averaging over the unitary group, \emph{e.g.} for a 2-point function $\Tr W(t)W$ we schematically will have equality's like

\begin{equation}
 \label{eq:Haar_two_points_introduction}
 \Tr{(W(t)W)}\xrightarrow[\text{times}]{\text{late}}\int_{U(d)}\Tr{(U^{\dagger}WUW)} [U^{\dagger}dU],
\end{equation}
where $[U^{\dagger}dU]$ denotes the normalized $U(d)$ Haar-Measure. Particularly, such an average agrees with the expected late time decay of the connected part of the two point correlation \cite{Alhambre/two_point_decay/2020} characteristic in chaotic many-body systems, \emph{i.e.} $\Tr (W(t)W)\sim \Tr{W}\Tr{W}$. However, the right side in \eqref{eq:Haar_two_points_introduction} has the failure that in order for such an average to be physical in the first place, each unitary inside the integral symbol must be generated by a Hamiltonian. Otherwise stated, the unitary average must match the average over some unknown ensemble of Hamiltonians $\mathcal{E}_{H}$, \emph{i.e.}

\begin{equation} 
\label{eq:Unitary_vs_hamiltonain_ensembles}
 \int_{U(d)}\Tr{(U^{\dagger}WUW)} [U^{\dagger}dU]\approx \int_{\mathcal{E}_{H}} \Tr\left(e^{itH}We^{-itH}W\right)d\mu[H],
\end{equation}
with $d\mu[H]$ the measure of such ensemble and the time variable must be interpreted to be evaluated at late times. For individual chaotic many-body systems, \eqref{eq:Unitary_vs_hamiltonain_ensembles} implies that the late time physics is captured by an ensemble of Hamiltonians, and even though this result is congruent with the random matrix theory (RMT) universality \cite{Atland/Holographic_quantum_chaos/2021}, it is also known that scars from the individual system must also appear. Concretely, for an individual system it is expected that the energy is conserved\footnote{This relation has a stronger counterpart by demanding that all the expected value of the traces are equal, \emph{i.e.} $\braket{\Tr{H^{n}}}_{\mathcal{E}_{H}}=\Tr{H^{n}}$. This can be also stated as saying that the energy is not being conserved \cite{Stanford/subleading_weingartens/2022}.}, at least on average, by the Hamiltonian ensemble that captures their chaotic properties, \emph{i.e.} $\braket{\Tr{H}}_{\mathcal{E}_{H}}=\Tr{H}$. For the uniform (Haar) ensemble in \eqref{eq:Haar_two_points_introduction} the corresponding unknown Hamiltonian ensemble implies $\braket{\Tr{H}}_{\mathcal{E}_{H}}=0$ and therefore the replacement $U(t) \to U$ does not conserve energy\footnote{For uniform distributed unitaries we have $\braket{\Tr H}_{\mathcal{E}_{H}}\sim \int_{U(d)}\Tr{\log U}\,[U^{\dagger}dU]=0$.}. 

Is worth mentioning that, in contrast to individual chaotic many-body systems, the authors in \cite{Cotler/spectral_decoupling/2020} 
show that the approximation \eqref{eq:Unitary_vs_hamiltonain_ensembles} holds rather broadly for ensemble-averaged ones.

\item
The second argument to go against the uniform (Haar) ensemble as a correct replacement for the unitary dynamics of individual chaotic many-body systems is to compare them at the level of spectral statistics. Concretely, if $e^{-itH}$ were a truly uniform distributed unitary, then its spectrum would follow a Wigner-Dyson distribution with complex eigenvalues lying on the unit circle. On the other hand, since $H$ is the Hamiltonian of a chaotic system, its eigenvalues lie on the real line and are expected to display level repulsion. The exponential $e^{-itH}$ maps the real line onto the unit circle by wrapping it multiple times, meaning that the final spectrum of $e^{-itH}$ will have many closed energy levels and will follow a Poisson distribution, rather than the Wigner-Dyson distribution of a Haar random distributed unitary. This argument, attributed to Michael Berry \cite{Cotler/spectral_decoupling/2020}, has been recently employed to argue against unitaries as a replacement for chaotic time evolution \cite{Stanford/subleading_weingartens/2022}.
\end{itemize}
However, even if the Haar ensemble is a rather crude approximation to the late-time physics of individual many-body chaotic systems, it has also allowed to unveil many features of many-body dynamics. Particularly, it has led to insights into the close relationship between chaos and scrambling in many-body systems \cite{Hosur/Chaos_Scrambling_Channels/2016} and also has predicted the leading terms of higher correlation functions for holographic duals \cite{Stanford/subleading_weingartens/2022}. In this direction, one of the strongest points of the Haar ensemble in favor of modeling late-time dynamics is its analytical simplicity at the moment of computing correlation functions. By simplicity, we refer to the ability to obtain the leading contribution of such correlations from the asymptotic of unitary-group integrals \cite{Brouwer/Asymptotic_weingarten/1996}.  

From this discussion an important question arises: Is the pseudo-randomness of the unitary time evolution of individual chaotic many-body systems exclusively captured by the Haar ensemble, or is there an additional ensemble, perhaps simpler (in terms of the number of necessary unitaries), that is capable of accurately reproduce/extent the Haar results which are expected to hold for chaotic many-body quantum systems?

If such an ideal ensemble that captures the pseudo-randomness of the chaotic-time evolution exists, then, as mentioned previously, a necessary condition is that certain scars from the Haar ensemble still remain, \emph{e.g.} we expect that the ensemble average two-point correlations between local operators still decay. Additionally, for higher-order correlations, such as four-point ones, we also expect a decay. However, as four-point correlations are more sensitive to the short-time physics characterized by the exponentially growth of the non-commutativity of the operators, we expect strong deviations from the Haar result, provided by the non-universal individual Hamiltonian. As shown in \cite{Yoshida/Chaos_design/2017,Hosur/Chaos_Scrambling_Channels/2016}, the concept of closeness or remoteness in  ensemble averaged $k$-order correlation functions can be diagnosed by unitary $k$-designs, \emph{i.e.} a smaller set of unitaries capable of reproducing the moments of the full unitary group. Therefore, we qualitatively would expect that such an ideal ensemble was closer enough to be an unitary $2$-design, \emph{i.e.} the averaged two-point correlations must decay in a largely universal manner. However, as soon as the order of the correlations begins to increase, then due the fine-details of the individual Hamiltonian, the forehead mentioned ideal ensemble must deviate from being an upper $k$-design, \emph{i.e.} averaged upper-point correlations must become more and more sensitivity to $H$ as the order increases.

In this paper, we present a candidate for such an ideal ensemble called the $C$-ensemble. The construction of such ensemble is based on a modification and re-interpretation of Deutsch's RMT approach to the foundations of the Eigenstate Thermalization Hypothesis \cite{Deutsch/ETH_famous/1991,Nation/Deutsh_ETH/2018} for single systems rather than ensembles of Hamiltonians. This deviation from the classical RMT landscape of Hamiltonian ensembles allows the $C$-ensemble to possess an intrinsic twofold appealing: first, it counts over a much smaller set of unitaries rather than the whole unitary group, making it more conceptually accurate. Second, it allows for the introduction of specific Hamiltonian dependence in correlation functions, thereby going beyond the Haar ensemble's predictions.

As a short-guide, we provide a concise list of the main results next to the particular section where they appear.

\begin{itemize}
  \item In Section. \ref{section:C_ensemble}, we introduce the $C$-ensemble and provide the physical motivation behind it. We start by revisiting the basic machinery from quantum chaos and quantum information necessary for the rest of the work. Particularly, we review the concepts of $k$-designs and spectral decoupling and define the notion of a $k$-spectral decoupled ensemble as one capable of producing $k$-spectral decoupled correlations. As the main results of the section, we show that the $C$-ensemble yields an unitary $1$-design, and for special classes of Hamiltonians, it yields an approximated $2$-design. Additionally, by highly constraining the form of two-point correlations we are able to provide a closed-form expression to the ensemble two-fold channel.
  
  \item In Section. \ref{Chapter:two_points} we probe the $C$-ensemble against the Haar-one at the level of finite and infinite temperature one- and two-point correlation functions. Concretely, we define a special object, called the plateau operator, which captures fine details of the individual Hamiltonian $H$ and gives rise to the deviations from the universal Haar result for the evaluation of correlation functions. At the end of this section, we conclude by providing some comments about the connection between the $C$-ensemble with the Eigenstate-Thermalization-Hypothesis and show that both statements are locally compatible, \emph{i.e.} over small energy windows.
  
  \item In Section. \ref{Chapter_chaos} we compute the $C$-ensemble average of four-point correlation functions. Concretely, we use a method to express high-order correlation functions as low-order ones but in a larger Hilbert-space. This method allows us to set the $C$-ensemble average of 
  four-point functions such as out-time-ordered correlations (OTOC) in terms of averaged two-point correlation functions previously discussed in Section. \ref{Chapter:two_points}. 
  
  We finish Section. \ref{Chapter_chaos} by providing a close expression for the size (\emph{i.e.} the number of unitaries) of the $C$-ensemble associated with a Hamiltonian $H$. Concretely, we comment about the role of the spectral statistics \emph{e.g.} the difference between the size of a $C$-ensemble associated with a system that presents level clustering and one that presents level repulsion. Additionally, as was done by the authors in \cite{Yoshida/Chaos_design/2017}, by interpreting the complexity of an ensemble of unitaries with minimum number of elementary gates needed to build the ensemble, we make some comments about the complexity of the $C$-ensemble as well.
  
  \item Finally, in Section. \ref{summary_outlook} we present a short summary of our results and also comments on further directions.

  \item The Appendices contain various formulas and derivations, that support the statements presented on the main part of the paper.
\end{itemize}

\section{The C-ensemble General Aspects}
\label{section:C_ensemble}

In this section, we introduce and motivate the main properties of the $C$-ensemble by posing the following question: \emph{For 
systems described by individual Hamiltonians, is there a suitable ensemble that accurately captures system-specific corrections to the universal RMT regime of late-time correlation functions?}. This question is subtle, in ensemble-averaged theories, the average in question is defined tautologically by sampling over the ensemble elements. However, for a fixed Hamiltonian we lose this type of ensemble-average\footnote{Let us clarify that we are interested in constructing an ensemble for a fixed Hamiltonian, this is not the same as representing exact correlation functions as some integral average, {\em i.e.} as in periodic orbit theory.}). Clearly, in any attempt to introduce an ensemble average for an individual system \footnote{This is not the case for certain Holographic duals, \emph{e.g.} JT-gravity \cite{Saad/Jt_gravity_matter/2019}, in where the bulk-physics is itself dual to an ensemble of Hamiltonian's. } system we expect to partially lose some information. The key point is then: What is the relevant information that we demand to retain by introducing such an ensemble?. We argue that the relevant information is in the spectral data of the individual Hamiltonian, \emph{i.e.} in the spectrum and the correlations between energy-eigenstates. Here, we will present the $C$-ensemble as an approach for introducing an ensemble average in systems with a fixed Hamiltonian.

\subsection{Eigenvector vs Hamiltonian ensembles}

Following \cite{Cotler/spectral_decoupling/2020}, let us address the concept ``disordered'' theory as one where physical observables are obtained by averaging over some ensemble of Hamiltonians $\mathcal{E}_{H}$ with probability measure $d\mu[H]$. The particular Hamiltonian ensemble might be the one of a system with quenched disorder, a random matrix ensemble, etc. As an example, for the Sachdev-Ye-Kitaev (SYK) model \cite{Cotler/Blackholes_SYK/2017,Maldacena/solution_SYK/2016}, this means that for every function $f$,
\begin{equation*}
  \Braket{f(H)}_{\mathcal{E}_{\text{SYK}}}=\int_{\text{couplings}} f(H_{SYK})[dJ].
\end{equation*}
Here the integral runs over the joint Gaussian distribution of random couplings $J$ defining the measure $[dJ]$. Another option instead of ``randomizing'' the Hamiltonian would be to randomize the change of basis matrix, \emph{e.g.}, by randomly select the transformation from the eigenbasis to any other local basis of the Hilbert space. 

The explicit construction of this ensemble proceeds as follows. Consider a $d$-dimensional Hilbert space $\mathcal{H}$ where we distinguish both a computational basis $\{\ket{1},\ket{2},\dots, \ket{d}\}$ (with respect to which we define local operators) and the basis of Hamiltonian energy-eigenstates $\{\ket{E_{1}},\ket{E_{2}},\dots, \ket{E_{d}}\}$. Over this Hilbert space we introduce the operator $C$ defined in the computational basis,

\begin{equation}
  \label{eq:C_operator}
  \braket{l|C|m}=\braket{E_{l}|m },
\end{equation}
as the matrix of overlaps between eigenvectors and computational basis states\footnote{Notice that with this definition the $C$ operator corresponds to the conjugate of the usual transition matrix between the computational and eigenbasis, {\em e.g.} $C^{*}_{lm}=\braket{m|E_{l}}$.}. An eigenvector ensemble is obtained by identifying the $C$ operator as a random variable of some matrix-ensemble $\mathcal{E}_{C}$. The motivation behind this is straightforward, for complex systems one expects that the overlaps between both bases are erratic variables. We want to highlight that an eigenvector ensemble always depends upon the specific Hamiltonian chosen. To begin with, the symmetry class of $H$ defines the group where $C$ belongs, {\em i.e.} for the three Dyson symmetry classes \cite{Dyson/symmetry_class/1962} $C$ can be either unitary, orthogonal or symplectic\footnote{For $H$ without anti-unitary symmetries $C$ will be unitary whereas for a conventional $T$ symmetry $C$ is an orthogonal matrix. For the non-standard symmetry classes \cite{Altland/Non_standard_symmetry/1997,Zirnbauer/symmetry_class/2010} the symmetry group will be a quotient one.}. In this work, we will focus on Hamiltonians of the unitary class and therefore work with the unitary eigenvector ensembles.

As a motivation for the analytical exploration we are about to undertake, let us present an example to show how averages over an eigenvector ensemble work and compare them against a standard Hamiltonian ensemble computation. Suppose that we want to average the thermal expected value of some local operator $W$ over the Gibbs state $\rho_{\beta}=Z(\beta)e^{-\beta H}$,

\begin{align*}
  \Braket{\frac{\Tr{e^{-\beta H}W}}{Z(\beta)}}_{\mathcal{E}_{H}}\, \xleftarrow[\text{ensemble}]{\text{Hamiltonian}} \frac{\Tr{e^{-\beta H}W}}{Z(\beta)}\xrightarrow[\text{ensemble}]{\text{eigenvector}}\, \frac{\braket{\Tr{e^{-\beta E}CWC^{\dagger}}}_{\mathcal{E}_{C}}}{Z(\beta)}\,,
 \end{align*}
 where $Z(\beta)=\Tr{e^{-\beta H}}$ denotes the canonical partition function at inverse temperature $\beta$. Notice that while in the Hamiltonian ensemble side we wash out the spectral degrees of freedom, in the eigenvector ensemble side those spectral degrees of freedom remain in the diagonal matrix of eigenvalues $H=C^{\dagger}EC$. For the special case of unitarily invariant Hamiltonian ensembles, there is an unique induced eigenvector ensemble, {\em i.e.} for any unitary invariant Hamiltonian ensemble, the measure invariance $d\mu[UHU^{\dagger}]=d\mu[H]$ for any unitary $U$ implies

\begin{equation}
  \int_{\mathcal{E}_{H}}f(H)d\mu[H]=\int_{\mathcal{E}_{H}}\int_{U(d)}f(UHU^{\dagger})[U^{\dagger}dU]d\mu[H].
\end{equation}
This particular example leads us to introduce the first explicit eigenvector ensemble, the \emph{Haar-distributed eigenvector ensemble} defined as the eigenvector ensemble with the $C$-operator sampled from the $U(d)$ Haar measure\footnote{This definition applies straightforward to orthogonal and symplectic invariant Hamiltonian ensembles by replacing the unitary Haar measure with the proper orthogonal or symplectic one.}. For the previous example of a thermal correlation, the integration over all possible unitaries $\Braket{\cdots}_{U(d)}$ (see \eqref{eq:2_moment_Haar} in the next section) yields,

\begin{equation}
  \frac{\braket{\Tr{e^{-\beta E}CWC^{\dagger}}}_{\mathcal{E}_{C}}}{Z(\beta)}=\frac{\braket{\Tr{e^{-\beta E}UWU^{\dagger}}}_{U(d)}}{Z(\beta)}=\frac{\Tr{e^{-\beta E}\braket{UWU^{\dagger}}_{U(d)}}}{Z(\beta)}=\frac{1}{d}\Tr{W}.
\end{equation}
The unitary Haar distributed ensemble incorporates the Hamiltonian dependence at the simplest level, that is, by only taking into account the symmetry class. However, the fine details of the Hamiltonian are somehow missed; in particular, it fails to encode the fact that simple operators do not change the energy by a huge amount \cite{Stanford/subleading_weingartens/2022}. A suggestive way to capture these fine details is by replacing the Haar ensemble with one explicitly dependent on the Hamiltonian in the first place. However, this is a difficult task: how can we statistically model the complex eigenvector structure of an arbitrary Hamiltonian?. Here, we provide a construction to accomplish this task and introduce the $C$-ensemble as an eigenvector ensemble that explicitly introduces the Hamiltonian dependence.

\begin{definition}[unitary $C$-ensemble]\label{def:C_ensemble}
  The unitary $C$-ensemble is the unitary eigenvector ensemble associated with the fixed Hamiltonian $H$, defined by the probability measure

  \begin{equation}
  \label{eq:C_ensemble_Measure}
    d\mu_{\mathcal{E}_{C}}[C] = \frac{1}{\Vol(H)} \delta_{\perp}(CHC^{\dagger})[C^{\dagger}dC],
  \end{equation}
where $\delta_{\perp}(\cdots)$ is the off-diagonal delta distribution of a matrix argument \cite{Alonso/dirac_delta/2020,Zhang/dirac_delta/2020} and $\Vol(H)$ is the ensemble partition function (see Appendix \ref{appendix:partition_function}). 
\end{definition}
As far as we know, this measure first appeared in the work of Deutsch \cite{Deutsch/great_idea/1991,Deutsch/ETH_famous/1991} and later in \cite{Nation/Deutsh_ETH/2018}. However, there is a crucial difference between the $C$-ensemble and aforementioned works. That is, in previous works, the Hamiltonian inside the measure \eqref{eq:C_ensemble_Measure} is not the one of an individual quantum system but rather a member of an ensemble. Concretely, as an attempt to take into account the complexity of the Many-body interactions, the authors chose to model their systems as a fixed diagonal Hamiltonian plus a perturbation sampled from the Gaussian orthogonal ensemble (GOE), meaning that on the top of the eigenvector ensemble averages they include RMT averages as well\footnote{An additional technical differences is that in order to impose orthogonality for the \emph{change} of basis matrix, they use a delta parametrization of the Haar measure instead (see Appendix \ref{Appendix:RMT_Haar}).}. Here we choose to do the complete opposite, \emph{i.e.} instead of trying to model the system Hamiltonian $H$, we will take the system Hamiltonian as input data and build an exclusive ensemble for it. Of course, if the input Hamiltonian is itself a member of an ensemble; we will not have one, but rather an ensemble of $C$-ensembles and (for the particular case when such Hamiltonian's are draw in from the Diagonal+GOE case) we expect to recover the results from \cite{Deutsch/great_idea/1991,Deutsch/ETH_famous/1991}.

\begin{figure}[ht]
  \centering
  \includegraphics[width=0.3\textwidth]{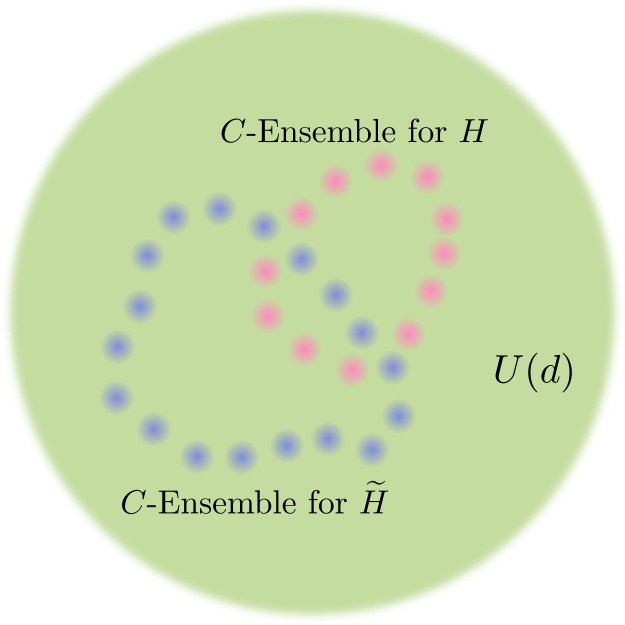}
  \caption{$C$-ensemble as a subset of the full unitary group. Given a unitary $U$ which diagonalizes the fixed Hamiltonian $H$, the $C$-ensemble can be identified as the orbit of $U$ generated by the $U(1)^{d} \times S_{d}$ group action. This is graphically represented by the sequence of blurred points, where disjoint points correspond to the discrete $S_{d}$ part of the orbit and the blur around each point denotes the family of ``close'' unitaries given by the $U(1)^{d}$ action.}
  \label{fig:C_ensemble_diagram}
\end{figure}

The measure \eqref{eq:C_ensemble_Measure} counts \emph{over all possible unitaries that ``diagonalize'' the fixed Hamiltonian $H$)} (see Fig. \ref{fig:C_ensemble_diagram}). Here the quotes over ``diagonalize'' have the following meaning: Inside \eqref{eq:C_ensemble_Measure} we are only fixing the off-diagonal part of $CHC^{\dagger}$ to be equal to zero whereas the diagonal part is left in principle arbitrary. We cannot fix the diagonal as well, {\em e.g.} by setting $\delta(CHC^{\dagger}-E)$ with $E$ diagonal, because the spectral theorem implies that there is a single solution up to an $U(1)^{d}$ global nonphysical phase and we would be left with no ensemble to average over at all!.

The obvious question is then, over what are we averaging in \eqref{eq:C_ensemble_Measure}? The upshot of this question is that the $C$-ensemble can be interpreted as an equal-probable average over the Hamiltonian eigenstates. To see this, let us review the symmetries of the measure \eqref{eq:C_ensemble_Measure}. First, $\delta_{\perp}(CHC^{\dagger})$ is invariant under global translations $H\to H + \mu \mathbb{I}$ with $\mu$ real, and transforms self-similarly under scalings, \emph{i.e.} $\delta_{\perp}(C \lambda HC^{\dagger}) \propto \delta_{\perp}(CHC^{\dagger})$ with $\lambda$ real, meaning that the ensemble is only concerned about energy differences and not about a global energy scale.\footnote{This must be a basic requirement for every eigenvector ensemble, otherwise we would end up with different ensembles by changing the energy units or shifting the ground state in the Hamiltonian. Notice that the Haar ensemble (trivially) accomplishes this requirement as well. } Second and even more important, the $C$-ensemble is left $U(1)^{d}\times S_{d}$ invariant, {\em e.g.}

\begin{equation}
  \delta_{\perp}(CHC^{\dagger})=\delta_{\perp}(e^{i\vec{\phi}}P_{\pi}CHC^{\dagger}P_{\pi^{-1}}e^{-i\vec{\phi}}),
\end{equation}
with $e^{i\vec{\phi}}$ and $P_{\pi}$ being $d\times d$ arbitrary diagonal and permutation matrices respectively\footnote{$
  \delta_{\perp}(P_{\pi}CHC^{\dagger}P_{\pi^{-1}})=\frac{\delta(P_{\pi}CHC^{\dagger}P_{\pi^{-1}})}{\delta_{\parallel}(P_{\pi}CHC^{\dagger}P_{\pi^{-1}})}=\frac{\delta(CHC^{\dagger})}{\delta_{\parallel}(P_{\pi}CHC^{\dagger}P_{\pi^{-1}})}.
$}. While the $U(1)^{d}$ part simply represents the global phase invariance of each one of the $d$ eigenvectors and is a symmetry that any physical eigenvector ensemble must have, it is the subtle $S_{d}$ symmetry which implies that there is no privileged ``eigenvector'' among the $d$ ones. As an example, consider the $d=2$ case, therefore the corresponding left $S_{2}$ symmetry implies that

\begin{equation}
   C=\begin{pmatrix}
    \braket{E_{1}|1} & \braket{E_{1}|2} \\
    \braket{E_{2}|1} & \braket{E_{2}|2} 
  \end{pmatrix} 
    \quad \text{and} \quad
  P_{(1\,2)}C=\begin{pmatrix}
    \braket{E_{2}|1} & \braket{E_{2}|2} \\
    \braket{E_{1}|1} & \braket{E_{1}|2} 
  \end{pmatrix},
\end{equation}
which shows that both $C$-matrices are equally probable in the context of the ensemble. It is worth clarifying that the left $S_{d}$ symmetry by no means implies an indistinguishable set of eigenvectors\footnote{An indistinguishable set of eigenvectors requires an invariant eigenvector ensemble under both, the left and right $S_{d}$ action. {\em e.g.} $C\rightarrow P_{\pi}CP_{\pi^{-1}}$ for all $\pi$ in $S_{d}$ just as the Haar distributed case.}, which cannot happen if we are fixing the Hamiltonian in the first place. What the symmetry does imply is that, although for different ``eigenvectors'' the specific overlaps over the computational basis depend upon the chosen Hamiltonian, labeling an eigenvector requires associating it with a corresponding eigenvalue. Since we are not fixing the diagonal part in $CHC^{\dagger}$ (the eigenvalues), therefore, there is no particular way to order this ``eigenvectors''. This is the reason why we mentioned that there was a non-privileged eigenvector in the first place.

A complementary perspective for this trade-off between the left $S_{d}$ symmetry and the ``diagonal freedom'' arises by interpreting the $C$-ensemble measure as the effective one after integrating out all diagonal degrees of freedom, {\em i.e.}

\begin{equation}
  \label{eq:C_coarse_graning}
  \delta_{\perp}(CHC^{\dagger})=\int_{\text{Diagonals}} \delta(CHC^{\dagger}-E)\, [dE],
\end{equation}
where again, the spectral theorem dictates that there is only one $C$ (up to an $U(1)^{d}$ global phase) in the expression under the integral on the right-hand side of \eqref{eq:C_coarse_graning} which fulfills the constraint for a fixed diagonal matrix of eigenvalues $E$. Integrating out over all possible diagonals is the same as adding up the $d!$ eigenvalue permutations (counting multiplicities) in the diagonal. Therefore the $S_{d}$ symmetry on the left-hand side of (\ref{eq:C_coarse_graning}) emerges as the average over all $d!$ matrices that diagonalize the fixed Hamiltonian, $CHC^{\dagger}=E$ for each $d!$ eigenvalue permutations. To summarize, we cannot fix the diagonal constraint in \eqref{eq:C_ensemble_Measure} because by the spectral theorem there will not be any nontrivial average. Therefore, one possibility is to drop this diagonal restriction, but by doing so, we automatically gain the $S_{d}$ symmetry.

Finally, it is worth noting that for systems with additional unitary symmetries, the $C$-ensemble can be used to induce a micro-canonical average within each symmetry sector. As an example, consider the simplest case of a single symmetry generator $J$, such that $[J,H]=0$, with eigenvalues ${j_{1},\dots,j_{d_{J}}}$. The Hilbert space decomposes into a direct sum of $d_{J}<d$ symmetry sectors,

\begin{equation}
\mathcal{H}=\bigoplus_{l=1}^{d_{J}}\mathcal{H}_{l},
\end{equation}
where each sector has dimension $d_{l}=\text{dim}(\mathcal{H}_{l})$. For each sector, one can construct a local $C$-ensemble $\mathcal{E}_{C^{l}}$ with an associated Hamiltonian $H_{l}$, by projecting the initial Hamiltonian onto the corresponding symmetry sector, \emph{i.e.} $H_{l}=\Pi_{l}H\Pi_{l}$ with $\Pi_{l}$ the projector on the $l$-th sector. Each local $C$-ensemble averages over all the $d_{l}!$ unitaries which diagonalize the projected Hamiltonian $H_{l}$ and preserve the $j_{l}$-sector. This idea will be further discussed in later chapters when four-point correlation functions are analyzed.

\subsection{Two and four moments}

Before we present the main results in this section, let us define some of the technology used in this work. First, we introduce the $2k$th-moment operator of an unitary matrix ensemble $\mathcal{E}$ with normalized measure $d\mu_{\mathcal{E}}[U]$ as the map $\hat{\Phi}_{2k}^{\mathcal{E}}:\mathcal{H}^{\otimes 2k}\rightarrow \mathcal{H}^{\otimes 2k}$,

\begin{equation}
  \label{eq:2k_moment}
  \hat{\Phi}_{2k}^{\mathcal{E}}=\braket{U^{\otimes k}\otimes U^{\dagger \otimes k}}_{\mathcal{E}}=\int_{\mathcal{E}}U^{\otimes k}\otimes U^{\dagger \otimes k}d\mu_{\mathcal{E}}[U].
\end{equation}
The relevance of (\ref{eq:2k_moment}) is that all correlation functions involving $2k$th ensemble moments can be obtained as specific matrix elements from it. As a particular example, by setting the ensemble $\mathcal{E}$ to be the Haar unitary one, the two moments are given by, 

\begin{equation*}
\label{eq:weingarten_formula_for_2}
  \bra{l \, m}\hat{\Phi}_{2}^{U(d)}\ket{k \, s}=\int_{U(d)}\braket{l|U|k}\braket{m|U^{\dagger}|s} [U^{\dagger}dU]=\int_{U(d)}U_{lk}U^{*}_{sm}[U^{\dagger}dU]=\frac{1}{d}\delta_{ls}\delta_{mk}.
\end{equation*}
Where the last unitary integral is given by the Weingarten functions (See Appendix \ref{appendix:unitary}). Diagrammatically the $2k$th-moment operators are suitably represented as follows,

\begin{equation}
  \begin{tikzpicture}
	\begin{pgfonlayer}{nodelayer}
		\node [style={big_box}] (0) at (-1.75, 0.25) {$\displaystyle{\hat{\Phi}^{\mathcal{E}}_{2k}}$};
		\node [style=none] (1) at (-0.75, 1.75) {};
		\node [style=none] (2) at (-2.75, 1.75) {};
		\node [style=none] (3) at (-0.75, 0.5) {};
		\node [style=none] (4) at (-2.75, 0.5) {};
		\node [style=none] (5) at (-2.75, -0.25) {};
		\node [style=none] (6) at (-0.75, -0.25) {};
		\node [style=none] (7) at (-2.75, -1.5) {};
		\node [style=none] (8) at (-0.75, -1.5) {};
		\node [style=none] (9) at (0.75, 1.75) {};
		\node [style=none] (10) at (0.75, 0.5) {};
		\node [style=none] (11) at (0.75, -0.25) {};
		\node [style=none] (12) at (0.75, -1.5) {};
		\node [style=none] (13) at (-4.25, -1.5) {};
		\node [style=none] (14) at (-4.25, -0.25) {};
		\node [style=none] (15) at (-4.25, 0.5) {};
		\node [style=none] (16) at (-4.25, 1.75) {};
		\node [style=none] (17) at (0.25, 1.5) {.};
		\node [style=none] (18) at (0.25, 1.25) {.};
		\node [style=none] (19) at (0.25, 1) {.};
		\node [style=none] (20) at (-3.75, 1.5) {.};
		\node [style=none] (21) at (-3.75, 1.25) {.};
		\node [style=none] (22) at (-3.75, 1) {.};
		\node [style=none] (24) at (6.25, 1.75) {};
		\node [style=none] (25) at (5.25, 1.75) {};
		\node [style=none] (26) at (6.25, 0.5) {};
		\node [style=none] (27) at (5.25, 0.5) {};
		\node [style=none] (28) at (5.25, -0.25) {};
		\node [style=none] (29) at (6.25, -0.25) {};
		\node [style=none] (30) at (5.25, -1.5) {};
		\node [style=none] (31) at (6.25, -1.5) {};
		\node [style=none] (32) at (7.5, 1.75) {};
		\node [style=none] (33) at (7.5, 0.5) {};
		\node [style=none] (34) at (7.5, -0.25) {};
		\node [style=none] (35) at (7.5, -1.5) {};
		\node [style=none] (36) at (4, -1.5) {};
		\node [style=none] (37) at (4, -0.25) {};
		\node [style=none] (38) at (4, 0.5) {};
		\node [style=none] (39) at (4, 1.75) {};
		\node [style=none] (43) at (4.25, -0.75) {.};
		\node [style=none] (44) at (4.25, -1) {.};
		\node [style=none] (45) at (4.25, -1.25) {.};
		\node [style=none] (46) at (5.75, 1.75) {$\displaystyle{U}$};
		\node [style=none] (48) at (5.75, 0.5) {$\displaystyle{U}$};
		\node [style=none] (49) at (5.75, -0.25) {$\displaystyle{U^{\dagger}}$};
		\node [style=none] (50) at (5.75, -1.5) {$\displaystyle{U^{\dagger}}$};
		\node [style=none] (51) at (3, 0.25) {$\displaystyle{=\int_{\mathcal{E}}}$};
		\node [style=none] (52) at (9, 0.25) {$d\mu_{\mathcal{E}}[U]$};
		\node [style=none] (53) at (-3.75, -0.75) {.};
		\node [style=none] (54) at (-3.75, -1) {.};
		\node [style=none] (55) at (-3.75, -1.25) {.};
		\node [style=none] (56) at (-5.5, 1.25) {$k$};
		\node [style=none] (57) at (0.25, -0.75) {.};
		\node [style=none] (58) at (0.25, -1) {.};
		\node [style=none] (59) at (0.25, -1.25) {.};
		\node [style=none] (60) at (7, -0.75) {.};
		\node [style=none] (61) at (7, -1) {.};
		\node [style=none] (62) at (7, -1.25) {.};
		\node [style=none] (63) at (4.25, 1.5) {.};
		\node [style=none] (64) at (4.25, 1.25) {.};
		\node [style=none] (65) at (4.25, 1) {.};
		\node [style=none] (66) at (7, 1.5) {.};
		\node [style=none] (67) at (7, 1.25) {.};
		\node [style=none] (68) at (7, 1) {.};
		\node [style=none] (70) at (-4.75, 1.5) {};
		\node [style=none] (71) at (-5, 1.25) {};
		\node [style=none] (72) at (-4.75, 1) {};
		\node [style=none] (73) at (-4.5, 2) {};
		\node [style=none] (74) at (-4.5, 0.5) {};
		\node [style=none] (75) at (1.225, -1.225) {};
		\node [style=none] (76) at (1.475, -0.975) {};
		\node [style=none] (77) at (1.225, -0.725) {};
		\node [style=none] (78) at (0.975, -1.725) {};
		\node [style=none] (79) at (0.975, -0.225) {};
		\node [style=none] (80) at (2, -1) {$k$};
	\end{pgfonlayer}
	\begin{pgfonlayer}{edgelayer}
		\draw (16.center) to (2.center);
		\draw (15.center) to (4.center);
		\draw (14.center) to (5.center);
		\draw (13.center) to (7.center);
		\draw (1.center) to (9.center);
		\draw (3.center) to (10.center);
		\draw (6.center) to (11.center);
		\draw (8.center) to (12.center);
		\draw (39.center) to (25.center);
		\draw (38.center) to (27.center);
		\draw (37.center) to (28.center);
		\draw (36.center) to (30.center);
		\draw (24.center) to (32.center);
		\draw (26.center) to (33.center);
		\draw (29.center) to (34.center);
		\draw (31.center) to (35.center);
		\draw [bend left] (70.center) to (71.center);
		\draw [bend left, looseness=1.25] (71.center) to (72.center);
		\draw [in=90, out=-180, looseness=1.25] (73.center) to (70.center);
		\draw [in=-180, out=-90, looseness=1.50] (72.center) to (74.center);
		\draw [bend left] (75.center) to (76.center);
		\draw [bend left, looseness=1.25] (76.center) to (77.center);
		\draw [in=-90, out=0, looseness=1.25] (78.center) to (75.center);
		\draw [in=0, out=90, looseness=1.50] (77.center) to (79.center);
	\end{pgfonlayer}
\end{tikzpicture}.
\end{equation}
Of particular importance are permutation operators $W_{\pi}$ over the $k$th fold Hilbert space $\mathcal{H}^{\otimes k}$ defined by their action over the local basis as,

\begin{equation}
    W_{\pi}\ket{i_{1} \, i_{2} \dots \, i_{k}}=\ket{\pi(i_{1})\, \pi(i_{2})\, \dots \pi(i_{k})}\quad \text{for every} \quad \pi \in S_{k}.
\end{equation}
They are naturally represented by decomposing the permutation $\pi$ into transpositions and assigning to each transposition the SWAP:$\,\mathcal{H}^{\otimes 2} \to \mathcal{H}^{\otimes 2}$ gate defined by $\text{SWAP}=W_{(12)}$\footnote{In the following we will use both, the cyclic and the transposition notation for permutations.} and graphically represented as

\begin{equation}
  \begin{tikzpicture}
	\begin{pgfonlayer}{nodelayer}
		\node [style=none] (9) at (4.25, 0) {$\displaystyle{\text{SWAP}=}$};
		\node [style=none] (10) at (6.25, -0.5) {};
		\node [style=none] (11) at (6.25, 0.5) {};
		\node [style=none] (12) at (7.75, 0.5) {};
		\node [style=none] (13) at (7.75, -0.5) {};
	\end{pgfonlayer}
	\begin{pgfonlayer}{edgelayer}
		\draw [in=-180, out=0] (11.center) to (13.center);
		\draw [style={white_edge}] (10.center) to (12.center);
		\draw [in=180, out=0] (10.center) to (12.center);
	\end{pgfonlayer}
\end{tikzpicture}\,.
\end{equation}
From the Weingarten formula \eqref{app:weingarten_formula} the 2nd-moment operator for the Haar unitary ensemble turns out to be proportional to a $\text{SWAP}$,

\begin{equation}
\label{eq:2_moment_Haar}
  \begin{tikzpicture}
	\begin{pgfonlayer}{nodelayer}
		\node [style={medium_box}] (0) at (-1.75, 0) {$\hat{\Phi}^{U(d)}_{2}$};
		\node [style=none] (1) at (-1, -0.5) {};
		\node [style=none] (2) at (-2.5, 0.5) {};
		\node [style=none] (3) at (-2.5, -0.5) {};
		\node [style=none] (4) at (-1, 0.5) {};
		\node [style=none] (5) at (-3.5, 0.5) {};
		\node [style=none] (6) at (-3.5, -0.5) {};
		\node [style=none] (7) at (0, 0.5) {};
		\node [style=none] (8) at (0, -0.5) {};
		\node [style=none] (9) at (1, 0) {$\displaystyle{=\frac{1}{d}}$};
		\node [style=none] (10) at (2, -0.5) {};
		\node [style=none] (11) at (2, 0.5) {};
		\node [style=none] (12) at (3.5, 0.5) {};
		\node [style=none] (13) at (3.5, -0.5) {};
	\end{pgfonlayer}
	\begin{pgfonlayer}{edgelayer}
		\draw (5.center) to (2.center);
		\draw (6.center) to (3.center);
		\draw (4.center) to (7.center);
		\draw (1.center) to (8.center);
		\draw [in=-180, out=0] (11.center) to (13.center);
		\draw [style={white_edge}] (10.center) to (12.center);
		\draw [in=180, out=0] (10.center) to (12.center);
	\end{pgfonlayer}
\end{tikzpicture}\,,
\end{equation}
while for the Haar fourth- moment operator $\hat{\Phi}^{U(d)}_{4}$ see \eqref{appendix:unitary}. 

Now, let us introduce the concept of a unitary $k$-design \cite{Chirtoph/two_design_thesis/2005,Chirtoph/2_design_origin/2009,Gross/2_designs_theorems/2008} as a unitary ensemble $\mathcal{E}$ for which the first $2k$th moments coincide with the ones of the Haar ensemble\footnote{The motivation behind designs is that they serve as statistical simulators of uniform distributed unitaries.}, {\em e.g.} $\hat{\Phi}_{2l}^{\mathcal{E}}=\hat{\Phi}_{2l}^{U(d)}$ for all $l\leq k$. Clearly, from Haar's theorem \cite{Pedersen/Simple_Haar_uniqueness}, the only ``$\infty$''-design corresponds to the Haar ensemble itself. Obtaining exact $k$th-designs is quite hard for $k>2$, some known examples of designs are the Clifford group as a 3-design \cite{Webb/clifford_3_design}, the Pauli group as a 1-design \cite{Roy/Pauli_design_bound/2009}, and as we will see later the $C$-ensemble as a 1-design. 

Intimately related to the $2k$th moment operator $\hat{\Phi}^{\mathcal{E}}_{2k}$, are the left and right $k$th fold channels $\Phi^{\mathcal{E}}_{k,L(R)}: \text{End}(\mathcal{H}^{\otimes k})\rightarrow \text{End}(\mathcal{H}^{\otimes k})$ respectively defined as,

\begin{equation}
  \Phi^{\mathcal{E}}_{k,L}(\cdots)=\int_{\mathcal{E}}U^{\otimes k}(\cdots)U^{\dagger \otimes k}d\mu_{\mathcal{E}}[U], \quad \Phi^{\mathcal{E}}_{k,R}(\cdots)=\int_{\mathcal{E}}U^{\dagger \otimes k}(\cdots)U^{\otimes k}d\mu_{\mathcal{E}}[U].
\end{equation}
The physical relationship between $k$th-fold channels and $k$-point Haar averaged correlation functions was established in \cite{Yoshida/Chaos_design/2017}. For an unitary ensemble with a measure invariant under conjugation, {\em i.e.} $d\mu_{\mathcal{E}}[U]=d\mu_{\mathcal{E}}[U^{\dagger}]$, the left and right $k$th-fold channels are indistinguishable. This is the case for Haar distributed unitaries. However, the measure \eqref{eq:C_ensemble_Measure}, due to the fixed Hamiltonian $H$ acting as an external source, is not invariant under conjugation. This symmetry breaking of the measure under conjugation is a manifestation of the distinction between the local and energy eigenbasis naturally imposed by the particular Hamiltonian. Notably, and as will be discussed in later sections, the right two-fold channel $\Phi^{\mathcal{E}_{C}}_{2,R}$ plays a crucial role in distinguishing the $C$-ensemble from the Haar one. An equivalent definition of $k$-design is given by demanding that $\Phi^{\mathcal{E}}_{k,L(R)}(A)=\Phi^{U(d)}_{k}(A)$ for every $A\in \text{End}(\mathcal{H}^{\otimes k})$  (see \cite{Gross/2_designs_theorems/2008,Yoshida/Chaos_design/2017}).

For the $C$-ensemble $k$th-moment operator the direct consequence of the $U(1)^{d}\times S_{d}$ symmetry is the following factorization,

\begin{equation}
  \label{eq:2kmoment_factorization}
  \begin{tikzpicture}
	\begin{pgfonlayer}{nodelayer}
		\node [style={big_box}] (0) at (-1.75, 0.25) {$\displaystyle{\hat{\Phi}^{\mathcal{E}_{C}}_{2k}}$};
		\node [style=none] (1) at (-0.75, 1.75) {};
		\node [style=none] (2) at (-2.75, 1.75) {};
		\node [style=none] (3) at (-0.75, 1) {};
		\node [style=none] (4) at (-2.75, 1) {};
		\node [style=none] (5) at (-2.75, -0.5) {};
		\node [style=none] (6) at (-0.75, -0.5) {};
		\node [style=none] (7) at (-2.75, -1.25) {};
		\node [style=none] (8) at (-0.75, -1.25) {};
		\node [style=none] (9) at (0.75, 1.75) {};
		\node [style=none] (10) at (0.75, 1) {};
		\node [style=none] (11) at (0.75, -0.5) {};
		\node [style=none] (12) at (0.75, -1.25) {};
		\node [style=none] (13) at (-4.25, -1.25) {};
		\node [style=none] (14) at (-4.25, -0.5) {};
		\node [style=none] (15) at (-4.25, 1) {};
		\node [style=none] (16) at (-4.25, 1.75) {};
		\node [style=none] (17) at (0.25, 0.5) {.};
		\node [style=none] (18) at (0.25, 0.25) {.};
		\node [style=none] (19) at (0.25, 0) {.};
		\node [style=none] (20) at (-3.75, 0.5) {.};
		\node [style=none] (21) at (-3.75, 0.25) {.};
		\node [style=none] (22) at (-3.75, 0) {.};
		\node [style=none] (51) at (2.25, 0.25) {$\displaystyle{=\int_{U(d)}}$};
		\node [style=none] (52) at (13, 0.25) {$d\mu_{\mathcal{E}_{C}}[U]$};
		\node [style={big_box}] (53) at (7.75, 0.25) {$\displaystyle{\hat{\Phi}^{U(1)^{d}\times S_{d}}_{2k}}$};
		\node [style=none] (54) at (8.75, 1.75) {};
		\node [style=none] (55) at (6.75, 1.75) {};
		\node [style=none] (56) at (8.75, 1) {};
		\node [style=none] (57) at (6.75, 1) {};
		\node [style=none] (58) at (6.75, -0.5) {};
		\node [style=none] (59) at (8.75, -0.5) {};
		\node [style=none] (60) at (6.75, -1.25) {};
		\node [style=none] (61) at (8.75, -1.25) {};
		\node [style=none] (62) at (10.25, 1.75) {};
		\node [style=none] (63) at (10.25, 1) {};
		\node [style=none] (64) at (11.75, -0.5) {};
		\node [style=none] (65) at (11.75, -1.25) {};
		\node [style=none] (66) at (5.25, -1.25) {};
		\node [style=none] (67) at (5.25, -0.5) {};
		\node [style=none] (68) at (3.75, 1) {};
		\node [style=none] (69) at (3.75, 1.75) {};
		\node [style=none] (70) at (10.5, 0.5) {.};
		\node [style=none] (71) at (10.5, 0.25) {.};
		\node [style=none] (72) at (10.5, 0) {.};
		\node [style=none] (73) at (5.25, 0.5) {.};
		\node [style=none] (74) at (5.25, 0.25) {.};
		\node [style=none] (75) at (5.25, 0) {.};
		\node [style=none] (76) at (10.75, 1.75) {$\displaystyle{U}$};
		\node [style=none] (77) at (11.25, 1.75) {};
		\node [style=none] (78) at (11.75, 1.75) {};
		\node [style=none] (79) at (10.75, 1) {$\displaystyle{U}$};
		\node [style=none] (80) at (11.25, 1) {};
		\node [style=none] (81) at (11.75, 1) {};
		\node [style=none] (82) at (4.75, -0.5) {$\displaystyle{U^{\dagger}}$};
		\node [style=none] (83) at (4.75, -1.25) {$\displaystyle{U^{\dagger}}$};
		\node [style=none] (84) at (3.75, -0.5) {};
		\node [style=none] (85) at (4.25, -0.5) {};
		\node [style=none] (86) at (3.75, -1.25) {};
		\node [style=none] (87) at (4.25, -1.25) {};
	\end{pgfonlayer}
	\begin{pgfonlayer}{edgelayer}
		\draw (16.center) to (2.center);
		\draw (15.center) to (4.center);
		\draw (14.center) to (5.center);
		\draw (13.center) to (7.center);
		\draw (1.center) to (9.center);
		\draw (3.center) to (10.center);
		\draw (6.center) to (11.center);
		\draw (8.center) to (12.center);
		\draw (69.center) to (55.center);
		\draw (68.center) to (57.center);
		\draw (67.center) to (58.center);
		\draw (66.center) to (60.center);
		\draw (54.center) to (62.center);
		\draw (56.center) to (63.center);
		\draw (59.center) to (64.center);
		\draw (61.center) to (65.center);
		\draw (77.center) to (78.center);
		\draw (80.center) to (81.center);
		\draw (84.center) to (85.center);
		\draw (86.center) to (87.center);
	\end{pgfonlayer}
\end{tikzpicture}
\end{equation}
with $\Phi^{U(1)^{d}\times S_{d}}$ the $U(1)^{d}\times S_{d}$ ensemble $2k$th moment operator (see Appendix \ref{Appendix:U1dSd}). By itself, as shown in Appendix \ref{Appendix:U1dSd}, the $U(1)^{d}\times S_{d}$ group is an unitary 1-design, therefore the integral (\ref{eq:2kmoment_factorization}) implies for $k=1$, that the $C$-ensemble also yields an unitary 1-design,

\begin{equation}
  \hat{\Phi}_{2}^{\mathcal{E}_{C}}=\hat{\Phi}_{2}^{U(d)}.
\end{equation}
The consequences of this result will be discussed in the next section. For higher moments the $U(1)^{d}\times S_{d}$ group does not yield into higher designs and for $k=2$, \eqref{eq:2kmoment_factorization} is constrained to be

\begin{equation}
  \label{eq:C_four_moment_operator}
  \input{C_fourmoment.tikz}
\end{equation}
where the 3-edge tensor corresponds to the $\text{COPY}:\mathcal{H}\rightarrow \mathcal{H}^{\otimes 2}$ map defined over the local (computational) basis as  $\text{COPY}=\sum_{l}\ket{l\, l}\bra{l}$. However, the object which plays a special role inside $\hat{\Phi}^{\mathcal{E}_{C}}$ is the s-tensor\footnote{The name $s$-tensor is straightforward because looks like the $s$-channel diagram in QFT. See Appendix \ref{Appendix:Scrambling}.} defined by joining two COPY tensors,

\begin{equation}
\label{eq:s_tensor_introduction}
  \begin{tikzpicture}
	\begin{pgfonlayer}{nodelayer}
		\node [style=none] (0) at (-6.5, 0) {$\displaystyle{s\text{-tensor}=\text{COPY}\,\text{COPY}^{\dagger}=}$};
		\node [style=none] (1) at (3, -1) {};
		\node [style=none] (2) at (3, 1) {};
		\node [style=none] (3) at (1, 0) {};
		\node [style=none] (4) at (2.25, 0) {};
		\node [style=none] (5) at (-1, 1) {};
		\node [style=none] (6) at (-1, -1) {};
		\node [style=none] (7) at (1, 0) {};
		\node [style=none] (8) at (-0.25, 0) {};
	\end{pgfonlayer}
	\begin{pgfonlayer}{edgelayer}
		\draw (1.center) to (4.center);
		\draw (4.center) to (2.center);
		\draw (4.center) to (3.center);
		\draw (5.center) to (8.center);
		\draw (8.center) to (6.center);
		\draw (8.center) to (7.center);
	\end{pgfonlayer}
\end{tikzpicture}
\end{equation}
and the action of the $C$-ensemble right 2-fold channel on the $s$-tensor in turn defines the plateau operator $G^{\mathcal{E}}$,
\begin{equation}
\label{eq:dephasing_operator_channel}
  G^{\mathcal{E}_{C}}=\Phi^{\mathcal{E}_{C}}_{2,R}(s\text{-tensor})=\int_{U(d)} C^{\dagger \otimes 2} (s\text{-tensor})C^{ \otimes 2} d\mu_{\mathcal{E}_{C}}[C]\,.
\end{equation}
The name will become clear when we show later that $G^{\mathcal{E}}$ corresponds to the long-time average of infinite-temperature two-point self correlation functions.

As a final comment, let us highlight the following points about this result. First, the tensor decomposition \eqref{eq:C_four_moment_operator} in terms of the plateau operator is not exclusive to the $C$-ensemble. In general, any unitary ensemble with a $U(1)^{d}\times S_{d}$ symmetry takes this form after integrating out the symmetric degrees of freedom \footnote{Notice that while $\Phi^{\mathcal{E}_{C}}_{2,R}(\text{s-tensor})$ is invariant under the left $U(1)^{d}\times S_{d}$ action, $\Phi^{\mathcal{E}_{C}}_{2,L}(\text{s-tensor})$ is not. This is the reason to distinguish between both left and right channels in first place.}, {\em e.g.} the Haar ensemble is obviously  $U(1)^{d}\times S_{d}$ invariant. Therefore, by replacing the Haar plateau operator by
\begin{equation}
  G^{U(d)}=\frac{1}{d+1}(\mathbb{I}^{\otimes 2}+\text{SWAP}),
\end{equation}
in \eqref{eq:C_four_moment_operator} we recover the well-known result for the Haar four moment operator $\hat{\Phi}^{U(d)}_{4}$ (see Appendix \ref{Appendix:RMT_Haar}). This observation in particular leads us to the second remark about the $C$-ensemble calculation, that is, \emph{for the C-ensemble all the information about the particular Hamiltonian $H$ is stored (at least for four-moments) in the plateau operator $G^{\mathcal{E}}$}. Then, we have a natural way to compare how much the specific $C$-ensemble generated from a particular Hamiltonian will differ from uniformly distributed unitaries. 

\subsection{C-ensemble and spectral decoupling}

The fact that unitary Haar averages of arbitrary  infinite temperature  correlation functions between local operators  factorize as,

\begin{equation}
  \Braket{\frac{1}{d}\Tr{W(t)V...}}_{U(d)}\sim \sum (\text{spectral form factor}(t)'s)\braket{WV...}\,,
\end{equation}
with $\braket{\cdots}=\frac{1}{d}\Tr{(\cdots)}$ was recently pointed out by Cotler and Jones in \cite{Cotler/spectral_decoupling/2020} under the name of ``spectral decoupling'' {\em i.e.} the time dependence in the correlation function decouples from the local operator. For two-points the explicit evaluation yields \cite{Reimann/Thermalization_Haar/2016,Cotler/spectral_decoupling/2020},

\begin{equation}
  \Braket{\frac{1}{d}\Tr{W(t)W(0)}}_{U(d)}=\braket{W}^{2}+\frac{|Z(it)|^{2}-1}{d^{2}-1}\left(\braket{W^{2}}-\braket{W}^{2} \right)\,
\end{equation}
where $Z(it)$ denotes the imaginary time partition function, and  $|Z(it)|^{2}$ the infinite temperature form factor\footnote{$| \Tr{e^{it H}}|^{2}=\sum_{l\, m}e^{it(E_{l}-E_{m})}$}. Motivated by this observation, in this section we further point out that spectral decoupling goes beyond the mathematical curiosity, rather , it corresponds to the expected late-time factorization of correlation functions in chaotic many-body systems. 

Concretely, in this section, we show that spectral decoupling arises as a consequence of the tensor structure of the $2k$th moment operators from the specific eigenvector-ensemble used, and show that in particular the \emph{C-ensemble provides apart from the Haar unitary ensemble another example of an unitary ensemble which accomplishes spectral decoupling}. 

\begin{figure}[ht]
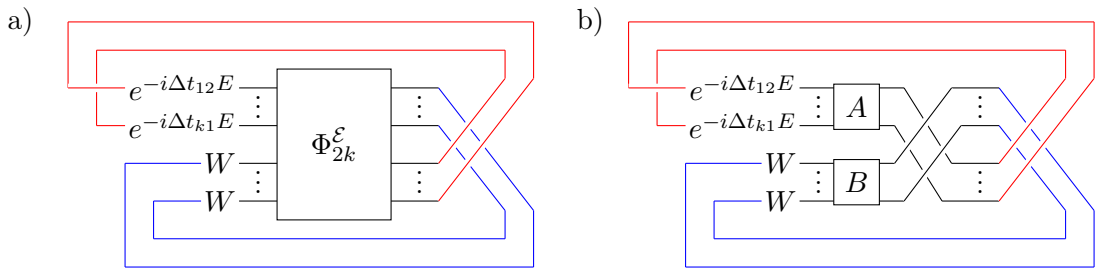

  \ctikzfig{spectral_decoupling_fig}
   \caption{Panel (a): Ensemble averaged $k$-point correlation $\braket{\Tr W(t_{1})W(t_{2})\dots W(t_{k})}_{\mathcal{E}}$.  Here $\Delta t_{i j} =t_{i}-t_{j}$ and we used the color code to distinguish contractions involving the $k$-first right Hilbert space replicas with the $k$-last left ones (blue) and the $k$-first left with the $k$-last right ones (red). Panel (b): For $2k$ moment operators which factorize as the product of two separated $k$-order tensors times a permutation between the first and second $k$-fold replicas, the time dependence in the correlation ``decouples'' from the operator $W$. }
   \label{fig:contraction_diagram}
  \end{figure}

Consider for the moment an arbitrary unitary ensemble $\mathcal{E}$. Infinite\footnote{This also holds for finite temperature by properly imaginary rotating the arguments.} temperature $k$-point out equilibrium averaged correlation functions require ensemble $2k$-moments (see Fig. \ref{fig:contraction_diagram} left panel). If the ensemble $2k$-moment operator $\hat{\Phi}^{\mathcal{E}}_{2k}$ is factorized as the product of two uncoupled $k$ order tensors then the time dependence ``decouples'' in the correlation (Fig. \ref{fig:contraction_diagram} right panel). This motivates us to define the notion of an uncoupled ensemble:

\begin{definition}[$k$-Spectral decoupled ensemble]
  \label{def:spectral_decoupling} 
  An unitary ensemble $\mathcal{E}$ with normalized probability measure $d\mu_{\mathcal{E}}[U]$ is said to be spectrally decoupled up to $k$, if for all $l \leq k$ the ensemble $2l$-moment operators factorize as

  \begin{equation}
    \hat{\Phi}_{2l}^{\mathcal{E}}=\sum_{\alpha=1}^{(l!)^2}(A_{\alpha} \otimes B_{\alpha}) W_{\pi_{\alpha}},
  \end{equation}
with $\{A_{\alpha}\}$ and $\{B_{\alpha}\}$ operators in $\mathcal{H}^{\otimes l}$, and $W_{\pi_{\alpha}}$ one of the $(l!)^2$ permutations which exchanges the first and second $l$-fold replicas. Additionally, $A_{\alpha}$ and $B_{\alpha}$ further respect the consistency constraints imposed by the unitarity of the moment operators, \emph{e.g.}, by pairing  one of the first $k$th-upper replicas, with one among the second $k$th-lower ones $(\hat{\Phi}_{2k}^{\mathcal{E}}) \to d\,\hat{\Phi}_{2(k-1)}^{\mathcal{E}}$ .
\end{definition}

\begin{figure}
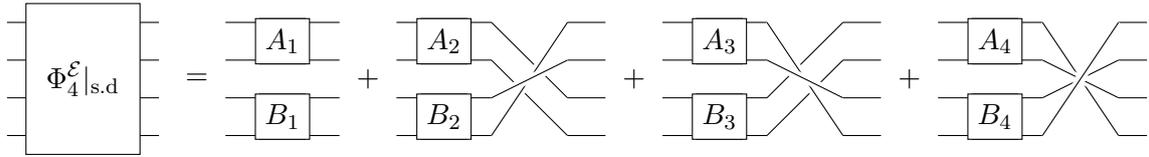

  \ctikzfig{2_spectral_decoupling_example}
   \caption{General form of the four-moment operator in a spectrally decoupled (s.p) unitary ensemble. For the case of the $C$-ensemble, the $A^{\text{'s}}$ and $B^{\text{'s}}$ are given in terms of the Plateau-operator $G^{\mathcal{E}_{C}}$, and the $s$-tensor.}
   \label{fig:two_spectral_decoupling_example}
  \end{figure}

From this definition, we arrive at the main result of this section. That is, the $C$-ensemble is a 2-spectral decoupled ensemble (see Fig. \ref{fig:two_spectral_decoupling_example}). In particular this means that, similarly to Haar distributed unitaries, the average of out-equilibrium (finite or infinite temperature) two-point correlation functions factorizes in such a way that the time dependence is decoupled from the operator dependence\footnote{We will show later in Section \ref{sec:OTOCS} that for four-point correlation functions, the $C$-ensemble also yields spectral decoupling.}.

Although we have shown two examples of spectral decoupled ensembles (\emph{i.e.} Haar unitaries, and the $C$-ensemble) we want to highlight that this is not a general feature of arbitrary unitary ensembles, in particular, for diagonally uniform distributed unitaries, {\em i.e.} $\mathcal{E}=U(1)^{d}$,

\begin{equation}
  \Phi^{U(1)^{d}}_{2}=\text{s-tensor},
\end{equation}
which is not of the form \eqref{def:spectral_decoupling} meaning that spectral decoupling is not a trivial property, \emph{i.e.} is not a direct consequence of unitarity. However, as a general result, notice that the only possible $k=1$ spectral decoupled ensemble must be equal to the Haar one. That is, at the level of two moments, spectral decoupling implies that the ensemble is an unitary $1$-design. For higher moments this equivalence no longer holds, and the Haar-distributed unitaries become an special case of spectral decoupling ensembles\footnote{Clearly the Haar ensemble is an spectral decoupled for every $k$, however it doesn't seem so straightforward the existence of additional unitary ensembles with the same property.}.

\subsection{Second Frame potential}\label{subsec:frame_potential}

 A natural way to compare the operational results between two different unitary ensembles $\mathcal{E}$ and $\mathcal{E}^{\prime}$ is by using the Hilbert-Schmidt operator norm\footnote{We can have used any other operator norm as well. However, as for finite dimension, operator norms are bounded by constant pre-factors between them, the final conclusion will be the same. See \cite{Hunter/Unitary_designs_statistical/2019}.} between the $k$th-moment operators,
 
\begin{equation}
  \label{eq:diff_operator}
  \Tr{\Delta S_{k}^{\dagger} \Delta S_{k} }\geq 0 ,\quad \text{with}\quad \Delta{S_{k}}=\hat{\Phi}_{2k}^{\mathcal{E}}-\hat{\Phi}_{2k}^{\mathcal{E}^{\prime}}.
\end{equation}
Clearly, if both ensembles yield the same moment operator up to some $k^{*}$,  $ \Tr{\Delta S_{k}^{\dagger} \Delta S_{k}}$ automatically vanishes for $k \leq k^{*}$. By choosing $\mathcal{E}^{\prime}$ to be the uniform Haar distributed ensemble, \eqref{eq:diff_operator} measures how much the ensemble $\mathcal{E}$ yields an unitary k-design. A closely related quantity to \eqref{eq:diff_operator} is the $k$th-frame potential $F^{\mathcal{E}}_{k}$ defined as,

\begin{equation}
  \Tr{\Delta S_{k}^{\dagger} \Delta S_{k} }=F_{k}^{\mathcal{E}}-F_{k}^{\text{Haar}},\quad \text{with} \quad F^{\mathcal{E}}_{k}=\int_{\mathcal{E}\times \mathcal{E}} |\Tr{U^{\dagger}V}|^{2k}d\mu_{\mathcal{E}}[U]d\mu_{\mathcal{E}}[V].
\end{equation}
Frame potentials first appeared in the literature of spherical designs \cite{Gross/2_designs_theorems/2008}, and recently have become a probe of strongly chaotic dynamics in quantum many-body systems \cite{Yoshida/Chaos_design/2017,Hunter/Unitary_designs_statistical/2019,Cotler/spectral_decoupling/2020}.

As the $C$-ensemble yields an unitary one-design, then $F^{\mathcal{E}_{C}}_{1}=F_{1}^{U(d)}=1$. However, is at the level of four-moments where both ensembles crucially differ. Let us recall that all the information regarding the specific Hamiltonian $H$ is stored (at the level of four-moments) in the $C$-ensemble by the plateau operator $G^{\mathcal{E}_{C}}$ \eqref{eq:dephasing_operator_channel}. In particular for $k=2$ the frame potential exactly yields,

\begin{equation}
\label{eq:2frame_potential}
  \begin{tikzpicture}
	\begin{pgfonlayer}{nodelayer}
		\node [style={medium_box}] (18) at (-6.25, 0) {$G^{\mathcal{E}_{C}}$};
		\node [style=none] (19) at (-5.25, -0.75) {};
		\node [style=none] (20) at (-5.25, 0.75) {};
		\node [style=none] (21) at (-3.75, -0.75) {};
		\node [style=none] (22) at (-3.75, 0.75) {};
		\node [style=none] (23) at (-4.25, 0) {};
		\node [style=none] (24) at (-4.75, 0) {};
		\node [style=none] (25) at (-5.25, -0.75) {};
		\node [style=none] (26) at (-7.25, -0.75) {};
		\node [style=none] (27) at (-7.25, 0.75) {};
		\node [style=none] (30) at (-7.25, 1.5) {};
		\node [style=none] (31) at (-7.25, -1.5) {};
		\node [style=none] (32) at (-8, 1.75) {};
		\node [style=none] (33) at (-8, -1.75) {};
		\node [style=none] (34) at (-2.25, 0) {$\displaystyle{-\frac{2d}{d+1}}$};
		\node [style=none] (35) at (-1, 1.75) {};
		\node [style=none] (36) at (-1, -1.75) {};
		\node [style=none] (42) at (-3.5, 0.75) {};
		\node [style=none] (43) at (-3.5, 1.5) {};
		\node [style=none] (44) at (-3.5, -0.75) {};
		\node [style=none] (45) at (-3.5, -1.5) {};
		\node [style=none] (62) at (-11.25, 0) {$\displaystyle{\left(\frac{d+1}{d(d-1)}\right)^{2}}$};
		\node [style=none] (63) at (-14.75, 0) {$\displaystyle{F_{2}^{\mathcal{E}_{C}}=}$};
		\node [style=none] (64) at (-0.25, 1.75) {2};
		\node [style=none] (65) at (1.25, 0) {$+F_{2}^{U(d)}$};
	\end{pgfonlayer}
	\begin{pgfonlayer}{edgelayer}
		\draw (21.center) to (23.center);
		\draw (23.center) to (22.center);
		\draw (23.center) to (24.center);
		\draw (24.center) to (25.center);
		\draw (24.center) to (20.center);
		\draw (19.center) to (26.center);
		\draw (20.center) to (27.center);
		\draw [bend left=90, looseness=1.25] (31.center) to (26.center);
		\draw [bend left=90, looseness=1.25] (27.center) to (30.center);
		\draw [bend right] (32.center) to (33.center);
		\draw [bend left] (35.center) to (36.center);
		\draw (30.center) to (43.center);
		\draw (22.center) to (42.center);
		\draw [bend left=90, looseness=1.25] (43.center) to (42.center);
		\draw (31.center) to (45.center);
		\draw (21.center) to (44.center);
		\draw [bend left=90, looseness=1.25] (44.center) to (45.center);
	\end{pgfonlayer}
\end{tikzpicture}
\end{equation}
Here, in advance of Section \ref{Chapter:two_points} and Appendix \ref{Appendix:Scrambling}, we claim that the contraction of the plateau operator with the $s$-tensor inside the frame potential \eqref{eq:2frame_potential} can be exactly identified as a sum over the overlaps between states in the local $\{\ket{l}\}$ and energy eigenbasis basis $\{\ket{E_{l}}\}$ of the $C$-ensemble Hamiltonian, \emph{i.e.}

\begin{equation}
  \label{eq:effective_dimension}
  \Tr\left((\text{s-tensor})G^{\mathcal{E}_{C}}\right) = \sum_{kl}|\braket{k|E_{l}}|^{4}.
\end{equation}
For a given initial state $\rho$, a measure of the localization over the energy eigenbasis is given by the so-called inverse participation ratio\footnote{In the literature sometimes the IPR is defined as the reciprocal, \emph{e.g.} $\left(\sum_{l}\braket{E_{l}|\rho|E_{l}}\right)^{-1}$. Here we adopt the convention of \cite{Haake/Quantum_chaos/2010} and define the IPR as in \eqref{eq:definition_IPR}.} (IPR) defined as,
\begin{equation}
\label{eq:definition_IPR}
  \text{IPR}(\rho)=\sum_{l}\Braket{E_{l}|\rho|E_{l}}^{2}.
\end{equation}
For initial highly localized states in the energy eigenbasis, the IPR is of order $\mathcal{O}(1)$ whereas for highly delocalized ones $\text{IPR}(\rho) = \mathcal{O}(d^{-1})$. By taking as the initial state every state in the local basis, the sum in \eqref{eq:effective_dimension} corresponds to the averaged inverse participation ratio $\overline{\text{IPR}}$ over the whole local basis, \emph{i.e.}
\begin{equation}
  \sum_{kl}|\braket{k|E_{l}}|^{4}=\sum_{l} \text{IPR}(\ket{l}\bra{l})=d\,\overline{\text{IPR}}.
\end{equation}
The averaged IPR characterizes the whole Hamiltonian rather than a particular energy eigenstate and solely depends on the local basis representation of the Hamiltonian $H$. Just as the IPR for a particular state, the averaged IPR is bounded by 
\begin{equation}
\frac{1}{d} \leq \overline{\text{IPR}} \leq 1,
\end{equation}
where the upper bound is saturated for strongly (on average) localized systems \emph{e.g.} $\Braket{k|E_{l}} \sim 1$, while the lower bound is saturated for strongly (on average) de-localized ones, \emph{e.g.} $\Braket{k|E_{l}} \sim \mathcal{O}(\sqrt{d^{-1}})$. As the $C$-ensemble second frame potential depends solely on the averaged inverse participation ratio, we can classify which types of Hamiltonian's will reproduce $C$-ensembles that are also unitary two-designs.

\begin{figure}
    \centering
    \includegraphics{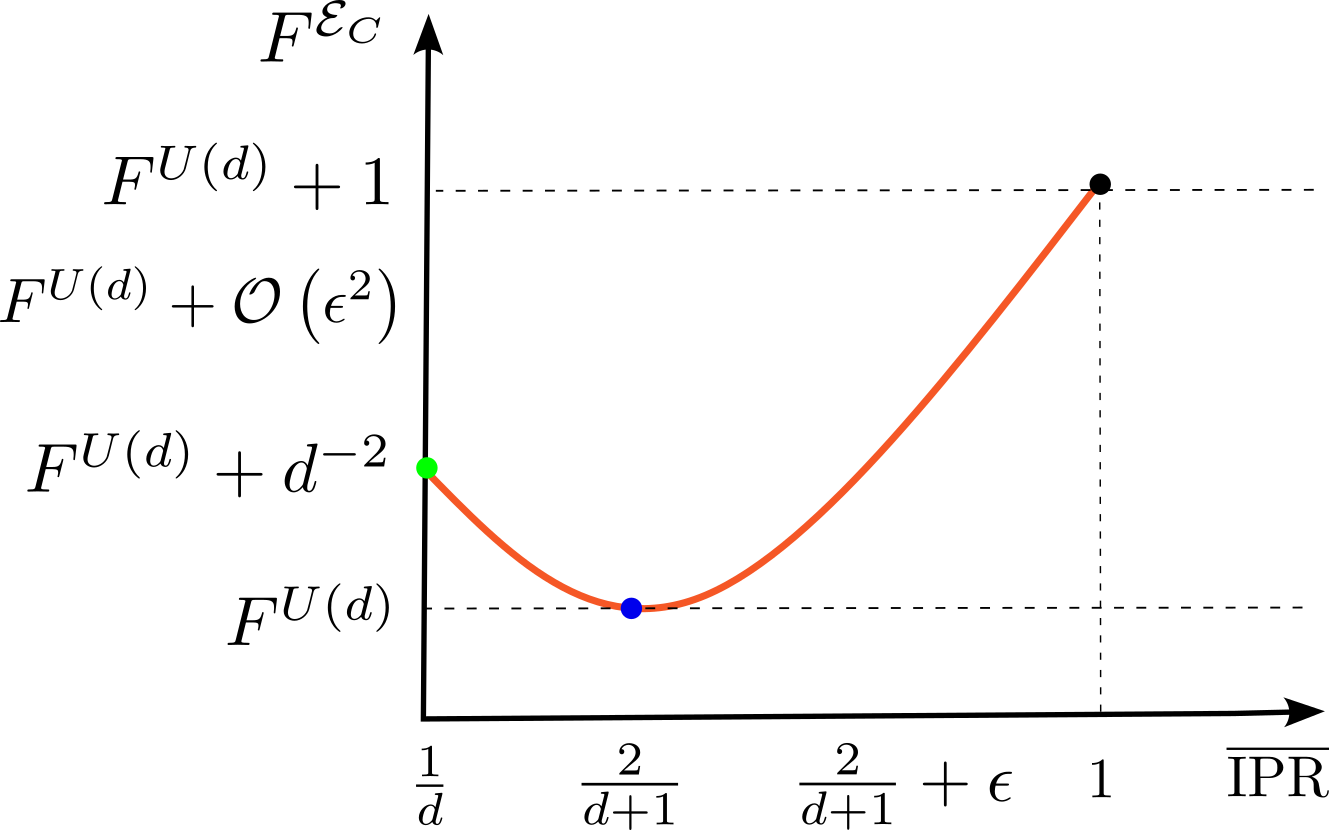}
    \caption{$C$-ensemble two-frame potential. For highly de-localized systems, where the average IPR lies between $\frac{1}{d} \leq \overline{\text{IPR}} < \frac{2}{d+1}$, the $C$-ensemble becomes an approximated two-design and the frame potential is bounded like $|F^{\mathcal{E}_{C}}_{2}-F^{U(d)}_{2}| \leq d^{-2}$. For the critical value $\overline{\text{IPR}}=\frac{2}{d+1}$, the $C$-ensemble becomes an exact two-design. For an average IPR greater than the critical value, \emph{e.g.} $\overline{\text{IPR}}=\frac{2}{d+1}+\epsilon$ with $\epsilon > 0$ and independent of the dimension $d$, the $C$-ensemble fails to be an exact two-design.}
    \label{fig:frame_potential_phase}
\end{figure}

\begin{theorem}[Condition for the $C$-ensemble to be an exactly two-design]\label{theorem:C-ensemble_two-design} 
Every $C$-ensemble generated by a Hamiltonian $H$ whose averaged IPR fulfills,

\begin{equation}
  \overline{\text{IPR}} = \frac{2}{d+1},
\end{equation}
yields an exactly unitary 2-design.
\end{theorem}
The proof of this theorem is straightforward from \eqref{eq:2frame_potential}. For Hamiltonians with averaged IPR shifted from the special value in \eqref{theorem:C-ensemble_two-design} by a factor $\epsilon$, \emph{e.g.} $IPR=\frac{2}{d+2}+\epsilon$, the resulting $C$-ensemble will yield (for large Hilbert-space dimensions) an \emph{approximate} two-design up to $\epsilon^{2}$-accuracy if the averaged IPR is strictly less than $1$ (see Fig.\ref{fig:frame_potential_phase}). 

The fact that the $C$-ensemble does not become an exactly two-design when the average IPR reaches its minimum value is a result of the underlying unitarity in the change of basis matrix \emph{i.e.} unitarity imposes correlations that cannot be reproduced by independently sampling the overlaps as $\braket{l|E_{l}}=\sqrt{d^{-1}}$. However, these correlations become sub-leading for large Hilbert spaces (in what to the design property concerns), and a maximum delocalized Hamiltonian can be used to build an approximately unitary $2$-designs.

The limit $\overline{\text{IPR}}=1$ is only reached when the local basis and the energy eigenbasis are the same, \emph{i.e.} when the Hamiltonian is already diagonal from the beginning. It was expected therefore that we cannot have a 2-design in this case. To summarize:

\begin{enumerate}
  \item For every Hamiltonian $H$, the associated $C$-ensemble is an exact unitary 1-design.
  \item For every (non-diagonal) Hamiltonian with $\overline{\text{IPR}}=\frac{2}{d+2}+\epsilon$ and sufficiently large Hilbert spaces, the associated $C$-ensemble deviates from a two-design by,

  \begin{equation}
      F^{\mathcal{E}_{C}}_{2}-F^{U(d)}_{2} = \epsilon^{2}<1.
  \end{equation}
  
  Particularly, for Hamiltonians with an averaged inverse participation ratio equal to the critical value in Theorem \ref{theorem:C-ensemble_two-design}, the $C$-ensemble becomes an exact unitary two-design.   
\end{enumerate}
\section{\texorpdfstring{$C$}{}-ensemble one- and two-point functions}\label{Chapter:two_points}

In the previous section, we introduced the concept of a spectral decoupled unitary ensemble $\mathcal{E}$ as the one for which moment operators $\hat{\Phi}_{2k}^{\mathcal{E}}$ factorize as in definition \ref{def:spectral_decoupling}. Additionally, we presented two particular examples, the well-known unitary Haar ensemble and the new $C$-ensemble. In this section, we compare the predictions made by the two ensembles on two-point functions. Additionally, we provide a physical interpretation of the plateau operator $G^{\mathcal{E}_{C}}$ as the $C$-ensemble averaged s-tensor \eqref{eq:s_tensor_introduction} in the energy eigenbasis.

Before starting, let us fix some convenient notation first. Infinite temperature averages $\frac{1}{d}\Tr(\cdots)$ will be denoted by brackets as $\braket{\cdots}$, where the normalization $d$ is the same as the carrying dimension of the argument, {\em i.e.} $\braket{W\otimes W}$ must be interpreted as $\frac{1}{d^{2}}\Tr{W\otimes W}$ unless otherwise stated. When dealing with finite temperature correlations we actually mean thermal regulated ones, {\em i.e.}, $\Braket{W}_{\beta}=\Tr{\rho_{\beta} W}$ for one point, $\Braket{W(t)V}_{\beta}=\Tr{\sqrt{\rho_{\beta}} W(t) \sqrt{\rho_{\beta}} V}$ for two-points and so on. Recurrently, we will perform eigenvector ensemble averages over the aforementioned correlation functions. Therefore, in order to avoid the double bracket annoying notation, {\em i.e.} $\langle \braket{\cdots} \rangle_{\mathcal{E}}$, when dealing with this kind of averages we will explicitly write the argument, {\em i.e.} $\langle \braket{\cdots} \rangle_{\mathcal{E}} \rightarrow \Braket{\frac{1}{d}\Tr(\cdots)}_{\mathcal{E}}$.

\subsection{Two-moments as 1-designs}

The fact that an eigenvector ensemble modeled by Haar random distributed unitaries leads (at the level of two moments) to independent eigenvector correlations is clear, as we are not choosing any preferred basis therefore there cannot be a preferred correlation. However, a more subtle question is why the $C$-ensemble, an ensemble which from the beginning has a basis-dependence given by the local basis where we know how to write down the Hamiltonian $H$, yields at the level of two moments the same results as if it were drawn from random Haar unitaries\footnote{This is the same as asking for a physical reason why the $C$-ensemble is a 1-design.}. 

To answer this question, let us avoid directly looking ahead to the ensemble two-moment operator $\hat{\Phi}^{\mathcal{E}_{C}}_{2}$ and instead ask why it should match $\hat{\Phi}^{U(d)}_{2}$. Let us take a more pragmatic point of view and ask the following: once we know all the two-moments for an arbitrary eigenvector ensemble $\mathcal{E}$, then, which type of observables of interest can we evaluate? The answer is simple, only one-Hamiltonian-valued functions, {\em i.e.} all $f(H):\mathcal{H}\to \mathcal{H}$ such that
$f(U^{\dagger}EU)=U^{\dagger}f(E)U$, with $E$ the diagonal of $H$ eigenvalues,

\begin{equation}
f(H) \rightarrow \braket{f(H)}_{\mathcal{E}}=\braket{U^{\dagger}f(E)U}_{\mathcal{E}}.
\end{equation}
Now, in the specific case of the $C$-ensemble, and from the discussion in Section \ref{section:C_ensemble}, averaging over the ensemble is the same as averaging over the equivalence class $\{C\sim U(1)^{d} \times S_{d}\,C \,\}$ of unitaries that diagonalize a given Hamiltonian subjected to no particular ordering of the eigenvalues. Within this interpretation, the $C$-ensemble average of Hamiltonian functions must be equivalent to an average over every possible eigenvalue ordering, \emph{i.e.}
\begin{equation}
  \label{eq:equivalence_class}
  \Braket{f(CEC^{\dagger})}_{\mathcal{E}_{C}}\xrightarrow{\text{equivalent}}\frac{1}{d!}\sum_{\sigma \in S_{d}}f(P_{\sigma^{-1}}EP_{\sigma}).
\end{equation}
As a proof of such equivalence notice that the right-hand side of \eqref{eq:equivalence_class} is a simple sum of diagonal matrices that yields $\frac{\Tr{f(E)}}{d} \mathbb{I}$ which turns out to be exactly the same result as the one obtained by replacing the $C$-ensemble 2-moment operator $\hat{\Phi}_{2}^{\mathcal{E}_{C}}$ on the left-hand side in the first place, {\em i.e.}

\begin{equation}
  \label{eq:1design_example}
  \begin{tikzpicture}
	\begin{pgfonlayer}{nodelayer}
		\node [style={medium_box}] (0) at (-1.75, 0) {$\hat{\Phi}_{2}^{\mathcal{E}^{C}}$};
		\node [style=none] (1) at (-1, -0.5) {};
		\node [style=none] (2) at (-2.5, 0.5) {};
		\node [style=none] (3) at (-2.5, -0.5) {};
		\node [style=none] (4) at (-1, 0.5) {};
		\node [style=none] (5) at (-3.5, 0.5) {};
		\node [style=none] (6) at (-3.5, -0.5) {};
		\node [style=none] (7) at (0, 0.5) {};
		\node [style=none] (8) at (0, -0.5) {};
		\node [style=none] (10) at (1.5, -0.5) {};
		\node [style=none] (11) at (1.5, 0.5) {};
		\node [style=none] (12) at (3, 0.5) {};
		\node [style=none] (13) at (3, -0.5) {};
		\node [style=none] (14) at (0.75, 0.5) {$f(E)$};
		\node [style=none] (15) at (-3.5, 1.25) {};
		\node [style=none] (16) at (3, 1.25) {};
		\node [style=none] (17) at (-6.5, 0) {$\langle{ C^{\dagger}f(E)C}\rangle_{\mathcal{E}_{C}} =$};
		\node [style=none] (18) at (5, 0) {$=\displaystyle{\frac{\text{Tr} f(E)}{d}}$};
	\end{pgfonlayer}
	\begin{pgfonlayer}{edgelayer}
		\draw (5.center) to (2.center);
		\draw (6.center) to (3.center);
		\draw (4.center) to (7.center);
		\draw (1.center) to (8.center);
		\draw [in=-180, out=0] (11.center) to (13.center);
		\draw [style={white_edge}] (10.center) to (12.center);
		\draw [in=180, out=0] (10.center) to (12.center);
		\draw (8.center) to (10.center);
		\draw [bend left=90, looseness=1.50] (5.center) to (15.center);
		\draw (15.center) to (16.center);
		\draw [bend left=90, looseness=1.50] (16.center) to (12.center);
	\end{pgfonlayer}
\end{tikzpicture}
\end{equation}
%
Therefore the upshot of why for every Hamiltonian the $C$-ensemble becomes a 1-design turns out to be that for every function $f(H)$ averaging over the $C$-ensemble mirrors averaging over the whole set of permutations of $H$ eigenvectors\footnote{For a sufficiently small energy window this is the same as a micro-canonical average.}, and as $\braket{f}_{U(d)}=\braket{f}_{\mathcal{E}_{C}}$ for all $f:\mathcal{H}\rightarrow \mathcal{H}$ or $\hat{\Phi}^{\mathcal{E}}_{2}=\hat{\Phi}^{U(d)}_{2}$ are equivalent definitions of design, then the $C$-ensemble must be an unitary 1-design. 

\subsection{Two-point functions and the plateau operator}

\begin{figure}[ht]
   \begin{subfigure}{.5\textwidth}
    \centering
    \includegraphics[width=.8\linewidth]{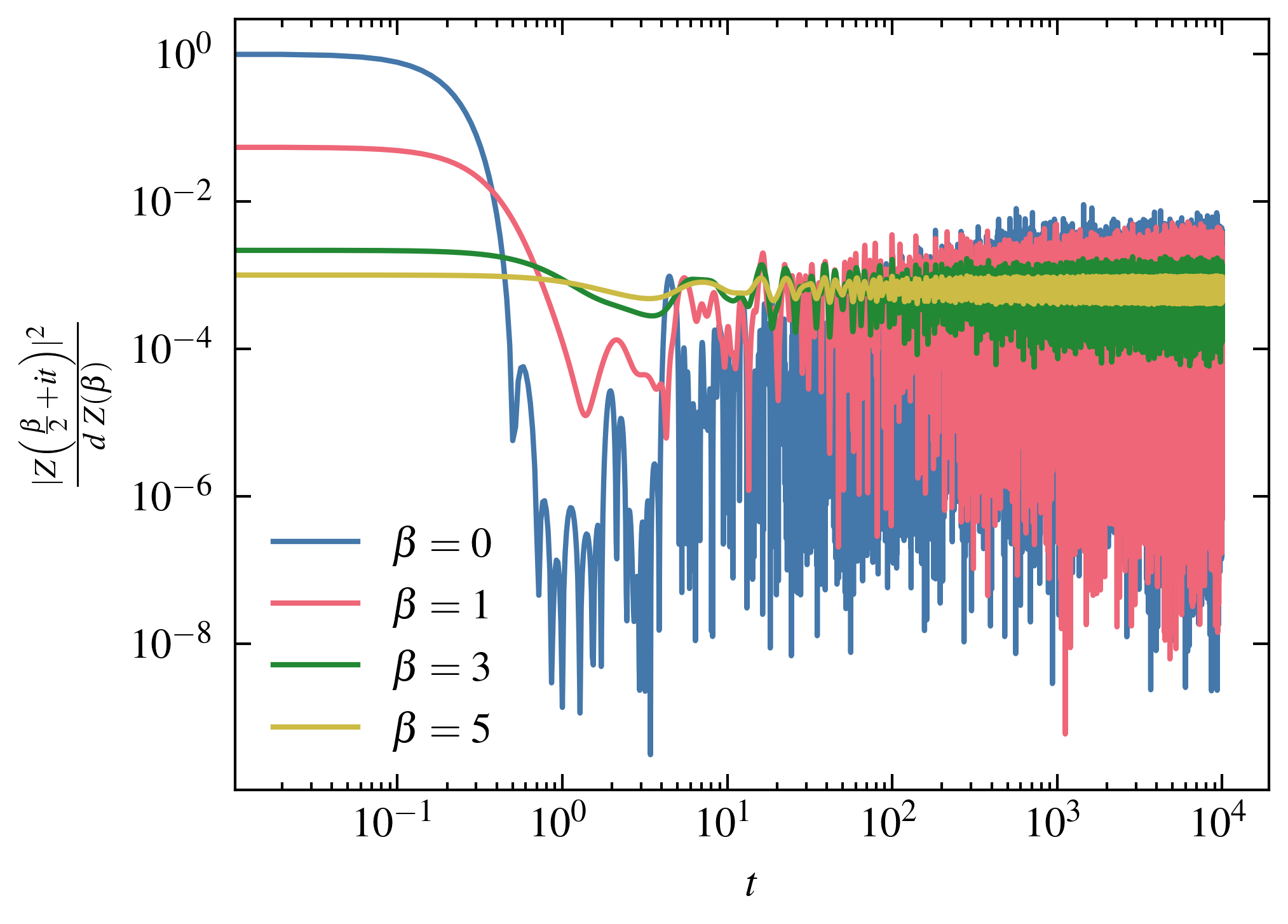}  
    \caption{Finite temperature form factor for a system of interacting Bosons.}
    \label{fig:sub-first}
  \end{subfigure}
  \begin{subfigure}{.5\textwidth}
    \centering
    \includegraphics[width=.8\linewidth]{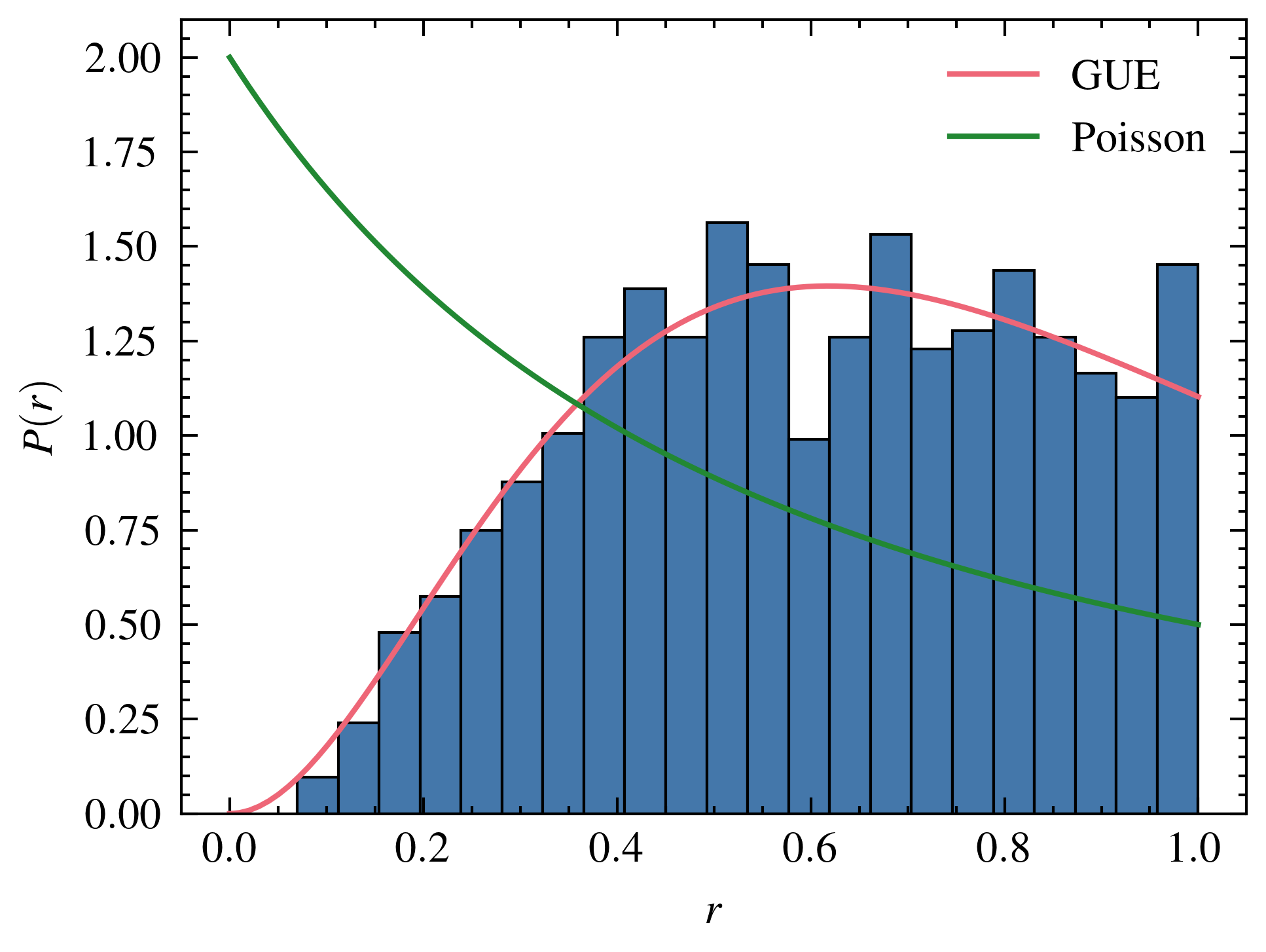}  
    \caption{Consecutive level spacing ratio distribution. }
    \label{fig:sub-second}
  \end{subfigure}
  
  \caption{(a) Example of the form factor for a non-disordered system of $N$ locally interacting Bosons over $L$ lattice sites described by a complex hopping Bose-Hubbard type Hamiltonian, $H=-\frac{J}{2}\sum_{l=1}^{L}\left(e^{i\theta}a^{\dagger}_{l+1}a_{l}+\text{h.c}\right)+\frac{U}{2}\sum_{l=1}^{L}n_{l}(n_{l}-1)$ with closed boundaries. For $\theta \in (0,\frac{\pi}{2})$ the model belongs to the unitary symmetry class. (b) The ratio of consecutive level spacings \cite{Atas/ratio_distribution/2013} follows the GUE distribution. The Hamiltonian $H$ is invariant under lattice reflections and we fix the parity sector to be $+1$. }
    \label{fig:Bose_hubbard_form_factor}
\end{figure}

Let us now pay attention to two-point functions. Due to spectral decoupling \cite{Cotler/spectral_decoupling/2020}, the time dependence for Haar-averaged, infinite temperature two-point functions,

\begin{equation}
  \label{eq:sec_2-Haar_two_points}
  \Braket{\frac{1}{d}\Tr{W(t)V}}_{U(d)}=\braket{W}\braket{V}+\frac{|Z(it)|^{2}-1}{d^{2}-1}\left(\braket{WV}-\braket{W}\braket{V} \right),
\end{equation}
is given by the infinite temperature spectral form factor $|Z(it)|^{2}=\sum_{l\,m}e^{it(E_{l}-E_{m})}$. For chaotic systems of the unitary symmetry class, an example of the typical behavior of the form factor is depicted in (Fig. \ref{fig:Bose_hubbard_form_factor}). While for short times $|Z(it)|^{2}\sim \mathcal{O}(d^{2})$, at times greater than the Heisenberg time $t_{H}$, \emph{i.e.}, the inverse of the mean-level spacing $\frac{2 \pi}{\overline{\Delta E}}$, it becomes a highly oscillating function around its infinite time averaged value\footnote{This can be naively seen as follows: By splitting the eigenvalue collisions from the form factor, we have $|Z(it_{H})|^{2}=d+\sum_{l\neq m}e^{i2 \pi 
 \overline{\Delta E} (E_{l}-E_{m})}$, where in the case of complex quantum systems, the last sum can be interpreted as a nearly uniform random walk of phases over the complex plane with $d^{2}-d$-steps. As $2D$ walks are recurrent, then the sum is expected to contribute only sub-leading with oscillations for large-$d$.},

\begin{equation}
  \lim_{T \to \infty}\frac{1}{T}\int_{0}^{T}|Z(it)|^{2}=d.
\end{equation}
In the case of a degenerate spectrum $d$ must be replaced by $\sum_{l}g_{l}$, with $g_{l}$ the multiplicity of the $l$th-level.  The $\mathcal{O}(d^{2})$ to oscillating behavior transition between short and late times in the form factor is mediated by an initial non-universal decay followed by a universal ramp \cite{Cotler/Blackholes_SYK/2017,Liu/Spectral_form_factor:lecture_notes/2018} (see also Appendix \ref{Appendix:RMT_Haar}). The transition between the decay and the ramp occurs at the dip (thermalization) time $t_{d}$ that for two-point correlation functions sets the scale for the connected to disconnected transition, {\em i.e.} $\braket{WV(t)}\rightarrow \braket{W}\braket{V}$.

For disordered chaotic many-body systems, where one has an ensemble of Hamiltonians instead of a single one, the disordered average suppresses the strong fluctuations after the dip time, and the resulting ensemble-averaged form-factor is mirrored by the Random Matrix Theory universality result (see Fig.\ref{App_fig:GUE_form_factor} in Appendix \ref{Appendix:RMT_Haar}). A typical example of this behavior is provided by the SYK model in the GUE phase (see \cite{Cotler/Blackholes_SYK/2017}). For non-disordered chaotic systems, the spectral form factor is not self-averaging \cite{Prange/Not_self_averaging_form_factor/1997}, meaning that the disordered average result does not coincide with the one of a single realization. This is again visible in the strong fluctuations, but the underlying dip, ramp, and plateau are still clearly visible.

Consider now the same two-point correlation but this time averaged over the $C$-ensemble instead. As the $C$-ensemble is, apart of the unitary Haar one, another example of a spectral decoupled ensemble (see definition \eqref{def:spectral_decoupling}), the four moment operator $\hat{\Phi}_{4}^{\mathcal{E}_{C}}$ \eqref{eq:C_four_moment_operator} yields spectral decoupled two-point functions,

\begin{align}
  \label{eq:C_two_points}
  \Braket{\frac{1}{d}\Tr W(t)V}_{\mathcal{E}_{C}} =\,&\frac{1}{d-1} \left( \Tr(G^{\mathcal{E}_{C}} W\otimes V)-\braket{WV}\right) \\ \notag
  +\,&\frac{|Z(it)|^{2}}{(d-1)d}\left( \braket{WV}-\frac{1}{d}\Tr(G^{\mathcal{E}_{C}} W\otimes V)\right).
\end{align}
Where the time dependence is stored again in the spectral form factor. Therefore, we expect that two-point functions follow similar transitions as the ones predicted by the Haar ensemble, {\em i.e.} for short times the correlations must remain connected $\braket{W(t)V}\sim \braket{WV}$, whereas for later times after the thermalization time, they must become roughly disconnected $\braket{W(t)V}\sim \braket{W}\braket{V}$. It is easy to show that the $C$-ensemble correctly reproduces the short-time behavior, 

\begin{equation}
  \Braket{\frac{1}{d}\Tr{W(t)V}}_{\mathcal{E}_{C}}\sim \braket{WV} \quad \text{for} \quad t \ll t_{d}.
\end{equation}
However, to show the connected-disconnected transition after the thermalization time is much more subtle. The reason is that the $C$-ensemble predicts\footnote{We use that at the dip time $|Z(it_{d})|^{2}\sim \mathcal{O}(\sqrt{d})$.},

\begin{equation}
\label{eq:dip_two_points}
  \braket{\Tr{W(t)V}}_{\mathcal{E}_{C}}\sim \frac{1}{d}\left( \Tr{G^{\mathcal{E}_{C}}W \otimes V}-\braket{WV}\right)+\mathcal{O}(d^{-\frac{3}{2}}) \quad \text{for} \quad t\sim t_{d,}
\end{equation}
and to show that the leading term contributes with a disconnected part $\braket{W}\braket{V}$ one needs to find the plateau operator $G^{\mathcal{E}_{C}}$  first. Fortunately, in Appendix \ref{Appendix:Scrambling} we prove that for the $C$-ensemble, the plateau operator has the following natural factorization,

\begin{equation}
  \label{eq:dephasing_factorization}
  G^{\mathcal{E}_{C}}=G^{U(d)}+G[H],
\end{equation}
where the term $G[H]$ exclusively depends on the particular Hamiltonian $H$. The decomposition of the plateau operator \eqref{eq:dephasing_factorization} into a universal (Haar) $G^{U(d)}$ contribution and a non-universal  $G[H]$ term can be used to check how a fixed Hamiltonian $H$ will yield into a purely Haar random unitary time evolution. Concretely, for strongly chaotic quantum many-body systems we expect relations like

\begin{equation}
  G[H] \sim 0,
\end{equation}
where the comparison must be interpreted within some operator norm. The argument behind this claim is the following: For the special case of an unitarily invariant Hamiltonian ensemble $\mathcal{E}_{H}$, we show (in Appendix \ref{Appendix:Scrambling}) that the non-universal contribution of the plateau operator identically vanishes\footnote{This is equivalent to say that for those systems the $C$-ensemble becomes an exactly two-design (see \ref{eq:C_four_moment_operator}).}, \emph{e.g.} $\braket{G[H]}_{\mathcal{E}_{H}}=0$. 
Although, strictly speaking, unitarily invariant ensembles only capture the chaotic features of highly non-local many-body systems \emph{i.e.} systems with $k$-body interactions with $k\to \infty$, we expect that for strongly chaotic (and not-necessarily non-local) many-body systems, the smallness of $G[H]$ still holds up to some extent. Particularly, the factorization \eqref{eq:dephasing_factorization} implies that for arbitrary two-point functions, the ensemble average decomposes as, 

\begin{align}
  \Braket{\frac{1}{d}\Tr{W(t)V}}_{\mathcal{E}_{C}}  = & \Braket{\frac{1}{d}\Tr{W(t)V}}_{U(d)}+\frac{1}{d-1} \left( \Tr(G[H] W\otimes V)-\braket{WV}\right) \\ \notag
  +\,&\frac{|Z(it)|^{2}}{(d-1)d}\left( \braket{WV}-\frac{1}{d}\Tr(G[H] W\otimes V)\right),
\end{align}
and the expected disconnected $\braket{W}\braket{V}$ term appears from the universal Haar contribution of the plateau operator. Schematically,

\begin{equation}
\frac{1}{d}\braket{\Tr{W(t)V}}_{\mathcal{E}_{C}}\sim \braket{W}\braket{V}+\mathcal{O}(\text{``individual Hamiltonian corrections''}),
\end{equation}
with $\mathcal{O}(\text{``individual Hamiltonian corrections''})$ the non-universal contribution coming from the non-universal part of the plateau operator $G[H]$.

\subsubsection{Long-time averages and late-time exactness}

So far one of the most attractive features of the $C$-ensemble lies in its specific Hamiltonian dependence. As argued in section \ref{section:C_ensemble}, among all the possible Hamiltonian-dependent eigenvector ensembles, the $C$-ensemble stands as the fairest one by considering all the Hamiltonian eigenvectors on the same footing. In this section, we show that, due to this fairness, the $C$-ensemble is late-time exact for two-point functions at infinite temperature, \emph{i.e.} the $C$-ensemble average exactly matches the time-average.

To see this, let us consider the long-time average of infinite temperature two-point functions given by the so-called diagonal ensemble \cite{DAlession/ETH_review/2016},

\begin{equation}
  \label{eq:sec3_diag_ensemble}
  \overline{\braket{A(t)A}}=\lim_{T \to \infty}\frac{1}{T}\int_{0}^{T} \frac{1}{d}\Tr( A(t)A) \,dt=\frac{1}{d}\sum_{l}\braket{E_{l}|A|E_{l}}\braket{E_{l}|A|E_{l}}. 
\end{equation}
Now, suppose we are to repeat the same calculation but this time averaging first over the Haar distributed eigenvector ensemble. The straightforward motivation for doing so is the following: If the Haar eigenvector ensemble is an accurate model for late-time chaotic dynamics, then one would expect that replacing the diagonal ensemble with the long-time average of the Haar eigenvector ensemble will yield similar results. 

From previous sections we know that Haar distributed unitaries constitute a spectral decoupled eigenvector ensemble (and from definition \ref{def:spectral_decoupling} turns out to be the simplest among any other spectral decoupled eigenvector ensemble). For spectral decoupled ensembles, long-time averages are quite simple to evaluate, because all time dependence of the correlation functions is contained in terms of spectral form factors \footnote{We want to remark that although for both, uniform distributed unitaries and the C-ensemble, this is the case. There could be more ``exotic'' eigenvector ensembles for the ones this may not hold.}, $e.g.$

\begin{equation}
  \label{eq:sec3_longtime_haar}
  \lim_{T \to \infty} \frac{1}{T}\int_{0}^{T}dt \Braket{\frac{1}{d}\Tr{A(t)A}}_{U(d)}= \frac{d}{d+1}\left(\braket{A}^{2}+\frac{1}{d}\braket{A^{2}}\right).
\end{equation}
However, when we try to reconcile this result with \eqref{eq:sec3_diag_ensemble} immediately arises the problem that although the diagonal ensemble is by definition $H$-dependent, the right-hand side in \eqref{eq:sec3_longtime_haar} is not. In particular, for any spectral decoupled ensemble without any explicit dependence on a particular Hamiltonian $H$, we will conclude the same. A standard approach to reconcile both results is by arguing that the observable $A$ fulfills the Eigenstate Thermalization Hypothesis and interpret $\braket{A}$ as the micro-canonical average over a small energy window. We will discuss more about this further in the following section. Here, we will adopt a different route. Instead of trying to reconcile both results, the Haar average one with the diagonal ensemble one, we will improve the average by using the $C$-ensemble in the first place, {\em i.e.}

\begin{equation}
  \label{eq:sec_2-dephasing_and_long_time}
  \lim_{T \to \infty} \frac{1}{T}\int_{0}^{T}dt \Braket{\frac{1}{d}\Tr{A(t)A}}_{\mathcal{E}_{C}}= \frac{1}{d}\Tr{ \left(G^{\mathcal{E}_{C}}A^{\otimes 2} \right)}.
\end{equation}
By doing so, we are able to: First, identify the $C$-ensemble plateau operator $G^{\mathcal{E}_{C}}$ with the ensemble-averaged $s$-tensor in the energy eigenbasis, \emph{i.e.}

\begin{equation}
\label{eq:plateu-to-s_tensor}
  G^{\mathcal{E}_{C}} =\left\langle \sum_{l}\ket{E_{l}\, E_{l}}\bra{E_{l} \, E_{l}} \right\rangle_{\mathcal{E}_{C}}.
\end{equation}

Second, we recover an explicit Hamiltonian contribution for the late-time average of two-point functions, initially absent in the Haar case. However, the key point is that as we show in Appendix \ref{Appendix:Scrambling}, the identification \eqref{eq:plateu-to-s_tensor} is not only at the average level but rather an exact one, \emph{i.e.}

\begin{equation}
  \left\langle \sum_{l}\ket{E_{l}\, E_{l}}\bra{E_{l} \, E_{l}} \right\rangle_{\mathcal{E}_{C}}=\sum_{l}\ket{E_{l}\, E_{l}}\bra{E_{l} \, E_{l}} .
\end{equation}
This means that the $C$-ensemble not only provides a Hamiltonian contribution, but rather gives the right one, \emph{i.e.} it  \emph{exactly} computes infinite temperature two-point functions at late times! We emphasize that this is a remarkable feature. Among all the possible choices of Hamiltonian-dependent ensembles, the $C$-ensemble stands out due to its ability to exactly capture the (infinity-temperature) late-time physics of correlation functions, and thus providing the right corrections beyond the universal RMT-regime. 

Apart from arbitrary two-point functions between pairs of local operators, the $C$-ensemble can also be used to find average out-of-equilibrium expectation values of local observables over arbitrary initially prepared states, {\em e.g.}

\begin{align}
\Braket{\Tr{A(t)\rho}}_{\mathcal{E}_{C}}&=\frac{d}{d-1}\left(\Tr{\left(G^{\mathcal{E}_{C}}A\otimes \rho\right)} -\frac{1}{d}\braket{A}_{\rho}\right)\\
  &+\frac{|Z(it)|^{2}}{d(d-1)}\left( \braket{A}_{\rho}-\Tr{\left(G^{\mathcal{E}_{C}} A\otimes \rho\right)}\right),
\end{align}
where $\braket{A}_{\rho}$ denotes the initial time expectation value $\braket{A}_{\rho}=\Tr{A\rho}$. By identifying the action of the plateau operator with the long-time average $\overline{\braket{A}_{\rho}}$ as in \eqref{eq:sec_2-dephasing_and_long_time} and taking the large-$d$ limit,

\begin{align}
  \label{eq:Out_equilimbrium_expected_value}
  \Braket{\Tr A(t)\rho}_{\mathcal{E}_{C}}=\frac{|Z(it)|^{2}}{d^{2}}\braket{A}_{\rho}+\left(1-\frac{|Z(it)|^{2}}{d^{2}}\right)\overline{\braket{A}_{\rho}},
\end{align}
the out-equilibrium expectation value then becomes the competition between two terms. On the one hand, the initial conditions stored in $\braket{A}_{\rho}$ and on the other the late-time behavior given by $\overline{\braket{A}_{\rho}}$. We note that a similar result to the one in \eqref{eq:Out_equilimbrium_expected_value} was obtained earlier in \cite{Reimann/Relaxation_many_body/2020}. In their calculation, the authors model the many-body Hamiltonian as a ``diagonal part plus a random matrix perturbation'', and perform a RMT average. However, the key point turns out to be that the $C$-ensemble addresses the conceptual failures of previous results which either do not use a fixed Hamiltonian, {\em i.e.} \cite{Reimann/Relaxation_many_body/2020}, or do not use the best model for an eigenvector ensemble \cite{Reimann/Thermalization_Haar/2016}.

\subsubsection{Finite temperature case}

In this section, we consider the ensemble average of (regulated) two-point functions at finite temperature, {\em i.e. }
\begin{equation}
  \braket{W(t)V}_{\beta}=\Tr\left(\rho^{1/2}_{\beta}W(t)\rho^{1/2}_{\beta}V\right),
\end{equation}
with $\rho_{\beta}=Z(\beta)^{-1}e^{-\beta H}$ the Gibbs state. For the $C$-ensemble, due to spectral decoupling, the average of thermal correlation functions factorizes as in Section. \ref{eq:sec_2-Haar_two_points},

\begin{align}
  \label{eq:C_ensemble_regulated_tow_points}
  \Braket{\Tr \rho_{\beta}^{\frac{1}{2}}W(t)\rho_{\beta}^{\frac{1}{2}}V}_{\mathcal{E}_{C}} =\,&\frac{1}{d-1} \left( \Tr(G^{\mathcal{E}_{C}} W\otimes V)-\braket{WV}\right) \\ \notag
  +\,&\frac{|Z(\frac{\beta}{2}-it)|^{2}}{(d-1)Z(\beta)}\left( \braket{WV}-\frac{1}{d}\Tr(G^{\mathcal{E}_{C}} W\otimes V)\right),
\end{align}
where the spectral form factor $|Z(it)|^{2}$ is now replaced by its analytically continued version $|Z(\frac{\beta}{2}+it)|^{2}$. Due to finite-temperature corrections, both, the plateau height and the initial decay become temperature-dependent. At short-times, 

\begin{equation}
  \left| Z\left(\frac{\beta}{2}-it\right) \right|^{2} \sim  Z\left(\frac{\beta}{2}\right)^{2}
\end{equation}
whereas for late-times the plateau height reads,

\begin{equation}
  \lim_{T \to \infty }\frac{1}{T}\int_{0}^{T}\left| Z\left(\frac{\beta}{2}-it\right) \right|^{2} dt \sim  Z\left(\beta\right)
\end{equation}
The transition between the (now temperature-dependent) initial-value and plateau-value for the finite-temperature form-factor is again mediated by a non-universal decay and a universal ramp. However, due to the exponential $\beta$ weight, the strong fluctuations after the dip time present in the infinite temperature case become more suppressed as $\beta$ increases\footnote{As $\beta$ increases, the leading contribution to the form factor comes from the Many-body ground-state rather than the Bulk of the spectrum. Therefore, the thermal-fluctuations in the ramp must decrease.}. For the $C$-ensemble what appears in the two-point functions is not the form factor, but rather the normalized version, \emph{e.g.}

\begin{equation}
  \frac{|Z(\frac{\beta}{2}-it)|^{2}}{(d-1)Z(\beta)} \leq \frac{Z\left(\frac{\beta}{2}\right)^{2}}{(d-1)Z(\beta)}.
\end{equation}
For the Haar ensemble, due to spectral decoupling, the thermal correlation function factorizes in a disconnected $\braket{W}\braket{V}$ term plus the connected term weighted by the finite temperature form factor, \emph{e.g.}

\begin{align}
  \label{eq:Haar_regulated_tow_points}
  \Braket{\Tr \rho_{\beta}^{\frac{1}{2}}W(t)\rho_{\beta}^{\frac{1}{2}}V}_{U(d)} = & \braket{W}\braket{V} \nonumber \\
  +&\left(\frac{|Z(\frac{\beta}{2}-it)|^{2}d}{Z(\beta)(d^{2}-1)}-\frac{1}{d^{2}-1}\right)\left(\braket{WV}-\braket{W}\braket{V}\right).
\end{align}
It is instructive to compare both the short- and late-time limits of regulated two-points functions with the ones obtained in the non-regulated case, {\em i.e.} by replacing $t \to t+i\frac{\beta}{2}$ in \eqref{eq:C_ensemble_regulated_tow_points}. First, at short times, the spectral dependence is $Z(\beta-it)Z(it)\sim \mathcal{O}(d\, Z(\beta))$ for the non-regulated case, and the ensemble averaged two-points become nearly temperature independent, \emph{e.g.}

\begin{align}
  \Braket{\Tr \rho_{\beta} W(t) V}_{\mathcal{E}_{C}} \sim \braket{WV}
\end{align}
However, for late-times, $Z(\beta-it)Z(it)\sim \mathcal{O}(Z(\beta))$, and we recover the same late-time behaviour as in the regulated case \eqref{eq:C_ensemble_regulated_tow_points}. 
\begin{figure}[ht]
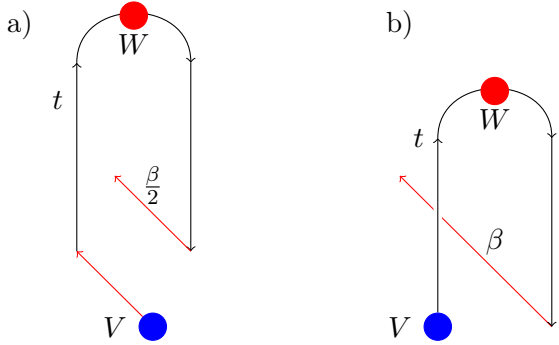

  \ctikzfig{Two_point_ordering}
   \caption{Regulated vs. non-regulated two-point functions. Unitary time evolution is depicted by black arrows, whereas thermal evolution is represented by perpendicular red arrows. For the regulated case (a), the operator insertions correspond to $e^{-\frac{\beta}{2}H}e^{itH}We^{-itH}e^{-\frac{\beta}{2}H}V$ and for all times the local operators $W,V$ are separated by a distance $\frac{-\beta}{2}$ in euclidean time. On the other hand, for the non-regulated case (b) at $t=0$ both $V$ and $W$ are close even in Euclidean time.} 
   \label{fig:two_point_orderings}
\end{figure}
The reason for this discrepancy at short times between the regulated and non-regulated ensemble-averaged two-point correlations is that, in the non-regulated ordering the pair of local operators $V,W$ become closer as time approaches zero and the two-point correlation progressively reduces to a one-point one of the composite $WV$ operator. In the $C$-ensemble\footnote{This also holds for the unitary Haar ensemble.} one-point functions are averages over the whole Hilbert space (a simple trace) and therefore the temperature dependence disappears. In contrast to the non-regulated case, in the regulated ordering the local operators are always separated at every time and therefore they never yield a one-point function (see Fig. \ref{fig:two_point_orderings}).

\subsection{ETH signatures}

A related framework to explain the universal signatures of correlation functions for a broad class of systems, and in particular, the emergence of thermal behavior in complex closed quantum systems is given by the Eigenstate Thermalization Hypothesis ETH \cite{Foini/Eth_OTOC/2022,Pappalardi/free_eth/2022}. In it current form, the ETH is a claim about the form of the
\emph{matrix elements} for local operator in the energy eigenbasis, {\em i.e.}

\begin{equation}
  \label{eq:sec_4_ETH}
  \braket{E_{l}|A|E_{m}}=\mathcal{A}(E_{lm})\delta_{lm}+e^{-\frac{S(E_{lm})}{2}}f^{A}(E_{lm},\omega_{lm})R_{lm},
\end{equation}
where $E_{lm}=\frac{E_{l}+E_{m}}{2}$, $\omega_{lm}=E_{l}-E_{m}$ are the mean energy and frequency, respectively. $S$ is the thermodynamic entropy, $\mathcal{A}$ is a smooth function of the mean energy\footnote{This function is usually identified with either the micro-canonical or canonical ensemble expectation value of $A$. In the canonical case, the temperature of the corresponding Gibbs state $\rho_{\beta}$ gets fixed by the requirement $\mathcal{A}(E)=\Tr{\rho A}$.}, $f^{A}(E_{lm},-\omega_{lm})=f^{A}(E_{lm},\omega_{lm})$ is an even smooth function dependent on the particular observable $A$ and finally $R_{lm}$ are zero mean and unit variance random variables. The key point in relating the ETH with eigenvector ensembles is to notice that any eigenvector ensemble $\mathcal{E}$ predicts a particular form for the matrix elements of local operators in the energy basis, \emph{i.e.}, $\braket{E_{l}|A|E_{m}}$ can be replaced by the particular ensemble average $\braket{l|UAU^{\dagger}|m}_{\mathcal{E}}$. Whereas the inverse is not true, that is, from the ETH we cannot infer a particular structure for the eigenvectors rather than the trivial one: that they must be sufficiently complex to fulfill the ETH in the first place. Here we claim that the specific Hamiltonian dependence of the $C$-ensemble can be used to recover some very specific features of the ETH. In particular, by looking at the variance of non-diagonal matrix elements, 

\begin{equation}
  \frac{1}{d}\left( \braket{A^{2}}-\frac{1}{d}\Tr{G^{\mathcal{E}_{C}} A^{\otimes 2}}\right)\xleftarrow[]{\text{\, C-ensemble \,}} |\braket{E_{l}|A|E_{m}}|^{2}\xrightarrow[]{\text{\, ETH \,}} e^{-S(E_{lm})}|f^{A}(E_{lm},\omega_{lm})|^{2}
\end{equation}
and by focusing on a small energy window, where one identifies the thermodynamical smoothed entropy as the logarithm of the number of states, \emph{i.e.}, $e^{-S}\sim d^{-1}$, the compatibility between both approaches is clear. Furthermore, the $C$-ensemble predicts a particular form for the ETH undetermined $|f^{A}|^{2}$ in terms of the plateau operator, \emph{i.e.},

\begin{equation}
  \label{eq:f_function_identification}
  |f^{A}(E_{lm},\omega_{lm})|^{2} \rightarrow \braket{A^{2}}-\frac{1}{d}\Tr{G^{\mathcal{E}_{C}} A^{\otimes 2}}=\braket{A^{2}}-\overline{\braket{A(t)A}},
\end{equation}
where we also used the identification of the plateau operator with the diagonal ensemble presented in the last section. However, for non-integrable many-body systems the ETH $f$-function is expected to decay as the frequency increases \cite{Sdreniky/ETH/1998,Murthy/ETH_bound/2019}, {\em i.e.} $|f^{A}(E,w)|^{2}\xrightarrow[\text{increases}]{w} 0$. For the $C$-ensemble one can prove that the predicted form of the $f^{A}$ function indeed decays, but trivially as $\delta_{w,0}$ due to the $S_{d}$ symmetry of the ensemble. 

\subsubsection{Requirements for an ETH eigenvector ensemble}

The fact that the $C$-ensemble correctly reproduces specific features of the ETH such as the exponentially decay of the non-diagonal correlations, or the long-time average value of two-point correlation functions mentioned in the previous section, suggests to promote eigenvector ensembles as the fundamental object instead of the plain ETH ansatz to explain the emergence of thermal behavior in complex systems and to seek for an ``ETH eigenvector ensemble $\mathcal{E}_{\text{ETH}}$'' which yields into ETH correlations (\ref{eq:sec_4_ETH}). 

\begin{itemize}
  \item \emph{Underlying unitary}: The first requirement and perhaps the most important is the unitarity of the ensemble $\mathcal{E}_{\text{ETH}}$. This requirement will not only constrain all the ensemble $k$th moment operators $\hat{\Phi}_{2k}^{\mathcal{E}_{\text{ETH}}}$ but additionally demands that the ETH ansatz (\ref{eq:sec_4_ETH}) must be modified by dropping the $R_{lm}$ random variable\footnote{{\em i.e.} by replacing (\ref{eq:sec_4_ETH}) with $\braket{E_{l}|A|E_{m}}=\mathcal{A}(E_{lm})\delta_{lm}+e^{-\frac{S(E_{lm})}{2}}f(E_{lm},\omega_{lm})$ instead. This incompatibility with unitarity was implicitly pointed out in \cite{Foini/Eth_OTOC/2022,Fava/ETH_designs/2023}.} in order to be compatible with the underlying unitarity of the eigenvector ensemble.
  
  \item Hamiltonian dependence: Any eigenvector ensemble that yields ETH matrix elements obviously must explicitly depend on a fixed Hamiltonian $H$. This requirement, in particular, trivially rules out the unitary Haar as a candidate for $\mathcal{E}_\text{ETH}$.
  
  \item  No left $S_{d}$ invariance: As mentioned before, a left $S_{d}$ symmetry\footnote{ A right $S_{d}$ symmetry is even worse, as it implies averages over the computational basis as well.} in the eigenvector ensemble implies that we are implicitly averaging over the whole set of Hamiltonian eigenvectors. This average destroys the local dependence of individual eigenvectors in ensemble-averaged correlation functions and will not yield an acceptable ETH ansatz. The $C$-ensemble fails to meet this requirement and, in particularly, this is the reason why the variance $\langle |\braket{E_{l}|A|E_{m}}|^{2} \rangle_{\mathcal{E}_{C}}$ is independent of $l,m$, and why both the $C$-ensemble and $ETH$ were only compatible within small energy windows.  
\end{itemize}

\newpage
\section{Chaos and Complexity within the \texorpdfstring{$C$}{}-Ensemble}\label{Chapter_chaos}

Although two-point self correlations such as the ones evaluated in Section \ref{Chapter:two_points} are good diagnostics of the thermalization or equilibration processes\footnote{This is clear from the naive classical point of view, where the exponential decay of connected two-point correlations means that the system has ``forgotten'' the initial conditions which in some sense can be interpreted as ergodicity.}, they fail to capture the sensitivity to initial conditions characteristic of the chaotic unitary Hamiltonian time evolution inherent in many-body complex systems. This is where four-point correlations (more accurately the out of order ones) come into play. However, ensemble-averaged four-point correlation functions would require the eight-moment operator $\hat{\Phi}^{\mathcal{E}}_{8}$ in every eigenvector ensemble $\mathcal{E}$. For Haar distributed unitaries the number of terms contributing to eight moments is $(4!)^{2}=576$\footnote{This comes from the Weingarten formula, see Appendix \ref{Appendix:RMT_Haar}.}, and to try to directly find the eight-moment operator for the $C$-ensemble is far more difficult than in the Haar case because of the Hamiltonian playing the role of an external source inside the measure.

In this section, we first we present a solution on how to circumvent the ``problem'' of explicitly requiring eight-moments to evaluate out-ordered four-point correlations by directly representing the correlation as a two-point one in a bigger space. Subsequently, we move to the evaluation of $C$-ensemble averaged out-time-ordered correlations by representing them as two-point correlation functions in proper twofold Hilbert spaces. Finally, motivated by \cite{Yoshida/Chaos_design/2017} we explore the complexity of building the $C$-ensemble from some elementary gate set and provide a system dependent bound.

\subsection{Replica space method}\label{sec:replica_method}

Similarly to what was done in reference \cite{Pappalardi/Replica_trick/2022}, let us briefly explain how the method works by considering the following four-point correlation $\Tr{A(t)BC(t)D}$. Diagrammatically, the correlation can be represented as,

\begin{equation*}
  \begin{tikzpicture}
	\begin{pgfonlayer}{nodelayer}
		\node [style={small_box}] (0) at (-7, 0) {$e^{iHt}$};
		\node [style={small_box}] (1) at (-5, 0) {$A$};
		\node [style={small_box}] (2) at (-2.75, 0) {$e^{-iHt}$};
		\node [style={small_box}] (3) at (-0.5, 0) {$B$};
		\node [style={small_box}] (4) at (1.5, 0) {$e^{iHt}$};
		\node [style={small_box}] (5) at (3.5, 0) {$C$};
		\node [style={small_box}] (6) at (5.75, 0) {$e^{-iHt}$};
		\node [style={small_box}] (7) at (8, 0) {$D$};
		\node [style=none] (8) at (-7.75, 1) {};
		\node [style=none] (9) at (8.5, 1) {};
		\node [style=none] (10) at (-7.75, 0) {};
		\node [style=none] (11) at (8.5, 0) {};
		\node [style={small_box}] (12) at (11.75, 1) {$e^{iHt}$};
		\node [style={small_box}] (13) at (13.75, 1) {$A$};
		\node [style={small_box}] (14) at (16, 1) {$e^{-iHt}$};
		\node [style={small_box}] (15) at (18.25, 1) {$B$};
		\node [style=none] (22) at (10.75, 1) {};
		\node [style={small_box}] (24) at (11.75, -0.5) {$e^{iHt}$};
		\node [style={small_box}] (25) at (13.75, -0.5) {$C$};
		\node [style={small_box}] (26) at (16, -0.5) {$e^{-iHt}$};
		\node [style={small_box}] (27) at (18.25, -0.5) {$D$};
		\node [style=none] (28) at (18.75, -0.5) {};
		\node [style=none] (29) at (18.75, 1) {};
		\node [style=none] (30) at (10.75, -0.5) {};
		\node [style=none] (31) at (20.25, 1) {};
		\node [style=none] (32) at (20.25, -0.5) {};
		\node [style=none] (33) at (10.75, 2) {};
		\node [style=none] (34) at (20.25, 2) {};
		\node [style=none] (35) at (20.25, -1.5) {};
		\node [style=none] (36) at (10.75, -1.5) {};
		\node [style=none] (37) at (9.75, 0.25) {$\displaystyle{=}$};
	\end{pgfonlayer}
	\begin{pgfonlayer}{edgelayer}
		\draw (0) to (1);
		\draw (1) to (2);
		\draw (2) to (3);
		\draw (4) to (5);
		\draw (5) to (6);
		\draw (6) to (7);
		\draw (3) to (4);
		\draw [bend left=90, looseness=1.25] (9.center) to (11.center);
		\draw [bend left=270, looseness=1.25] (8.center) to (10.center);
		\draw (10.center) to (0);
		\draw (8.center) to (9.center);
		\draw (7) to (11.center);
		\draw (12) to (13);
		\draw (13) to (14);
		\draw (14) to (15);
		\draw (22.center) to (12);
		\draw (24) to (25);
		\draw (25) to (26);
		\draw (26) to (27);
		\draw (27) to (28.center);
		\draw (15) to (29.center);
		\draw (30.center) to (24);
		\draw [in=180, out=0, looseness=1.25] (29.center) to (32.center);
		\draw [style={white_edge}, in=180, out=0, looseness=1.25] (28.center) to (31.center);
		\draw [in=180, out=0] (28.center) to (31.center);
		\draw [bend right=90, looseness=1.25] (33.center) to (22.center);
		\draw [bend left=270, looseness=1.25] (30.center) to (36.center);
		\draw [bend left=90, looseness=1.25] (32.center) to (35.center);
		\draw [bend left=90, looseness=1.25] (34.center) to (31.center);
		\draw (33.center) to (34.center);
		\draw (36.center) to (35.center);
	\end{pgfonlayer}
\end{tikzpicture}
\end{equation*}
where the right hand side is simply obtained by twisting the wire. Although both, left and right, hand sides are topologically equivalent, there is a subtle (and in fact crucial) difference between both representations in their interpretation, that is, we can identify the right hand side as a trace over $\mathcal{H}\otimes\mathcal{H}$ while in the left hand side we are tracing only over $\mathcal{H}$. Therefore, by defining the ``upgraded'' Hamiltonian in the two-fold space, $\hat{H}=H \otimes \mathbb{I} + \mathbb{I}\otimes H$ \footnote{ This definition arises naturally from the identity $e^{A\oplus B}=e^{A}\otimes e^{B}$, with $A\oplus B = A \otimes \mathbb{I}+ \mathbb{I}\otimes B$ the Kronecker sum of two operators of the same dimensions.}, the four-point correlation in $\mathcal{H}$ can be set as a two-point correlation in the replicated space $\mathcal{H}\otimes \mathcal{H}$, \emph{e.g.},

\begin{equation}
  \label{eq:replica_space_method}
  \Tr_{\mathcal{H}}{\left(e^{iHt}Ae^{-iHt}Be^{iHt}Ce^{-iHt}D\right)}=\Tr_{\mathcal{H}\otimes \mathcal{H}}{\left(e^{it \hat{H}} (A\otimes C) e^{-it \hat{H}}(B\otimes D)\,  \text{SWAP}\right)}.
\end{equation}
In fact the result in \eqref{eq:replica_space_method} can be further generalized to higher-order correlations where we highlight the interpretation of the replica space method\footnote{Although similar in motivation, it should not be confused with the replica trick in the Spin-Glass literature \cite{Dotsenko/Spin_Glasses/2000}.} as a duality relating higher order correlation functions in small systems to low-order correlations in larger ones.

As a remark, we wish to point out that this method also works for disconnected products of $n$-point correlations. Particularly, for the product of disconnected two-points $Tr{A(t)B}\Tr{C(t)D}$, the equivalent lower correlations in the twofold space correspond to a disconnected pair of one points, {\em e.g.}

\begin{equation*}
  \begin{tikzpicture}
	\begin{pgfonlayer}{nodelayer}
		\node [style={small_box}] (12) at (8.5, 1) {$e^{iHt}$};
		\node [style={small_box}] (13) at (10.5, 1) {$A$};
		\node [style={small_box}] (14) at (12.75, 1) {$e^{-iHt}$};
		\node [style={small_box}] (15) at (15, 1) {$B$};
		\node [style=none] (22) at (7.25, 1) {};
		\node [style={small_box}] (24) at (8.5, -0.5) {$e^{iHt}$};
		\node [style={small_box}] (25) at (10.5, -0.5) {$C$};
		\node [style={small_box}] (26) at (12.75, -0.5) {$e^{-iHt}$};
		\node [style={small_box}] (27) at (15, -0.5) {$D$};
		\node [style=none] (30) at (7.25, -0.5) {};
		\node [style=none] (33) at (7.25, 2) {};
		\node [style=none] (34) at (15.5, 2) {};
		\node [style=none] (35) at (15.5, -1.5) {};
		\node [style=none] (36) at (7.25, -1.5) {};
		\node [style=none] (44) at (15.5, 1) {};
		\node [style=none] (45) at (15.5, -0.5) {};
		\node [style=none] (46) at (16.75, 0.25) {$\displaystyle{=}$};
		\node [style={small_box}] (47) at (19, 1) {$e^{iHt}$};
		\node [style={small_box}] (48) at (21, 1) {$A$};
		\node [style=none] (51) at (17.75, 1) {};
		\node [style=none] (52) at (17.75, 2) {};
		\node [style=none] (53) at (22.5, 2) {};
		\node [style=none] (54) at (22.5, 1) {};
		\node [style={small_box}] (55) at (19, -0.5) {$e^{-iHt}$};
		\node [style={small_box}] (56) at (21, -0.5) {$B$};
		\node [style=none] (57) at (21.75, 1) {};
		\node [style=none] (58) at (21.75, -0.5) {};
		\node [style=none] (59) at (22.5, -0.5) {};
		\node [style={small_box}] (60) at (25, 1) {$e^{iHt}$};
		\node [style={small_box}] (61) at (27, 1) {$C$};
		\node [style=none] (62) at (23.75, 1) {};
		\node [style=none] (63) at (23.75, 2) {};
		\node [style=none] (64) at (28.5, 2) {};
		\node [style=none] (65) at (28.5, 1) {};
		\node [style={small_box}] (66) at (25, -0.5) {$e^{-iHt}$};
		\node [style={small_box}] (67) at (27, -0.5) {$D$};
		\node [style=none] (68) at (27.75, 1) {};
		\node [style=none] (69) at (27.75, -0.5) {};
		\node [style=none] (70) at (28.5, -0.5) {};
		\node [style=none] (71) at (17.75, -0.5) {};
		\node [style=none] (72) at (23.75, -1.5) {};
		\node [style=none] (73) at (23.75, -0.5) {};
		\node [style=none] (74) at (28.5, -1.5) {};
		\node [style=none] (75) at (17.75, -1.5) {};
		\node [style=none] (76) at (22.5, -1.5) {};
	\end{pgfonlayer}
	\begin{pgfonlayer}{edgelayer}
		\draw (12) to (13);
		\draw (13) to (14);
		\draw (14) to (15);
		\draw (22.center) to (12);
		\draw (24) to (25);
		\draw (25) to (26);
		\draw (26) to (27);
		\draw (30.center) to (24);
		\draw [bend right=90, looseness=1.25] (33.center) to (22.center);
		\draw [bend left=270, looseness=1.25] (30.center) to (36.center);
		\draw (33.center) to (34.center);
		\draw (36.center) to (35.center);
		\draw [bend left=90, looseness=1.50] (34.center) to (44.center);
		\draw (15) to (44.center);
		\draw (27) to (45.center);
		\draw [bend left=90, looseness=1.50] (45.center) to (35.center);
		\draw (47) to (48);
		\draw (51.center) to (47);
		\draw [bend right=90, looseness=1.25] (52.center) to (51.center);
		\draw (52.center) to (53.center);
		\draw [bend left=90, looseness=1.25] (53.center) to (54.center);
		\draw (58.center) to (54.center);
		\draw [style={white_edge}] (57.center) to (59.center);
		\draw (57.center) to (59.center);
		\draw (60) to (61);
		\draw (62.center) to (60);
		\draw [bend right=90, looseness=1.25] (63.center) to (62.center);
		\draw (63.center) to (64.center);
		\draw [bend left=90, looseness=1.25] (64.center) to (65.center);
		\draw (69.center) to (65.center);
		\draw [style={white_edge}] (68.center) to (70.center);
		\draw (68.center) to (70.center);
		\draw [bend right=90, looseness=1.25] (73.center) to (72.center);
		\draw (73.center) to (69.center);
		\draw (72.center) to (74.center);
		\draw [bend left=90, looseness=1.25] (70.center) to (74.center);
		\draw (61) to (68.center);
		\draw (75.center) to (76.center);
		\draw [bend left=270, looseness=1.25] (71.center) to (75.center);
		\draw [bend left=90, looseness=1.25] (59.center) to (76.center);
		\draw (48) to (57.center);
		\draw (56) to (58.center);
		\draw (71.center) to (55);
		\draw (55) to (56);
	\end{pgfonlayer}
\end{tikzpicture} 
\end{equation*}
where the Hamiltonian in the twofold Hilbert space will now be $\hat{H}=H \otimes \mathbb{I} - \mathbb{I}\otimes H$.

\subsubsection{A warm-up example: Haar plus replica }

Before applying the replica method directly in combination with the $C$-ensemble let us first present a warm-up example for the case of Haar distributed unitaries. For simplicity, let us consider an arbitrary two-point self correlation function $\braket{W(t)W}$. For this correlation, the replica method gives a one-point correlation in the two-fold Hilbert space,

\begin{equation}
  \Tr_{\mathcal{H}}\left( W(t)W\right)=\Tr_{\mathcal{H}\otimes \mathcal{H}}\left( e^{it\hat{H}_{(-)}} \hat{W} \,\text{SWAP}\right)
\end{equation}
with $\hat{H}_{(-)}=H\otimes \mathbb{I}-\mathbb{I}\otimes H$. However, the true power of the method emerges when in addition we average over some particular ensemble\footnote{For most ensembles lower moments are always easier to find than higher ones. Therefore, the replica method provides a recipe to require fewer moments at the cost of increasing the dimensionality.}, {\em e.g.} if the chosen ensemble is the unitary Haar distributed one, then the replica method suggests

\begin{equation}
  \Braket{\Tr\left(e^{itE}UWU^{\dagger}e^{-itE}UWU^{\dagger}\right)}_{U(d)}\rightarrow \Braket{\Tr_{\mathcal{H}\otimes \mathcal{H}}\left(e^{it\hat{E}_{(-)}}\hat{U}\hat{W}\hat{U}^{\dagger}\text{SWAP}\right)}_{U(d^{2})}.
\end{equation}
The right hand side involves only two-moments (see Appendix \ref{appendix:unitary}) and the ``naive'' average over $U(d^{2})$ simply yields

\begin{equation}
  \label{eq:replica_trick_bad_example}
  \frac{\Tr_{\mathcal{H}^{\otimes 2} }e^{it \hat{E}_{(-)}}}{d^2}\Tr_{\mathcal{H}^{\otimes 2} }\hat{W}\,\text{SWAP}=\frac{|Z(it)|^{2}}{d^{2}}\Tr{W^{2}}.
\end{equation}
Although one is tempted to think that \eqref{eq:replica_trick_bad_example} corresponds to $\braket{\Tr W(t)W}_{U(d)}$, this is not true. In fact the correct answer is \eqref{eq:C_two_points},

\begin{equation}
  \braket{\Tr{W(t)W}}_{U(d)}=\Tr{W^{2}}\left(\frac{|Z(it)|^{2}-1}{d^{2}-1} \right) +\frac{(\Tr W)^{2}}{d}\left(\frac{d^{2}-|Z(it)|^{2}}{d^{2}-1}\right),
\end{equation} 
and therefore we must be doing something wrong. Here we claim that indeed we are making a mistake, but it is an small one\footnote{It is an small one because the result \eqref{eq:replica_trick_bad_example} contains the correct leading form factor $|Z(it)|^{2}$ dependence expected for two-point correlations (see \cite{Cotler/spectral_decoupling/2020}). This can also be seen as an interesting approximation for two-points.}. The upshot is that in the replicated space we are doing the ensemble average badly by averaging over the whole $d^{2}$ twofold space $\mathcal{H}^{\otimes 2}=\mathcal{S}(\mathcal{H}^{\otimes 2})\oplus \mathcal{A}(\mathcal{H}^{\otimes 2})$ rather than averaging over the symmetric $\mathcal{S}(\mathcal{H}^{\otimes 2})$ and anti-symmetric $\mathcal{A}(\mathcal{H}^{\otimes 2})$ \emph{irreducible} sub-spaces separately. In fact, by doing the latter, we are able to recover the exact Haar averaged self correlator. This particular example for the unitary Haar ensemble motivates us to extrapolate the identification, 

\begin{equation}
\label{eq:Replica_correct_identification}
  \Braket{\Tr_{\mathcal{H}\otimes \mathcal{H}}\left(\cdots\right)}_{U(d^{2})} = \Braket{\Tr_{\mathcal{S}(\mathcal{H}^{\otimes 2})}\left(\cdots\right)}_{U\left(\frac{d(d+1)}{2}\right)} + \Braket{\Tr_{\mathcal{A}(\mathcal{H}^{\otimes 2})}\left(\cdots\right)}_{U\left(\frac{d(d-1)}{2}\right)}
\end{equation}
as the proper recipe to combine the replica space method with arbitrary unitarily eigenvector ensemble averages.

\subsection{Evaluating Out-Time-Ordered Correlators}
\label{sec:OTOCS}

\begin{figure}[ht]
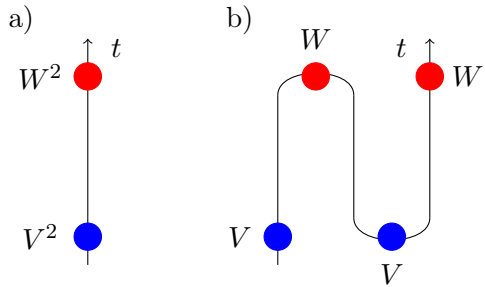

  \ctikzfig{Time_ordered_vs_OTOC}
    \caption{Time ordered vs out-time order correlation. For a time-ordered correlation (a) the operator insertions are in chronological order, {\em e.g.} $\braket{VVW(t)W(t)}$, whereas for the out-ordered case (b) the operator insertions wraps succeed in a non-chronological way, $\braket{VW(t)VW(t)}$. Notice that for the OTOC as time increases all operators become farther apart, whereas for the time-ordered, this does not happen.}
  \label{fig:OTOC_ordering}
\end{figure}

A standard fingerprint of classical chaotic systems is their sensitivity to small changes in the initial conditions. This phenomena, commonly referred as the ``butterfly effect'', is suitably captured by the Lyapunov exponent \cite{Lichtenberg/chaos_books/1992} which quantifies the divergence between nearly close trajectories in phase space. For quantum systems, because of unitary time evolution, one cannot characterize ``quantum chaos'' by looking into the overlap between two different initial states. However, one can extrapolate this expected dependence on the initial conditions by looking instead at the squared commutator between two local operators\footnote{Here we choose to probe the square commutator over the maximally mixed state. For finite temperature the discussion is similar.},

\begin{equation}
  \label{eq:square_commutator}
  C_{WV}(t)=-\Braket{[W(t),V]^{2}}
\end{equation}
The standard motivation behind this identification comes from systems with a classical analog, where in the semi-classical description the commutator is replaced by a Poisson bracket, {\em i.e.} $[,] \to \frac{1}{i\hbar}\{,\}+\mathcal{O}(\hbar)$ and in particular for position and momentum operators $-[q(t),p]^{2} \to \hbar^{2}e^{2\lambda_{L}t}$ (see \cite{Maldacena/bound_chaos/2015,Larkin/Original_OTOC/1969,Cotler/OTOC_wigner/2018}). By extrapolating the $p,q$ example to an arbitrary local pair of operators in chaotic quantum systems one would suspect a similar growth in \eqref{eq:square_commutator}. Even for systems without an obvious semi-classical limit, the interpretation of the square commutator as a probe of the butterfly effect arises naturally from the quantum information perspective where during recent years \eqref{eq:square_commutator} has become a direct link tool to identify chaos with information scrambling \cite{Swingle/Scrambling_otoc/2018,Swingle/Scrambling_tutorial/2022,Swingle/scrambling_initial_nature/2018,Couch/Chaotic_info_spreadign/2020,Mezei/entenglament_spreading_fields/2017}. After explicitly expanding the square in \eqref{eq:square_commutator} we get,

\begin{align}
\label{eq:square_commutator_expanded}
   C_{WV}(t)&=\braket{W(t)W(t)VV}+\braket{VVW(t)W(t)}\\
   &-\braket{W(t)VW(t)V}-\braket{VW(t)VW(t)}, \nonumber
\end{align}
where two of the four resulting terms yield time-ordered correlations whereas the remaining two yield out-time-order correlations (OTOC) such as $\braket{VW(t)VW(t)}$ (see Fig.\ref{fig:OTOC_ordering}). 

To explain the role of such terms consider that the $V,W$ operators are those of a local system, {\em e.g.} such as in a lattice model. For short times, and $W$ sufficiently spatially separated from $V$, both operators approximately commute, $[W(t),V]^{2}\approx 0$, and $C_{WV}(t)$ is small. As time increases, due to chaotic time evolution, all the operators (right figure in Fig.\ref{fig:OTOC_ordering}) in the out-ordered case become progressively farther and farther apart making $\braket{W(t)VW(t)V}$ to decrease. Similarly, for the time-ordered case,  as time increases the correlators $\Braket{W(t)W(t)VV}$ also decay due to the separation between the operators. However, in contrast to the out-ordered case, this decay is much slower because only $W^{2}$ and $V^{2}$ are separated (left figure in Fig.\ref{fig:OTOC_ordering}). The imbalance between both, the out-ordered and the ordered contributions in \eqref{eq:square_commutator_expanded} makes $C_{WV}(t)$  increase.

For the OTOC, as it decays with time, the late time major contribution is expected to be given by the leading Wick contraction $\braket{W^{2}}\braket{V^{2}}$, because the sub-leading one $\Braket{WV}\Braket{WV}$ is suppressed at late times due to the separation between $W$ and $V$.

\subsubsection{Ensemble average OTOC}

Having gained insights about the expected OTOC behavior, let us consider the infinite temperature OTOC average over the $C$-ensemble,

\begin{equation}
  \Braket{\Tr\left( W(t)VW(t)V\right)}_{\mathcal{E}_{C}}
\end{equation}
By the replica space method, the non-averaged OTOC can be set as a two-point correlation over the twofold Hilbert space,

\begin{equation*}
  \begin{tikzpicture}
	\begin{pgfonlayer}{nodelayer}
		\node [style={small_box}] (0) at (-7, 0) {$e^{iHt}$};
		\node [style={small_box}] (1) at (-5, 0) {$W$};
		\node [style={small_box}] (2) at (-2.75, 0) {$e^{-iHt}$};
		\node [style={small_box}] (3) at (-0.5, 0) {$V$};
		\node [style={small_box}] (4) at (1.5, 0) {$e^{iHt}$};
		\node [style={small_box}] (5) at (3.5, 0) {$W$};
		\node [style={small_box}] (6) at (5.75, 0) {$e^{-iHt}$};
		\node [style={small_box}] (7) at (8, 0) {$V$};
		\node [style=none] (8) at (-7.75, 1) {};
		\node [style=none] (9) at (8.5, 1) {};
		\node [style=none] (10) at (-7.75, 0) {};
		\node [style=none] (11) at (8.5, 0) {};
		\node [style={small_box}] (12) at (11.75, 1) {$e^{iHt}$};
		\node [style={small_box}] (13) at (13.75, 1) {$W$};
		\node [style={small_box}] (14) at (16, 1) {$e^{-iHt}$};
		\node [style={small_box}] (15) at (18.25, 1) {$V$};
		\node [style=none] (22) at (10.75, 1) {};
		\node [style={small_box}] (24) at (11.75, -0.5) {$e^{iHt}$};
		\node [style={small_box}] (25) at (13.75, -0.5) {$W$};
		\node [style={small_box}] (26) at (16, -0.5) {$e^{-iHt}$};
		\node [style={small_box}] (27) at (18.25, -0.5) {$V$};
		\node [style=none] (28) at (18.75, -0.5) {};
		\node [style=none] (29) at (18.75, 1) {};
		\node [style=none] (30) at (10.75, -0.5) {};
		\node [style=none] (31) at (20.25, 1) {};
		\node [style=none] (32) at (20.25, -0.5) {};
		\node [style=none] (33) at (10.75, 2) {};
		\node [style=none] (34) at (20.25, 2) {};
		\node [style=none] (35) at (20.25, -1.5) {};
		\node [style=none] (36) at (10.75, -1.5) {};
		\node [style=none] (37) at (9.75, 0.25) {$\displaystyle{=}$};
	\end{pgfonlayer}
	\begin{pgfonlayer}{edgelayer}
		\draw (0) to (1);
		\draw (1) to (2);
		\draw (2) to (3);
		\draw (4) to (5);
		\draw (5) to (6);
		\draw (6) to (7);
		\draw (3) to (4);
		\draw [bend left=90, looseness=1.25] (9.center) to (11.center);
		\draw [bend left=270, looseness=1.25] (8.center) to (10.center);
		\draw (10.center) to (0);
		\draw (8.center) to (9.center);
		\draw (7) to (11.center);
		\draw (12) to (13);
		\draw (13) to (14);
		\draw (14) to (15);
		\draw (22.center) to (12);
		\draw (24) to (25);
		\draw (25) to (26);
		\draw (26) to (27);
		\draw (27) to (28.center);
		\draw (15) to (29.center);
		\draw (30.center) to (24);
		\draw [in=180, out=0, looseness=1.25] (29.center) to (32.center);
		\draw [style={white_edge}, in=180, out=0, looseness=1.25] (28.center) to (31.center);
		\draw [in=180, out=0] (28.center) to (31.center);
		\draw [bend right=90, looseness=1.25] (33.center) to (22.center);
		\draw [bend left=270, looseness=1.25] (30.center) to (36.center);
		\draw [bend left=90, looseness=1.25] (32.center) to (35.center);
		\draw [bend left=90, looseness=1.25] (34.center) to (31.center);
		\draw (33.center) to (34.center);
		\draw (36.center) to (35.center);
	\end{pgfonlayer}
\end{tikzpicture}
\end{equation*}
However, when one combines the replica method with some ensemble average, we learned in Section \ref{sec:replica_method} that we must apply it carefully and perform the ensemble average over the symmetric $\mathcal{S}(\mathcal{H}^{\otimes 2})$ and anti-symmetric $\mathcal{A}(\mathcal{H}^{\otimes 2})$ subspaces separately,

\begin{align*}
  \braket{\Tr{W(t)V W(t)V}}_{\mathcal{E}_{C}}&= \\
  \Braket{\Tr_{\mathcal{S}(\mathcal{H}^{\otimes 2})}\left( e^{it \hat{H}}\hat{W}e^{-it \hat{H}} \hat{V}\,\text{SWAP}\right)}_{\mathcal{E}_{C}^{+}}&+
  \Braket{\Tr_{\mathcal{A}(\mathcal{H}^{\otimes 2})}\left( e^{it \hat{H}}\hat{W}e^{-it \hat{H}} \hat{V}\,\text{SWAP}\right)}_{\mathcal{E}_{C}^{-}}.
\end{align*}
Similarly, as in the Haar unitary example in Section \ref{sec:replica_method} every $C$-ensemble in the above identification is different. While $\mathcal{E}_{C}$ corresponds to the $d$-dimensional $C$-ensemble associated with the Hamiltonian $H$ in $\mathcal{H}$, $\mathcal{E}_{C}^{+}$ and 
$\mathcal{E}_{C}^{-}$ are the two $C$-ensembles associated with the projected Hamiltonians $\hat{H}_{+}=S_{+}HS_{+}$ for the $d(d+1)/2$-dimensional symmetric subspace $\mathcal{S}(\mathcal{H}^{\otimes 2})$ and $\hat{H}_{-}=S_{-}HS_{-}$ $d(d-1)/2$-dimensional anti-symmetric subspace $\mathcal{A}(\mathcal{H}^{\otimes 2})$. It is worth remarking that the relevance of this result comes from  noticing that we can use all our previously gained knowledge on averaged two-point functions (see Section \ref{Chapter:two_points}) to examine the behavior of OTOC's. 

From the single Hilbert space perspective, two-point functions in the twofold Hilbert space contribute with four-point terms as,

\begin{align}
\label{eq:Full_OTOC}
   \Braket{\Tr W(t) V W(t) V}_{\mathcal{E}_{C}} &= \frac{\Tr{WVWV}}{2}\left( \frac{|Z_{+}(it)|^{2}}{D_{+}(D_{+}-1)}+\frac{|Z_{-}(it)|^{2}}{D_{-}(D_{-}-1)}-\frac{1}{D_{+}-1}-\frac{1}{D_{-}-1} \right)\\ \notag
   &+\frac{\Tr{WV}\Tr{WV}}{2}\left( \frac{|Z_{+}(it)|^{2}}{D_{+}(D_{+}-1)}-\frac{|Z_{-}(it)|^{2}}{D_{-}(D_{-}-1)}-\frac{1}{D_{+}-1}+\frac{1}{D_{-}-1} \right)\\
   &+\Tr_{\mathcal{S}(\mathcal{H}^{\otimes 2})}\left(G^{\mathcal{E}_{C}}[\hat{H}_{+}]\left(\hat{W}\otimes \hat{V}\text{SWAP}\right)\right)\left( \frac{D_{+}}{D_{+}-1}-\frac{|Z_{+}(it)|^{2}}{D_{+}(D_{+}-1)}\right)\\ \notag
   &+\Tr_{\mathcal{A}(\mathcal{H}^{\otimes 2})}\left(G^{\mathcal{E}_{C}}[\hat{H}_{-}]\left(\hat{W}\otimes \hat{V}\text{SWAP}\right)\right)\left( \frac{D_{-}}{D_{-}-1}-\frac{|Z_{-}(it)|^{2}}{D_{-}(D_{-}-1)}\right).
\end{align}
For either $\mathcal{H}$ or $\mathcal{H}^{\otimes 2}$, the $C$-ensemble average will always lead to spectral decoupled two-point correlation functions with the time dependence fixed by some form factor. From the twofold Hilbert space perspective, the form factor is either $|Z_{+}(it)|^{2}=\Tr_{\mathcal{S}(\mathcal{H}^{\otimes 2})} e^{i t \hat{H}}$, for the symmetric subspace or, $|Z_{-}(it)|^{2}=\Tr_{\mathcal{A}(\mathcal{H}^{\otimes 2})} e^{i t \hat{H}}$, for the anti-symmetric one. However, from the single Hilbert space point of view, each of those form factors is indeed function of the single partition function $Z(it)=\Tr_{\mathcal{H}}e^{itH}$, \emph{i.e.},

\begin{equation}
\label{eq:sym_anti_partition_functions}
    |Z_{\pm}(it)|^{2}=\left|\frac{Z(it)^{2}\pm Z(2it)}{2}\right|^{2}=\frac{|Z(it)|^{4}+|Z(2it)|^{2}\pm2Re[Z(it)^{2}Z(-2it)]}{4},
\end{equation}
and therefore one expects to retrieve the much more fine-tuned time scales involved in the OTOC\footnote{As a sanity check, notice that indeed from \eqref{eq:sym_anti_partition_functions} the first contribution is given by the ``square" form factor $|Z(it)|^{4}$ just as the Haar average of OTOCS \cite{Cotler/spectral_decoupling/2020}.}. Indeed, from \eqref{eq:Full_OTOC} we can identify in advance the expected short time limit given by the connected term $\Tr{WVWV}$ contribution. To gain insight into the further time scales involved in the OTOC (and also for simplicity) let us consider from now on the large-$d$ limit,

\begin{align}
   \Braket{\Tr W(t) V W(t) V}_{\mathcal{E}_{C}} &= \Tr{WVWV}\left( \frac{|Z_{+}(it)|^{2}+|Z_{-}(it)|^{2}}{2d^{4}}\right)\\ \nonumber
   &+\Tr{WV}\Tr{WV}\left( \frac{|Z_{+}(it)|^{2}-|Z_{-}(it)|^{2}}{2d^{4}}\right)\\ \nonumber
   &+\Tr_{\mathcal{S}(\mathcal{H}^{\otimes 2})}\left(G^{\mathcal{E}_{C}}[\hat{H}_{+}]\left(\hat{W}\otimes \hat{V}\text{SWAP}\right)\right)\left( 1-\frac{|Z_{+}(it)|^{2}}{d^{4}}\right)\\ \nonumber
   &+\Tr_{\mathcal{A}(\mathcal{H}^{\otimes 2})}\left(G^{\mathcal{E}_{C}}[\hat{H}_{-}]\left(\hat{W}\otimes \hat{V}\text{SWAP}\right)\right)\left( 1-\frac{|Z_{-}(it)|^{2}}{d^{4}}\right).
\end{align}
In this limit, the OTOC displays similarities with the two-point case at the late time behavior (see Section \ref{Chapter:two_points}), \emph{i.e.}, as for late times the normalized form factor $|Z(it)|^{2}/d^{2}$ decays to zero, the two-point correlations progressively approach to their infinite time average given by the action of the plateau operator.  Now for the averaged OTOC case, both, symmetric and anti-symmetric form factors are bounded by $d^{4}$ (as well as in Section \ref{Chapter:two_points} one has to take care of degeneracies) and therefore $|Z_{\pm}(it)|^{2}/d^{4}$ decays towards zero (clearly not in the same way as $|Z(it)|^{2}/d^{2}$) contributing to a late time behavior of the OTOC given by\footnote{Recall that for two-points the action of the plateau operator can be identified with the long time average of such correlations, in that sense it was expected that the twofold plateau operators will also be related with long time behaviors.},

\begin{align}
     \overline{\Braket{\Tr W(t) V W(t) V}_{\mathcal{E}_{C}}}& \approx \Tr_{\mathcal{S}(\mathcal{H}^{\otimes 2})}\left(G^{\mathcal{E}_{C}}[\hat{H}_{+}]\left(\hat{W}\otimes \hat{V}\text{SWAP}\right)\right)\left( 1-\frac{|Z_{+}(it)|^{2}}{d^{4}}\right)\\ \nonumber
   &+\Tr_{\mathcal{A}(\mathcal{H}^{\otimes 2})}\left(G^{\mathcal{E}_{C}}[\hat{H}_{-}]\left(\hat{W}\otimes \hat{V}\text{SWAP}\right)\right)\left( 1-\frac{|Z_{-}(it)|^{2}}{d^{4}}\right).
\end{align}
For short times on the other hand, the leading contribution is $\Tr{WVWV}$, while the sub-leading disconnected one $\Tr{WV} \Tr{WV}$. The distinction between leading and sub-leading contributions at earlier times is physically justified because for strongly correlated quantum systems, one expects the latter contribution to decay\footnote{For the $C$-ensemble, this becomes clear from the beginning because as $d$ increase, both $|Z_{+}(it)|^{2}$ and $|Z_{-}(it)|^{2}$ become closer to each other, and the minus sign makes the contraction $\Tr{WV}^{2}$ contribution diminish.}.

\begin{figure}[ht]
  \begin{subfigure}{.5\textwidth}
    \centering
    \includegraphics[width=.85\linewidth]{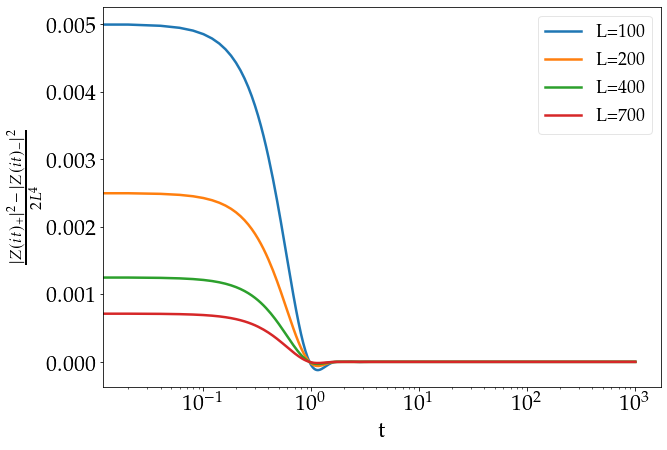}  
    \caption{Symmetric $|Z_{+}(it)|^{2}$.}
    \end{subfigure}
  \begin{subfigure}{.5\textwidth}
    \centering
    \includegraphics[width=.85\linewidth]{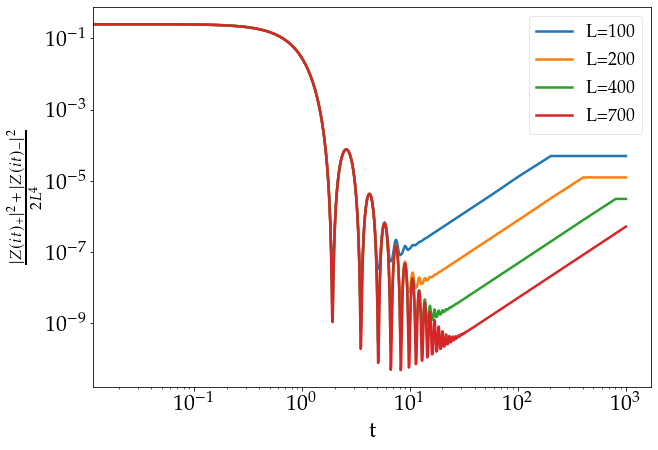}  
    \caption{Anti-Symmetric $|Z_{+}(it)|^{2}$.}
   \end{subfigure}
  \caption{Symmetric $|Z_{+}(it)|^{2}$ (a) and anti-symmetric $|Z_{-}(it)|^{2}$ (b) twofold form factors for the Gaussian unitary ensemble. As $d$-increases, the difference between symmetric and anti-symmetric form factors (left panel), $|Z_{+}(it)|^{2}-|Z_{-}(it)|^{2}$ ,decreases, and the contribution from the contraction $\Tr{WV} \Tr{WV}$ to the OTOC becomes negligible at all times. On the other hand, the short time behavior for the OTOC is dominated by the $\Tr{WVWV}$ term which contributes with a time dependence given by the sum between symmetric and anti-symmetric form factors $|Z_{+}(it)|^{2}+|Z_{-}(it)|^{2}$(right panel). In the large-$d$ limit, the leading contribution to the symmetric (anti-symmetric) form factors is provided by $|Z(it)|^{4}$. Therefore, we expect a similar dissipation time scale as the one in $|Z(it)|^{2}$ \cite{Cotler/spectral_decoupling/2020}.   }
  \label{fig:sym_antisym_form_factors}
  \end{figure}
Here we want to emphasize that $G^{\mathcal{E}_{C}}[\hat{H}_{+}]$ and $G^{\mathcal{E}_{C}}[\hat{H}_{-}]$ are far more complex than their single Hilbert space analog $G^{\mathcal{E}_{C}}[H]$. First, they act as maps over the twofold symmetric sub-spaces instead the twofold Hilbert space, {\em i.e.,} $G^{\mathcal{E}_{C}}[\hat{H}_{+}]:\mathcal{S}(\mathcal{H}^{\otimes 2})^{\otimes 2}\to \mathcal{S}(\mathcal{H}^{\otimes 2})^{\otimes 2}$ and $G^{\mathcal{E}_{C}}[\hat{H}_{-}]:\mathcal{A}(\mathcal{H}^{\otimes 2})^{\otimes 2}\to \mathcal{A}(\mathcal{H}^{\otimes 2})^{\otimes 2}$. Second, from the interpretation of the plateau operator as the $s$-state in energy-eigenbasis for the bipartite system $\mathcal{H}\otimes \mathcal{H}$ ( see Appendix \ref{Appendix:Scrambling}), the twofold symmetric plateau operators (from the single Hilbert space point of view) must then correspond to a $s$-state in the four-fold space $\mathcal{H}^{\otimes 4}$.

A non-straightforward computation when dealing with OTOCs is the ability to identify the expected leading Wick contraction $\Tr{W^{2}}\Tr{V^{2}}$ \footnote{This is closely related to the chaos bound \cite{Maldacena/bound_chaos/2015}, and will be discussed in a forthcoming paper.}. This is an straightforward task when the unitary dynamics is modeled by the Haar-ensemble because from the beginning one is imposing highly non-local interactions. However, being able to proof that a given $C$-ensemble generated by the fixed (and local) Hamiltonian $H$ will yield the expected leading-contraction contributing to the disconnected part of the OTOC is more subtle. To address this, we will make use again of the natural decomposition of the plateau operator \eqref{eq:dephasing_factorization}, but this time used on the symmetric and anti-symmetric sub-spaces independently, \emph{i.e.,}

\begin{align}
 G^{\mathcal{E}_{C}}_{+}[\hat{H}_{+}] & =G^{U(\frac{d(d+1)}{2})}+G[\hat{H}_{+}], \\
 G^{\mathcal{E}_{C}}_{-}[\hat{H}_{-}] & =G^{U(\frac{d(d-1)}{2})}+G[\hat{H}_{-}].
\end{align}
Similar to the two-point case, the universal $G^{U(\frac{d(d+1)}{2})}$, and $G^{U(\frac{d(d-1)}{2})}$ parts of the plateau operator will yield the Haar contribution for the OTOC; while the non-universal contribution to the plateau operator due to the local $H$ will split, \emph{e.g.},

\begin{equation}
 \Braket{W(t)V W(t)V}_{\mathcal{E}^{C}}=\Braket{W(t)VW(t)V}_{U(d)}+\Braket{W(t)VW(t)V}_{H}.
\end{equation}

\subsection{\texorpdfstring{$C$}{}-ensemble Complexity}\label{sub:C_ensemble_complexity}

Until now, we have mainly focused our attention on comparing the $C$-ensemble with the unitary Haar one but only at the statistical level, that is, by looking at their respective moment operators and comparing the predictions made for correlation functions by both. However, we have still not made any comments about what would be resource difference needed in order to build (at least theoretically) both ensembles. By identifying the resources with an available ``easy to implement'' set unitaries, a particular way to capture this resource dependence is given by the ensemble complexity (see \cite{Yoshida/Chaos_design/2017}) defined as the number of ``easy to implement'' unitaries needed for build up the whole ensemble.

\begin{figure}[ht]
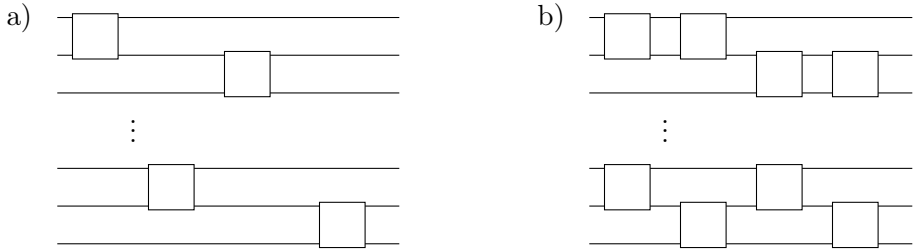

  \ctikzfig{Circuit_example}
  \caption{Equivalent notions of ensemble complexity in unitary circuits. For a circuit architecture where only a single unitary is applied per step (a), there are $|\mathcal{G}|\binom{n}{q}$ choices which after $\mathcal{C}$ steps yield in $\left(|\mathcal{G}|\binom{N}{q}\right)^{\mathcal{C}}$ different circuits. For a circuit architecture that allows gates to run in parallel (b) there are $|\mathcal{G}|^{l}\binom{N}{q}\binom{N-q}{q}\binom{N-2q}{q}\dots\binom{N-lq}{q}$ choices each step, where $l$ is the number of parallel gates.
The parallel architecture can be emulated by the non-parallel one by increasing the circuit length of the latter. Therefore, the ensemble complexities defined either with one underlying architecture or the other are related.}
  \label{fig:circuit_architectures}
\end{figure}

Following the setup in \cite{Yoshida/Chaos_design/2017}, suppose we have a universal $q$-qubit local gate set $\mathcal{G}=\{g_{1},g_{2},\dots,g_{|\mathcal{G}|}\}$\footnote{For $q=2$ this can be the 2-qubit Cliffords and the $T$-gate.} acting on a system of $N$ qubits, \emph{i.e}, $\mathcal{H}\cong \mathbb{C}^{N}$ and $d=2^{N}$. If we only apply one possible gate at each step (see Fig.\ref{fig:circuit_architectures}), the number of different circuits of length $\mathcal{C}$ is given by,

\begin{equation}
 \label{eq:number_circuits}
 \# \text{circuits}=\left(|\mathcal{G}|\binom{N}{q}\right)^{\mathcal{C}},
\end{equation}
where $|\mathcal{G}|$ denotes the cardinality of $\mathcal{G}$. For a finite ensemble of unitaries, the ensemble complexity is defined as the number of steps $\mathcal{C}$ necessary to obtain the ensemble $\mathcal{E}$. From a simple (but clever) counting argument, generating all the ensemble elements $|\mathcal{E}|$ requires that the number of circuits \eqref{eq:number_circuits} must be greater than $|\mathcal{E}|$ and therefore the ensemble complexity is lower bounded by,

\begin{equation}
 \label{eq:complexity_bound}
 \mathcal{C}(\mathcal{E}) \geq \frac{\log{|\mathcal{E}|}}{\log{|\mathcal{G}|\binom{N}{q}}},
\end{equation}
with $\log{|\mathcal{E}|}$ identified as the thermodynamic entropy associated with the ensemble of unitaries. For the $C$-ensemble, identified as the equivalence class $\{ C\sim (U(1)^{d} \times S_{d})\,\,C \}$ of unitaries, the ensemble complexity will be bounded in the large-$d$ limit by\footnote{Here we omit the $U(1)^{d}$ term which contributes with an additional $(2\pi)^{d}$ inside the logarithm. However this factor is subleading and will not change the final conclusion.},

\begin{equation}
 \label{eq:dclass_complexity}
 \mathcal{C}(\mathcal{E}_{C}) \geq \frac{d\log{d}-d}{\log{|\mathcal{G}|\binom{N}{q}}}.
\end{equation}
However, there is something suspicious about this result. First, the fact that apparently holds for every Hamiltonian. Second, even if \eqref{eq:complexity_bound} is not a tight lower bound, one expects at least a distinction between the complexity of a $C$-ensemble associated to a many-body chaotic Hamiltonian and the complexity of the $C$-ensemble for a many-body integrable one (provided the appropriate gate set $\mathcal{G}$), which is clearly not reflected by \eqref{eq:dclass_complexity}.
The reason behind why this bound is not completely accurate goes as follows: As mentioned in Section \ref{section:C_ensemble} (see Fig. \ref{fig:C_ensemble_diagram}), the $C$-ensemble can be identified with the orbit generated by the $U(1)^{d} \times S_{d}$ group action on a unitary $C$, such that $CHC^{\dagger}$ is diagonal.
However, not all the $U(1)^{d} \times S_{d}$ orbits of unitaries are equal, and we expect that some of them will be more complex to generate than others. This distinction between the complexity of different sets of orbits is originated form the individual Hamiltonian $H$ itself, \emph{i.e.}, once an ``initial'' unitary $C$ that makes $H$ diagonal is fixed, the corresponding $U(1)^{d} \times S_{d}$- orbit will be completely determined. Therefore, by properly taking into account the complexity of this initial unitary, it is expected that the bound \eqref{eq:dclass_complexity} becomes system dependent.

To include the system dependence in the complexity bound we have to count the number of unitaries $|\mathcal{E}_{C}(H)|$ in the $C$-ensemble, \emph{i.e.}, the ensemble cardinality. Formally, the $C$-ensemble has an infinite number of unitaries, due to the continuous $U(1)^{d}$ part, and in order to get a meaningful result we will need to regularize this value by doing some epsilon-ball counting. Here, we will follow two different approaches: First, we will use the result presented in \cite{Yoshida/Chaos_design/2017} to bound the ensemble cardinality in terms of the frame potential (see \eqref{subsec:frame_potential}). Second, we will make use of the $C$-ensemble partition function to count the cardinality.

\subsubsection*{$C$-ensemble complexity and two-frame potential}
 
Following \cite{Yoshida/Chaos_design/2017}, for continuous ensembles such as the $C$-ensemble or Haar-random distributed unitaries, the complexity lower bound \eqref{eq:complexity_bound} has to be replaced with,

\begin{equation}
 \label{eq:bound_frame_potential}
 \mathcal{C}(\mathcal{E}) \geq \frac{2k \log{d}-\log{F^{\mathcal{E}}_{k}}}{\log{|\mathcal{G}|\binom{N}{q}}}.
\end{equation}
The proof behind this relies in the fact that the frame potential of a unitary ensemble serves as a lower-bound of the ensemble-cardinality (see \cite{Yoshida/Chaos_design/2017}). For the particular $C$-ensemble case, we already know the exact two-frame potential \eqref{eq:2frame_potential} and the complexity bound reads,

\begin{equation}
 \label{eq:pre_complexity_bound}
 \mathcal{C}(\mathcal{E}_{C}) \geq \frac{4 \log{d}-\log{\left(F_{2}^{U(d)}+\left(\frac{d+1}{(d-1)}(\overline{\text{IPR}}-\frac{2}{d+1})\right)^{2}\right)}}{\log{|\mathcal{G}|\binom{N}{q}}},
\end{equation}
with $\overline{\text{IPR}}$ the average inverse participation ratio introduced in Section \ref{subsec:frame_potential} \footnote{As the average IPR is bounded by $d^{-1}\leq \overline{\text{IPR}}\leq 1$, the lower bound inside \eqref{eq:pre_complexity_bound} is itself bounded.}. Although the bound \eqref{eq:pre_complexity_bound} explicitly depends on the Hamiltonian $H$, in the large-$d$ limit this bound is only able to recover a sub-leading $\mathcal{O}(\log d)$ complexity growth term compared with the expected $\mathcal{O}(d\log d)$ growth in \eqref{eq:dclass_complexity}, the original bound obtained due to the contribution from the $C$-ensemble $S_{d}$ symmetry. The fact that \eqref{eq:pre_complexity_bound} is much less tight compared to \eqref{eq:dclass_complexity}, can be traced back to \eqref{eq:bound_frame_potential}, where in order to obtain the $\mathcal{O}(d\log(d))$ complexity growth term, we must have used the $\frac{d}{2}$th $C$-ensemble frame potential instead. For large Hilbert spaces, attempting to give an analytical estimate of such a higher frame potential is out of the scope of this work\footnote{We would need the $\mathcal{O}(d)$-moment operators, and we struggle to exactly find the $8$th ones. Even for the Haar unitary ensemble case, as we mentioned before, $\mathcal{O}(d)$ moments contribute with $\mathcal{O}((d!)^{2})$ terms!}.

\subsubsection*{$C$-Complexity and partition function}

As mentioned in the last section, a good lower complexity bound for the $C$-ensemble must depend explicitly on the Hamiltonian, and also must include the $\mathcal{O}(d\log{d})$ complexity growth due the ensemble $S_{d}$ symmetry. In particular, this is where the \eqref{eq:pre_complexity_bound} bound fails. The whole problem as been that the bound \eqref{eq:complexity_bound} (the one in which the $S_{d}$ term appears) is proportional to the cardinality $|\mathcal{E}|$ of the unitary ensemble, and in the frame potential based one \eqref{eq:bound_frame_potential} we are implicitly replacing this cardinality with a lower estimation of it.\footnote{In fact, this lower estimation is so poor that we would need a frame potential of the order of the Hilbert space dimension term in order to provide a good estimate of the bound. } To solve this issue, we argue here, that the cardinality $|\mathcal{E}_{C}|$ of the $C$-ensemble associated to an individual Hamiltonian $H$ can be obtained directly from the ensemble volume $\Vol(H)$, a.k.a the partition function (see Appendix \ref{appendix:partition_function}) via an epsilon counting, \emph{i.e.}

\begin{equation}
 \label{eq:C_ensemble_states}
 |\mathcal{E}_{C}(H)|_{\epsilon} \propto \frac{\Vol(H)}{\Vol(B_{\epsilon}(d^{2}))},
\end{equation}
where the proportionality factor will be fixed short after, and $\Vol(B_{\epsilon}(d^{2}))$ denotes the volume of an $\epsilon$-ball\footnote{For an $\epsilon$-ball of dimension $2n$, $\Vol(B_{\epsilon}(2n))=\frac{\pi^{n}}{n!}\epsilon^{2n}$.} of dimension $d^2$. This type of counting argument has been used in \cite{susskind/Three_lectures_Black_Holes/2020} to estimate the number of unitaries $|U(d)|_{\epsilon}$ in the Haar ensemble, \emph{i.e.}

\begin{equation}
 \label{eq:Vol_unitary_group}
 |U(d)|_{\epsilon} \coloneqq \frac{\Vol(U(d))}{\Vol(B_{\epsilon}(d^{2}))} = \frac{(2\pi)^{d+\binom{d}{2}}}{\prod_{l=1}^{d-1}l!}\frac{1}{\Vol(B_{\epsilon}(d^{2}))}\sim \mathcal{O}(e^{\frac{d^{2}}{2}\log{d}})
\end{equation}
where $\Vol(U(d))$ denotes the volume of the unitary group \cite{Zhang/Unitary_volume/2017,Zyczkowsky/Unitary_volume/2003}.

The motivation to interpret the ratio in \eqref{eq:C_ensemble_states} as the cardinality of the $C$-ensemble comes from the definition of the ensemble measure itself \emph{i.e.} the $\delta_{\perp} (CHC^{\dagger})$ term inside \eqref{eq:C_ensemble_Measure} is an uniform counting measure. Therefore, we expect that after integrating over the whole unitary group in \eqref{app_partition_func:eq_vol} we obtain something that is proportional to the total number of unitaries inside the $C$-ensemble associated with the Hamiltonian $H$. However, we desire to emphasize that although the previous counting argument is natural, \eqref{eq:C_ensemble_states} does not completely correspond to the number of unitaries inside the $C$-ensemble. The reason is that the partition function $\Vol(H)$ has dimensions of $[\text{Energy}]^{-2\binom{d}{2}}$ and a truly valid number of unitaries must be dimensionless. To solve this issue, we use the following trick: All the ensemble moments $\hat{\Phi}_{k}^{\mathcal{E}_{C}}$, and therefore all ensemble averaged correlation functions, are invariant under a global energy scaling, \emph{e.g.}, $H \to \lambda H$ with $\lambda$ real. Therefore, we can replace the Hamiltonian $H$ in \eqref{app_partition_func:eq_vol}
by a suitable dimensionless one, {\emph i.e.},

\begin{equation}
\label{eq:proper_energy}
 H\to \frac{H}{[\text{some energy scale}]},
\end{equation}
and all the $C$-ensemble results presented so far would remain unchanged. As the canonical energy scale for a Hamiltonian is given by the mean level spacing $\overline{\Delta E}$, we will use $\overline{\Delta E}$ to turn \eqref{eq:C_ensemble_states} dimensionless\footnote{For other choices of the energy scale, our final conclusions will still hold on up to some proportionality pre-factors.}.

Now, in order to fix the proportionality inside \eqref{eq:C_ensemble_states}, notice that the for arbitrary Hamiltonians we expect that the number of unitaries of the associated $C$-ensemble grows much slower than the number of Haar unitaries, \emph{i.e.}

\begin{equation}
\label{eq:C_ensemble_vs_unitary_ratio}
 \frac{|\mathcal{E}_{C}|_{\epsilon}}{|U(d)|_{\epsilon}} \ll 1.
\end{equation}
If this was not the case, then, we could use an arbitrary Hamiltonian to construct a $C$-ensemble comparable to the unitary group, and obviously, this cannot be true\footnote{Except for the Identity operator which is trivially diagonalized by all $U(d)$.}. For a non-degenerate\footnote{In Appendix \ref{appendix:partition_function} we make some comments about the degenerate case as well.} Hamiltonian, the $C$-ensemble partition function exactly yields (see Appendix \ref{appendix:partition_function}),

\begin{equation}
\label{eq:final_form_C_ensemble volume}
 \Vol(H)= 2^{\binom{d}{2}}\frac{\Vol{\left( U(1)^{d}\times S_{d}\right)}}{\Vol{U(d)}}\prod_{l< m}\left(\frac{\overline{\Delta E}}{E_{m}-E_{l}}\right)^{2},
\end{equation}
where $2^{\binom{d}{2}}$ is related to the universality class of the Hamiltonian, and the $\Vol(U(1)^{d}\times S_{d})$ term comes from the multiplicity of the $U(1)^{d}\times S_{d}$ -orbit that will give rise to the expected $\log(d!)$ contribution to the complexity bound present in \eqref{eq:dclass_complexity} but missing in \eqref{eq:pre_complexity_bound}. In a crude approximation, we can estimate the contribution from the inverse Vandermonde term inside \eqref{eq:final_form_C_ensemble volume} by neglecting all the level-fluctuations and setting the energy levels to be equally spaced, \emph{i.e.}

\begin{equation}
\label{eq:Harmonic_oscillator}
 E_{l}-E_{m}=(l-m)\overline{\Delta E}.
\end{equation}
For this particular case, the associated $C$-ensemble partition function exactly gives

\begin{equation}
 \Vol(H)|_{\text{Equally-spaced}}=\frac{d!}{\pi^{\binom{d}{2}} \prod_{l=1}^{d-1}l!}.
\end{equation}
If $|\mathcal{E}_{C}(H)|$ was directly equal to the ratio between $\Vol(H)$ and $\Vol(B_{\epsilon}(d^2))$, \emph{i.e.}, the proportionality factor inside \eqref{eq:C_ensemble_states} was one, then the number of unitaries of the associated $C$-ensemble would grow as $|\mathcal{E}_{C}|_{\text{Equally-spaced}}\sim \mathcal{O}(e^\frac{d^2}{2}\log d) \sim \mathcal{O}(|U(d)|)$ which contradicts \eqref{eq:C_ensemble_vs_unitary_ratio}. But, if we set instead 

\begin{equation}
\label{eq:Final_ensmeble_cardinality}
    |\mathcal{E}_{C}(H)|_{\epsilon}=\Vol(U(d))\, \frac{\Vol(H)}{\Vol(B_{\epsilon}(d^{2}))},
\end{equation}
then $|\mathcal{E}_{C}|_{\text{Equally-spaced}} \sim \mathcal{O}(e^{d^{2}})$ which seems to have more sense as far as it does not contradicts \eqref{eq:C_ensemble_vs_unitary_ratio}. The motivation for setting the proportionality pre-factor equal to the volume of the unitary group can be traced back to \eqref{app_partition_func:eq_vol}, where inside the definition of the $C$-ensemble volume we are using the \emph{normalized} Haar measure. Therefore, by integrating over the unitary group, we will, instead of counting the number of unitaries, compute the density with respect to the total number of Haar unitaries. We thus define the number of unitaries as \eqref{eq:Final_ensmeble_cardinality}.
Before continuing, let us remark some points about the particular Hamiltonian dependence inside the $C$-ensemble volume:

\begin{itemize}
\item First, the fact that the volume of the $C$-ensemble only cares about the level statistics rather than the local basis of the underlying Hamiltonian $H$ may seem strange at first. However, this can be easily explained as follows: Given two $C$-ensembles, one generated by $H$ and the other by $UHU^{\dagger}$ with $U \in U(d)$, the number of unitaries in both ensembles is the same (provided that $H$ is not diagonal in the first place) because there is a bijection between the $U(1)^{d}\times S_{d}$ orbits of both ensembles. Notice however, that in contrast to the volume which only depends on the level statistics, the moment-operators will yield different results for both ensembles, \emph{e.g.} the one generated by $H$ and the other by $UHU^{\dagger}$, as they depend not on the number but rather on the chosen unitaries themselves.

\item The second comment concerns the contribution of the square of the inverse Vandermonde in \eqref{eq:final_form_C_ensemble volume}, \emph{i.e.}, the term

\begin{equation}
\label{eq:inverse_vandermonde}
    \prod_{l<m} \left( \frac{\overline{\Delta E}}{E_{l}-E_{m}}. \right)^{2},
\end{equation}
where, out of the $\binom{d}{2}$ energy differences $E_{l}-E_{m}$ inside \eqref{eq:inverse_vandermonde}, $d-1$ will be of order of the mean level spacing, \emph{e.g.}, $E_{l+1}-E_{l}\sim \overline{\Delta E}$, and therefore they will not contribute significantly to the product. On the other hand, the remaining $\binom{d}{2}-(d-1)\sim \mathcal{O}(d^2)$ terms will be typically greater than the mean level spacing and therefore we expect that, independently of the choice of the particular many-body Hamiltonian $H$, \eqref{eq:inverse_vandermonde} will be small overall, \emph{i.e.},

\begin{equation}
 \overline{\Delta E}^{2 \binom{d}{2}} \Delta(H)^{-2} \ll 1.
\end{equation}
In particular, this was confirmed for the equally-level spaced case \eqref{eq:Harmonic_oscillator}, where we have $\Delta (H)^{-2} = \left(\prod_{l=1}^{d-1}l!\right)^{2}$. For the $C$-ensemble complexity bound \eqref{eq:complexity_bound}, the net effect in the smallness of $\Delta(H)^{-2}$ is to compensate the contribution coming from the $U(1)^{d} \times S_{d}$ -orbit, \emph{i.e.},

\begin{equation}
 \log{|\mathcal{E}_{C}(H)|_{\epsilon}}=\binom{d}{2}\log(2)+\log \left(\frac{\Vol(U(1)^{d} \times S_{d}) }{\Vol(B_{\epsilon}(d^{2}))}\right)-\log(\Delta(H)^{2}).
\end{equation}
However, we desire to emphasize that as we mentioned previously, the $C$-ensemble complexity associated with an individual Hamiltonian $H$ cannot grow faster than $\mathcal{O}(\frac{d^2}{2}\log d)$. Particularly, this means that $\log(\Delta(H)^{2})$ must contribute with a term that exactly cancels the leading $\mathcal{O}(d^2\log d)$ term coming from the entropy of the $U(1)^{d}\times S_{d}$ ensemble

\begin{equation}
\label{eq:orbit_entropy}
 \log\left( |\mathcal{E}_{U(1)^{d}\times S_{d}}|_{\epsilon}\right)=\log \left(\frac{\Vol(U(1)^{d} \times S_{d}) }{\Vol(B_{\epsilon}(d^{2}))}\right).
\end{equation}

\end{itemize}
Apart from the discussion on the overall smallness of the $\Delta(H)^{-2}$ term, we have not made any comments yet about the explicit dependence of this term on the level-correlation of the Hamiltonian, \emph{e.g.}, as \eqref{eq:inverse_vandermonde} depends on the level differences, we should expect a strong distinction
between a system that presents level clustering with respect to one having level repulsion \cite{Haake/Quantum_chaos/2010}. The distinction between the level-repulsion and level-clustering scenarios, can be made concrete by means of the normalized level spacings, $s_{l}=(E_{l+1}-E_{l})\overline{\Delta E}$, \emph{i.e.},

\begin{equation}
\label{eq:vandermonde_level_spacing}
 \Delta(H)^{-2}=\prod_{l<m}\left(\frac{1}{\sum_{k=l}^{d-1} s_{k}}\right)^{2},
\end{equation}
where, for Hamiltonians with level clustering the level spacings will follow a Poissonian distribution\footnote{For a distribution close to a Poissonian one, \emph{e.g.}, one with the probability peak in $s=0$, this argument will also hold.} $e^{-s}$ and, therefore, many terms of each sum inside \eqref{eq:vandermonde_level_spacing} will yield zero compared to the case where the Hamiltonian presents level repulsion and the level spacings have a vanishing probability of being zero. Therefore, even if the inverse Vandermonde is small, we expect a clear distinction between the contribution from a spectrum with energy spacings following a Wigner-Dyson distribution (paradigmatic of level repulsion) from the one with of a Poissonian (paradigmatic of level clustering). Schematically,

\begin{equation}
\left(\frac{1}{\Delta(H)}\right)^{2}\bigg|_{\text{Level repulsion}} \leq \left(\frac{1}{\Delta(H)}\right)^{2}\bigg|_{\text{Level clustering}},
\end{equation}
meaning that for systems with level clustering, the associated $C$-ensemble is expected to be larger in terms of unitaries compared to the $C$-ensemble associated with a system with level repulsion.

This last result about the size of the ensembles, and thereof the $C$-ensemble complexities for the level clustering and level repulsion cases, may seem counter-intuitive at first sight. Explicitly, one would be tempted to guess that for chaotic systems (the ones presenting energy repulsion) complexity should be higher. However, the result in \eqref{eq:final_form_C_ensemble volume} indicates the opposite, \emph{i.e.}, Hamiltonians with a high-degree of degeneracies are the ones that will yield $C$-ensembles with higher complexities.

\begin{figure}
     \centering
 \includegraphics[width=0.9\textwidth]{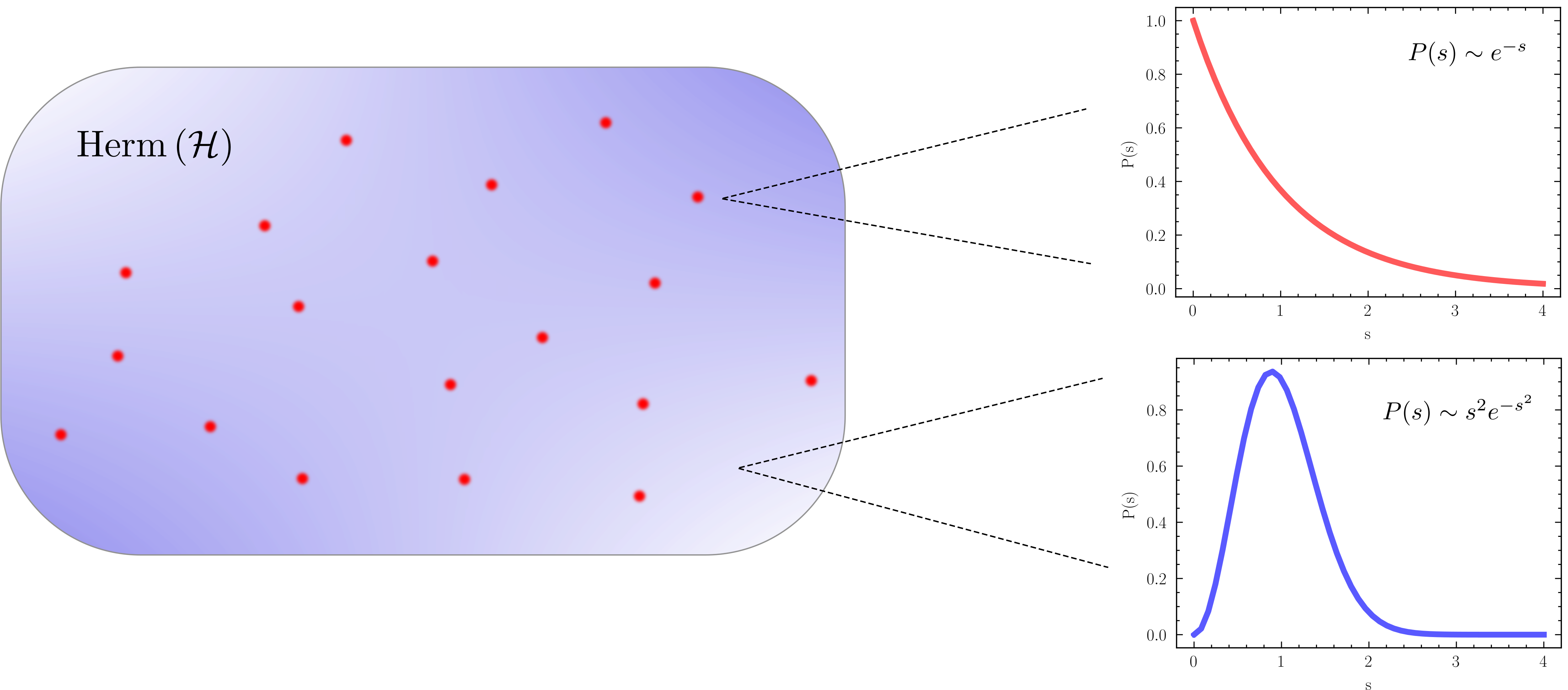}
    \caption{Space of all-possible many-body Hamiltonians of the unitary symmetry class identified with the space of Hermitian operators over the Hilbert space $\text{Herm}(\mathcal{H})\cong \mathbb{R}^{d=2+\binom{d}{2}}$. Integrable Hamiltonians (red blurred regions) are atypical over $\text{Herm}(\mathcal{H})$ and the probability of randomly picking an integrable Hamiltonian from $\text{Herm}(\mathcal{H})$ is exponentially small as the dimension of the many-body Hilbert space increases. For typical Hamiltonians, the spectrum presents level repulsion and the probability distribution for the normalized-energy spacings is close to a GUE Wigner-Dyson distribution. However, for atypical (integrable) models, the spectrum exhibits level-clustering and the probability distribution for normalized-energy spacings is close to a Poissonian one. }
    \label{fig:Hamiltonian_moduli_space}
\end{figure}
To argue in favor of this result, we provide the following typicality-type argument: Consider the space of all possible many-body Hamiltonians of the unitary symmetry class. This space is isomorphic to the space $\text{Herm}(\mathcal{H})$ of Hermitian operators over the many-body Hilbert space\footnote{In this identification we aren't imposing locality restrictions of any type, \emph{e.g.} we are considering two-body interacting and $k$-body interacting, with $k \to \infty$ as valid many-body Hamiltonians.}, and each Many-body Hamiltonian $H$ can be represented as

\begin{equation}
\label{eq:typicalyti_many_body}
 H=\sum_{l=1}^{d^{2}}\alpha_{l}T_{l},
\end{equation}
with $\{T_{l}\}_{l=1}^{d^2}$ any chosen basis for $\text{Herm}(\mathcal{H})$, \emph{e.g.}, the
$U(d)$ Lie-algebra generators. By randomly picking out a combination of the $d^{2}$-couplings inside \eqref{eq:typicalyti_many_body}, the most probable outcome is that the resulting Many-body Hamiltonian will be far from being an integrable one, \emph{i.e.}, it will not possess a considerable number of non-trivial integrals of motion\footnote{By non-trivial, we refer to globally defined integrals of motion, and not the large number of local integrals defined by projection operators.}. Concretely, integrable Many-body Hamiltonian's will be characterized by a very fine-tuned combination of the couplings which make them atypical inside $\text{Herm}(\mathcal{H})$. Now, in contrast to typical Hamiltonians (which tend to present a degree of level repulsion), the spectrum of an integrable Hamiltonian is characterized by an attraction between neighboring energy levels, \emph{i.e.}, level clustering. Therefore, as many-body Hamiltonians with level clustering are atypical inside $\text{Herm}(\mathcal{H})$, we expect that the $C$-complexity of the lattes to be grater\footnote{Although this last claim is (strictly speaking) about the $C$-ensemble complexity-lower bound rather than the actual value of the $C$-ensemble complexity, we are implicitly assuming that an increase/decrease of the previous implies an increase/decrease of the latter.} (see Fig. \ref{fig:Hamiltonian_moduli_space}).

\begin{figure}[ht]
  \begin{subfigure}{.5\textwidth}
    \centering
    \includegraphics[width=.95\linewidth]{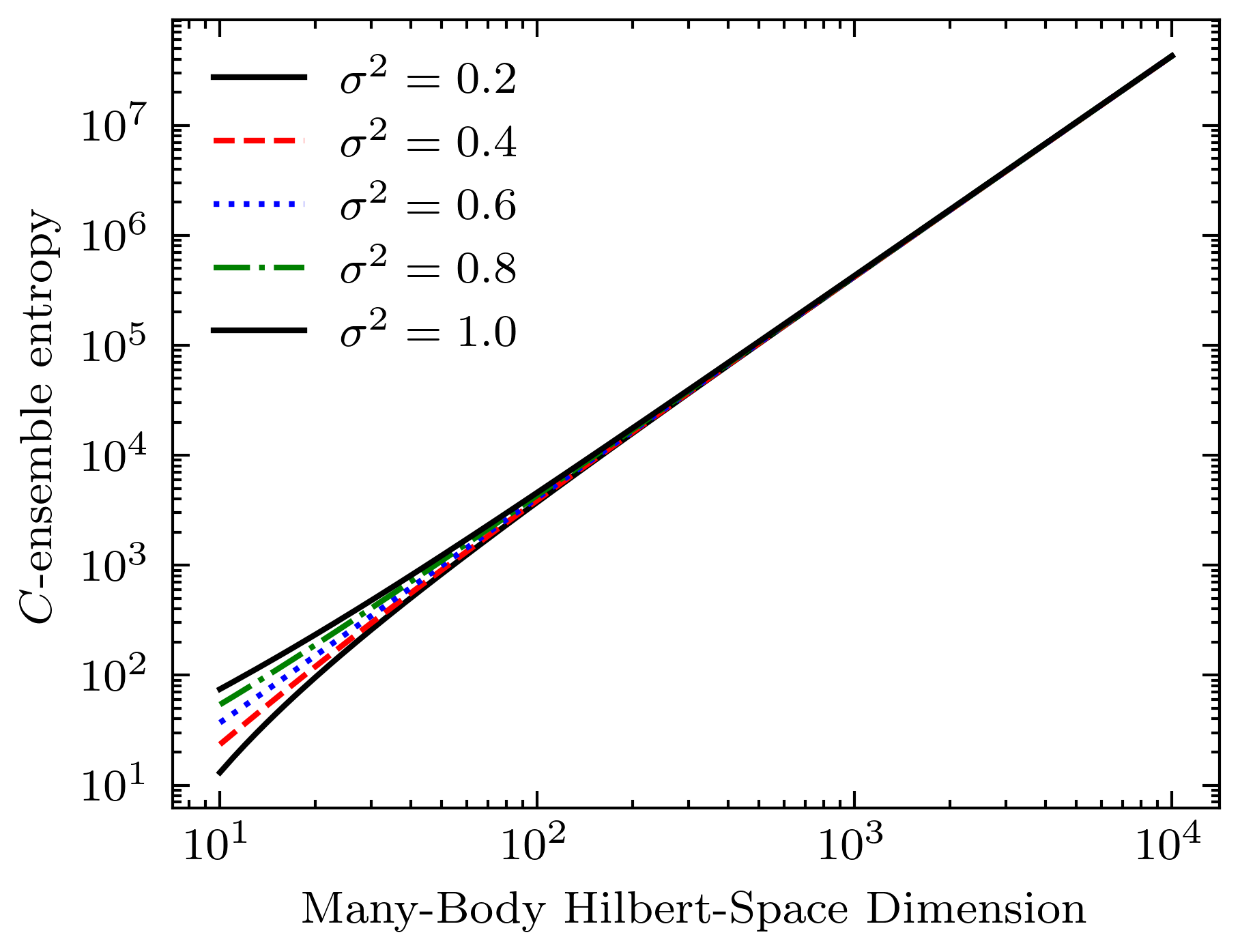}  
    \end{subfigure}
  \begin{subfigure}{.5\textwidth}
    \centering
    \includegraphics[width=.95\linewidth]{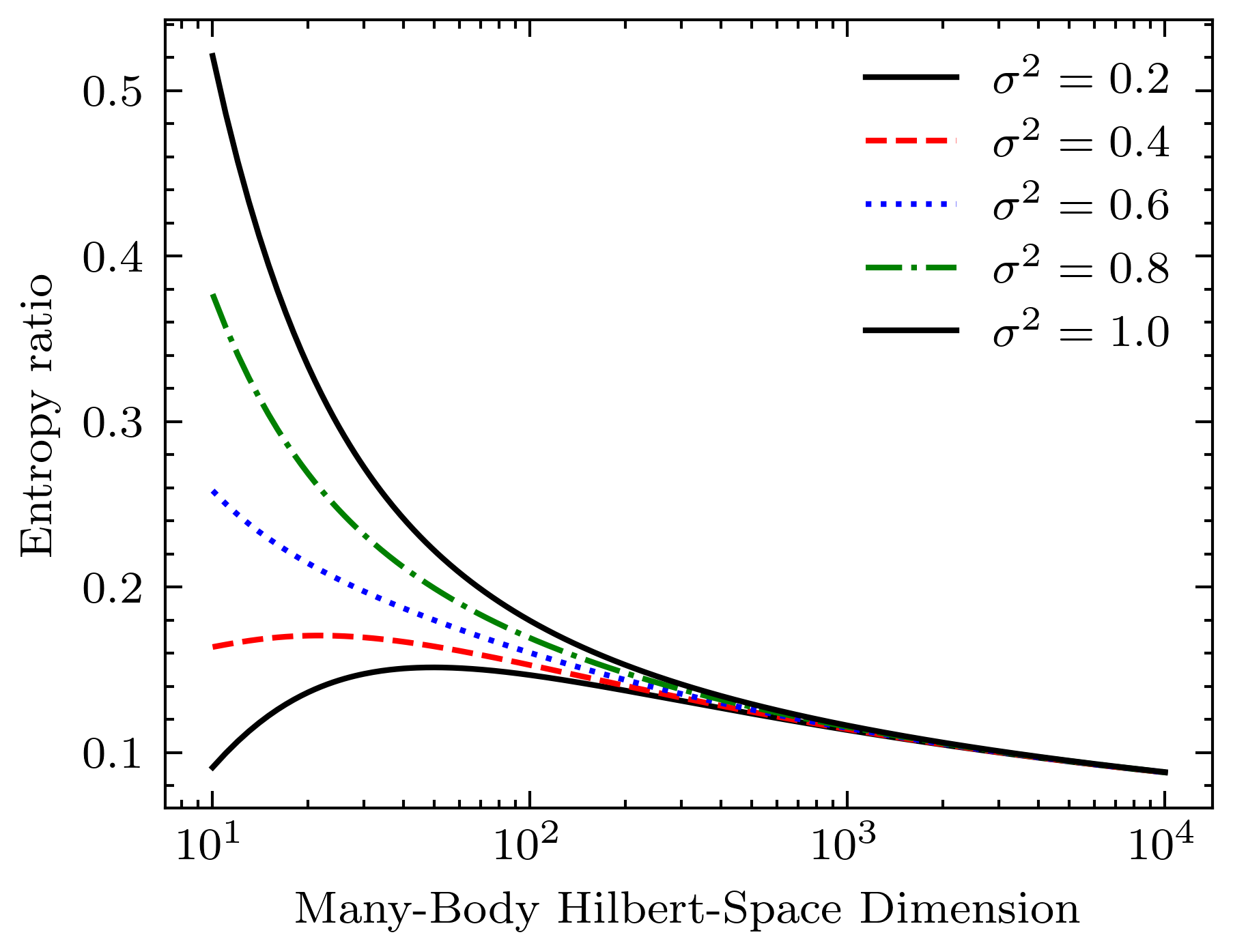}  
   \end{subfigure}
  \caption{Entropy $\log|\mathcal{E}_{C}(H)|_{\epsilon}$ of the $C$-ensemble as a function of the many-body Hilbert space dimension $d$. In the left-panel, we depict the dependence of the entropy in terms of the variance $\sigma^{2}$ of the level spacings $s_{l}=(E_{l+1}-E_{l})/\overline{\Delta E}$ associated with the particular many-body Hamiltonian $H$. As the Hilbert-space increases, the contribution from the level-statistics becomes sub-leading, and the entropy presents a universal growth. In the right-panel we depict the ratio between the $C$- and Haar-ensemble entropies $\frac{\log|\mathcal{E}_{C}(H)|_{\epsilon}}{\log|U(d)|_{\epsilon}}$. Again, as the dimension of the Hilbert space increases, the number of unitaries inside the $C$-ensemble, and thereof the entropy, becomes small compared with the number of unitaries inside the Haar-ensemble. For the computations involving large-factorials we use the \emph{mpmath} package \cite{mpmath/package/2023} and also set the regulator $\epsilon$ equal to one.}
  \label{fig:Complexity_and_variance}
  \end{figure}
Finally, as a toy computation to estimate the previous statements, in Appendix. \ref{appendix:partition_function}, we derive an approximation for the entropy of a $C$-ensemble generated by a many-body Hamiltonian $H$ with level-spacings following an arbitrary distribution $P(s)$ of variance $\sigma^{2}$,
\begin{align}
\label{eq:C-ensemble_entropy_leading_term}
 \log \left(|\mathcal{E}_{C}(H)|_{\epsilon}\right) &\approx \binom{d}{2}\log(2)+\log\left( |\mathcal{E}_{U(1)^{d}\times S_{d}}|_{\epsilon}\right)-2\log \left( \prod_{l=1}^{d-1}l^{d-l}\right)\\
 &+ \sigma^{2}\left(d(H^{(d-1)}-1)+1\right)+\text{sub-leading-terms} \nonumber.
\end{align}
Here, $H^{(n)}$ denotes the $n$th Harmonic-number, and the ``sub-leading-terms'' correspond to contributions exponentially small in $\frac{1}{\sigma^{2}}$ that are present in \eqref{eq:QED_vandermonde_claim}.
The approximation behind \eqref{eq:C-ensemble_entropy_leading_term} is based on using use the central limit theorem to estimate large energy-differences $E_{l}-E_{m}$ (\emph{i.e.}, the ones that significantly contribute to \eqref{eq:inverse_vandermonde}) by rewriting them as averages of a significant number of consecutive level spacings.

Apart from the $\binom{d}{2}\log(2)$ contribution to the $C$-ensemble entropy \eqref{eq:C-ensemble_entropy_leading_term} due to the universality class of the many-body Hamiltonian $H$, we desire to point out that the contribution from the (short-range) level correlations captured by the level spacing distribution $P(s)$ is rather weak (see left panel in Fig. \ref{fig:Complexity_and_variance}). Concretely, at leading order, the entropy does not cares about the particular distribution $P(s)$ but rather on the variance $\sigma^{2}$, \emph{i.e.}, on the term

\begin{equation}
   \sigma^{2}\left(d(H(d-1)-1)+1\right)\sim \sigma^{2}\left(d\log(d)-d(1-\gamma)\right),
\end{equation}
with $\gamma$ the Euler-Gamma constant. Notice that the weak contribution from the many-body spectral statistics inside \eqref{eq:C-ensemble_entropy_leading_term} does not  contradict our previous discussion about the role entropy in the $C$-ensemble for a system presenting either level-clustering or level-repulsion. For example, in the case of level-clustering $\sigma^{2}\approx 1$ (characteristic of a Poissonian distribution), whereas for level-repulsion $\sigma^{2} \approx 0.1781$ (characteristic of a Wigner-Dyson type). 

Additionally, and as pointed out previously in the discussion, the entropy of the $C$-ensemble cannot grow faster than the entropy of the unitary-group. This is confirmed from the asymptotics of \eqref{eq:C-ensemble_entropy_leading_term} where the leading $\mathcal{O}(d^{2}\log(d))$ contribution from the third term in the right-hand side of \eqref{eq:C-ensemble_entropy_leading_term} acts as a counterterm that exactly cancels the leading contribution from the $U(1)^{d}\times S_{d}$ -orbit entropy \eqref{eq:orbit_entropy}. However, what is more remarkable is the slow decay of the ratio between the $C$- and Haar- entropies as the dimension of the Hilbert space increases (Fig. \ref{fig:Complexity_and_variance} right panel). The explicit form of this slow decay can be indeed traced back to be generated by the universality class term $\binom{d}{2}\log(2)$, \emph{i.e},

\begin{equation}
\label{eq:entropy_ratios}
     \frac{\log|\mathcal{E}_{C}(H)|_{\epsilon}}{\log|U(d)|_{\epsilon}} \sim \frac{\binom{d}{2}\log(2)}{\log|U(d)|_{\epsilon}}\sim \mathcal{O}\left(\frac{1}{\log(d)}\right).
\end{equation}
The last result implies that we should expect a sharp distinction between the $C$-ensemble entropy and, therefore, the $C$-ensemble complexity associated with Many-body Hamiltonians selected from different universality classes. Particularly, if $H$ is taken from the orthogonal-class, the universality class term vanishes\footnote{\emph{e.g.} $\log(\beta)^{\binom{d}{2}}$ for $\beta=1$.}, and the ratio between the $C$-ensemble entropy and the entropy of the orthogonal group must decay faster compared to the decay of the unitary-class \eqref{eq:entropy_ratios}.

\section{Summary and Outlook}\label{summary_outlook}

To summarize, in this work we have presented an option on how to build an ensemble of unitary operators suitable for the description of the late-time dynamics of arbitrary many-body systems beyond the Universal Random Matrix Theory regime. This ensemble, named the $C$-ensemble\footnote{Upshot: $C$ can stand for either a) chaotic, or b) complex.}, takes as initial data the individual many-body Hamiltonian to later, and based on it, select a restricted fine-tuned amount of unitaries from the whole unitary group to sample the dynamics of many-body correlation functions. The fact that we are able to select a definite number of unitaries rather than the whole unitary group ensures that the individual details of the many-body Hamiltonian are not only present at late times, \emph{e.g.}, at the decay of the two-point correlation functions beyond the thermalization time but at short times as well, \emph{e.g.}, on the behavior of the Out-of-Time-Ordered correlators.
A brief list of key-points is the following:

\begin{itemize} 
    
    \item Even with an explicit Hamiltonian-dependence acting as an external source, the $C$-ensemble yields a unitary 1-design. However, already at the level of four-moments, and due to the explicit Hamiltonian dependence, the $C$-ensemble fails to yield higher order designs. The degree of deviation from the Haar-result (at the level of four-moments) was captured by the second-frame potential. Concretely,
    we present a connection with the participation ratio in the many-body-localization literature, and show why for chaotic many-body systems, the $C$-ensemble turns into an approximate $2$-design.
    
    \item By integrating out the symmetries of the $C$-ensemble, \emph{i.e.}, the $U(1)^{d}\times S_{d}$ ensemble, an expression for the $C$-ensemble averaged four moments, equivalently the $C$-ensemble twofold channel, was found in terms of an operator $G^{\mathcal{E}_{C}}$, named the \emph{plateau-operator}, living in the \emph{twofold} Hilbert-space $\mathcal{H}^{\otimes 2}$. The plateau operator contains all the late-time information of two-point correlation functions, and makes the $C$-ensemble self-averaging at late times, meaning that the ensemble average of two-point correlations at infinite temperatures exactly matches the correct value at late times. Additionally, we show that the plateau operator $G^{\mathcal{E}_{C}}$ is highly constrained, and provide an equation that by taking $H$ as input data, completely fixes $G^{\mathcal{E}_{C}}$.

    \item Motivated by the late-time splitting between spectral, and operator-parts in the evaluation of correlation functions for strongly chaotic many-body systems, which is captured by Haar-random unitaries (see \cite{Cotler/spectral_decoupling/2020}), we introduce the notion of spectral decoupled unitary ensembles. In particular, it was shown that, apart from Haar-random distributed unitaries, the $C$-ensemble also yields spectral decoupled correlation functions (at least for two-, and four-points). A closely related framework to explain the universal signatures of correlation functions in many-body complex systems is given by Eigenstate Thermalization Hypothesis (ETH), and in the scope of further connections, we discuss the possibility of an underlying ``ETH unitary eigenvector ensemble $\mathcal{E}_{\text{ETH}}$''  which reproduces the ETH claims. We provide some specific tight constraints for this ensemble and, in addition, show that over small energy windows, this ``ETH unitary eigenvector ensemble'' corresponds to the $C$-ensemble, \emph{i.e.} ,$\mathcal{E}_{\text{ETH}} = \mathcal{E}_{C}$. 
    
    \item Expressions for the $C$-ensemble average of (finite-, and infinite- temperature) two-point correlation functions were provided. Additionally, an expression for the $C$-ensemble averaged OTOC as a sum of two-point correlation functions in the symmetric and anti-symmetric sub-spaces of the twofold Hilbert space $\mathcal{H}^{\otimes 2}$ was provided. Finally, the $C$-ensemble partition function was found to be given in terms of the inverse Vandermonde determinant of the fixed Hamiltonian (\emph{i.e} the external source). By using this result, and based on  \cite{Yoshida/Chaos_design/2017} an upper-bound for the ensemble complexity was obtained, relating the $C$-ensemble complexity with the spectral statistics of the underlying many-body Hamiltonian $H$. 
\end{itemize}

In the rest of this section we make some comments about future perspectives.

\subsection{\texorpdfstring{$C$}{}-ensemble and operator growth}

\begin{figure}[ht]
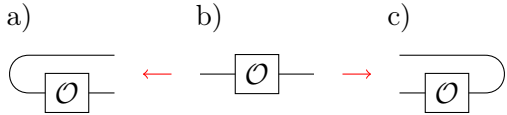

  \ctikzfig{Operator_to_state_map}
    \caption{Operator to state map. Due the isomorphism between a finite dimensional Hilbert space $\mathcal{H}$ and is dual $\overline{\mathcal{H}}$, every operator $\mathcal{O}$ in $\text{End}(\mathcal{H})$ (b) can be seen as a state $\ket{\mathcal{O}}$ in $\mathcal{H} \otimes \mathcal{H}$ (c) or as a dual state $\bra{\mathcal{O}}$ in $\overline{\mathcal{H}} \otimes \overline{\mathcal{H}}$ (a). The map naturally induces the Hilbert-Schmidt inner-product by diagrammatically contracting the right and left wires, i.e $\braket{ \mathcal{O}|\mathcal{O} }=\Tr_{\mathcal{H}}(\mathcal{O}^{2})$.}
  \label{fig:operator_to_state}
\end{figure}

Following \cite{Nick/operator_growth/2018}, let $\mathcal{O}$ be an arbitrary bounded linear operator in $\mathcal{H}$. By the operator to state map (see figure \ref{fig:operator_to_state}), $\mathcal{O}$ can be identified with an state $\ket{\mathcal{O}}$ in the twofold Hilbert space $\mathcal{H}\otimes \mathcal{H}$. For a system of $N$-qubits the whole set of $4^{N}$ Pauli strings\footnote{Unitaries of the form $\sigma^{i_{1}}\otimes \dots \otimes \sigma^{i_{N}}$, with $\sigma^{i_{k}}$ drawn from the single qubit Pauli matrices plus the identity, {\em e.g.} $\{\mathbb{I},X,Y,Z\}$.} $\{\mathcal{P}_{S}\}$ form a basis for $\mathcal{H}\otimes \mathcal{H}$, therefore at every time,

\begin{equation}
  \label{eq:operator_to_pauli}
  \ket{\mathcal{O}(t)}=\sum_{S \in \, \text{Pauli's}}\gamma_{S}^{\mathcal{O}}(t)\,\mathcal{P}_{S},
\end{equation}
where $\gamma_{S}^{\mathcal{O}}(t)$ denotes the time-dependent amplitude to be in the $S$ Pauli-string. Therefore, for an initially local operator $\mathcal{O}$, one can quantify the growth into a non-local operator $\mathcal{O}(t)$, tracking the unitary Hamiltonian time evolution of the string-to-string transition probabilities

\begin{equation}
\label{eq:string_to_string}
  |\gamma_{S}^{S^{\prime}}(t)|^{2}=\left(\frac{1}{4^{N}}\right)^{2}|\Tr{\sigma_{S^{\prime}}(t)\sigma_{S}}|^{2}.
\end{equation}
As a remark, notice that the Pauli strings, are only a particular (and in this case suitable) choice for the twofold space $\mathcal{H}^{\otimes 2}$ basis. An additional example of basis choice, which also works for arbitrary $d$-dimensional spaces, is given by $d^{2}$ independent Hermitian matrices $\{T_{l}\}_{l=1}^{d^2}$ normalized with the Hilbert-Schmidt inner product $\Tr(T_{l}T_{m})=\delta_{l m}$\footnote{$d$-dimensional Pauli matrices, \emph{e.g.} the generators of the canonical representation of $SU(d)$ plus the identity.}.
 
The string-to-string transition probability \eqref{eq:string_to_string} is a four-point function on the $N$-qubit Hilbert space $\mathcal{H}\sim \mathbb{C}^{N}$ and by a similar strategy as for the OTOC case, we can find its $C$-ensemble average. 

\subsection{Solutions for the plateau equation}

By itself the $C$-ensemble plateau operator $G^{\mathcal{E}_{C}}[H]$, viewed as an eigenvector $s$-state in the twofold-Hilbert space $\mathcal{H}^{\otimes 2}$ has proven to be a remarkable object. It contains, in each respective case, the expected late time contributions of two- and four-point functions plus additional non-universal dependence on the specific fixed Hamiltonian $H$. At first sight, it is surprising, that solely due to unitarity, it was possible to completely bootstrap the structure of $G^{\mathcal{E}_{C}}[H]$ up to a single-operator valued function $\phi(H)$. However, what is even more remarkable (at least for us) is that there is an universal equation which, once given the particular Hamiltonian $H$ as input data, gives $\phi(H)$ and therefore completely fixes $G^{\mathcal{E}_{C}}[H]$. A further direction will be search for, at least asymptotically (large-$d$), solutions to this equation, and thereof an asymptotic form for the plateau operator. Although we already made some comments in  Appendix \ref{Appendix:Scrambling} regarding on an strategy to solve the plateau equation, we desire to leave open the possibilities for searching for more efficient ways to attack this equation.

\subsection{\texorpdfstring{$C$}{}-ensemble and lower-dimensional quantum-gravity}

An additional interesting direction is to look at the $C$-ensemble from the holographic duality perspective. Particularly, in a recent paper \cite{Stanford/subleading_weingartens/2022} the authors study the subleading contributions from the $U(d)$ four-moment operators in OTOCs by identifying them term-by-term with the expected ones obtained from the bulk JT gravity \cite{Saad/Jt_gravity_matter/2019,Andreas/ETH_JT/2022} evaluation. At the end of the paper, they proposed the Haar unitary eigenvector ensemble as qualitative toy-model to identify the late-time dynamics in the JT OTOC. As the $C$-ensemble introduces the Hamiltonian fine-grained details as additional non-universal contributions apart from the universal Haar-contribution, it would be interesting to identify the non-universal contributions from the bulk perspective as well.

Additionally, recent proposals \cite{Daniel/JT_ETH_complete/2022,Daniel/JT_ETH_soft/2022} to couple matter-fields to bulk $JT$-gravity require the boundary theory being described not by a one-, but rather a two-matrix model (\emph{i.e.}, one for the standard JT-part and the other for the matter-fields). There, the authors point out that the two-matrix model can be tuned to give rise to generalized ETH-type correlations. As the $C$-ensemble, in a certain limit, gives rise to local ETH-type correlations, it would therefore be interesting to ask whether there exists a particular limit in which the Two-matrix model for JT-gravity with matter fields can be compared with the $C$-ensemble.

\pagebreak

\appendix

\section{On RMT Hamiltonian Ensembles and Haar Distributed Unitaries} 
\label{Appendix:RMT_Haar}

For the sake of completeness, in this appendix we present some standard results from random matrix theory, mainly the Gaussian unitary ensemble and integration over the unitary group. The first two sections introduce the basic machinery needed for quantum chaos and quantum information. This presentation is based on \cite{Liu/Spectral_form_factor:lecture_notes/2018,Mehta/RMT/2004,Yoshida/Chaos_design/2017}. The last part, on the other hand, contains the proof of the $C$-ensemble partition inversion formula as presented in (Appendix \ref{appendix:partition_function}).

\subsection{Gaussian Unitarity Ensemble}

RMT Hamiltonian ensembles are described by invariant measures under some Lie group $G$ action, {\em i.e.}, $d\mu_{\mathcal{E}}[H]=d\mu_{\mathcal{E}}[UHU^{\dagger}]$ for all $U$ in $G$. Here we will focus on the Gaussian unitary ensemble (GUE) defined as the random matrix ensemble of $L\times L$ Hermitian matrices\footnote{In this section we use $L$ instead of $d$ to denote the dimension of the matrices.} equipped with the unitary $U(L)$ invariant measure,

\begin{equation}
 d\mu_{\text{GUE}}[H]\propto e^{-\frac{L}{2}\Tr{H^{2}}}[dH].
\end{equation}
Here $[dH]=\prod_{l}dH_{ll}\prod_{l<m}dRe[H_{lm}]\,dIm[H_{lm}]$ is the Lebesgue (flat) measure over the $L+2 \binom{L}{2}$ independent matrix elements. In this definition, we set the mean, $\braket{H_{lm}}_{\text{GUE}}=0$, equal to zero and the variance, $\braket{|H_{lm}|^{2}}_{\text{GUE}}$, to be $\frac{1}{L}$. In particular this will yield an energy spectrum supported over the interval $[-2,2]$\footnote{For a GUE with unit variance distributed Hamiltonians the spectrum will be supported instead on $[-2\sqrt{L},2\sqrt{L}]$.}. Higher-order correlations between matrix elements are easily obtained by Wick contractions,

\begin{equation}
 \braket{H_{i\,j}H_{k\,s}\dots H_{l\,m}}_{\text{GUE}}=\sum_{P \in \text{pairings}} \braket{|H_{P(i)\,P(j)}|^{2}}_{\text{GUE}}\dots\braket{|H_{P(l)\, P(m)}|^{2}}_{\text{GUE}}.
\end{equation}
Every invariant RMT Hamiltonian ensemble induces a natural spectral measure for the eigenvalues through the polar decomposition, \emph{i.e.}, $H=UEU^{\dagger}$. In particular, for the GUE,

\begin{equation}
 d\mu_{\text{GUE}}[E]\propto e^{-\frac{L}{2}\Tr{E^{2}}}\Delta^{2}(E)dE.
\end{equation}
Here, $dE=\prod_{l=1}^{d}dE_{l}$ is the flat measure over the eigenvalues, and $\Delta(E)=\prod_{l<m}^{d}(E_{l}-E_{m})$ denotes the Vandermonde determinant. For the GUE, all $k$-point spectral correlations,

\begin{equation}
 \rho(x_{1},\dots,x_{k})=\Braket{\Tr(\delta(x_{1}-H))\dots \Tr(\delta(x_{k}-H))}_{\text{GUE}},
\end{equation}
are exactly solvable in terms of a determinant involving Hermite polynomials\footnote{From the pragmatic point of view the GUE is analogous to the ``harmonic oscillator'' in matrix models.} \cite{Mehta/RMT/2004}. These are rather complicated expressions, however, in the large-$L$ limit they highly simplify. In this limit the one-point function yields the famous Wigner semicircle distribution,

\begin{equation}
 \label{eq:Semi-circle}
 \rho(x)=\frac{L}{2\pi}\sqrt{4-x^{2}}.
\end{equation}
For a \emph{single} random Hermitian matrix $H$, the spectral density\footnote{We mean the physical spectral density, {\em i.e.} $\int dx \rho(x)=L$.} $\Tr{(x-H)}$ is a complicated non-smooth distribution of delta peaks located on its eigenvalues. The ensemble average smooths out this complicated distribution into \eqref{eq:Semi-circle}. For an arbitrary system, the spectral density sets the theoretical ``largest time scale'', \emph{i.e.}, the Heisenberg time, to be the inverse of the mean level spacing\footnote{The mean level spacing $\Delta \overline{E}$ is proportional to the inverse averaged spectral density $\Delta \overline{E} \propto \int \rho(E)^{-1}dE$.}. In particular, for the GUE, the average level spacing is of order $\frac{1}{L}$ and defines a Heisenberg time to be $t_{H} \sim L$.

\begin{figure}[ht]
  \begin{subfigure}{.5\textwidth}
    \centering
    \includegraphics[width=.9\linewidth]{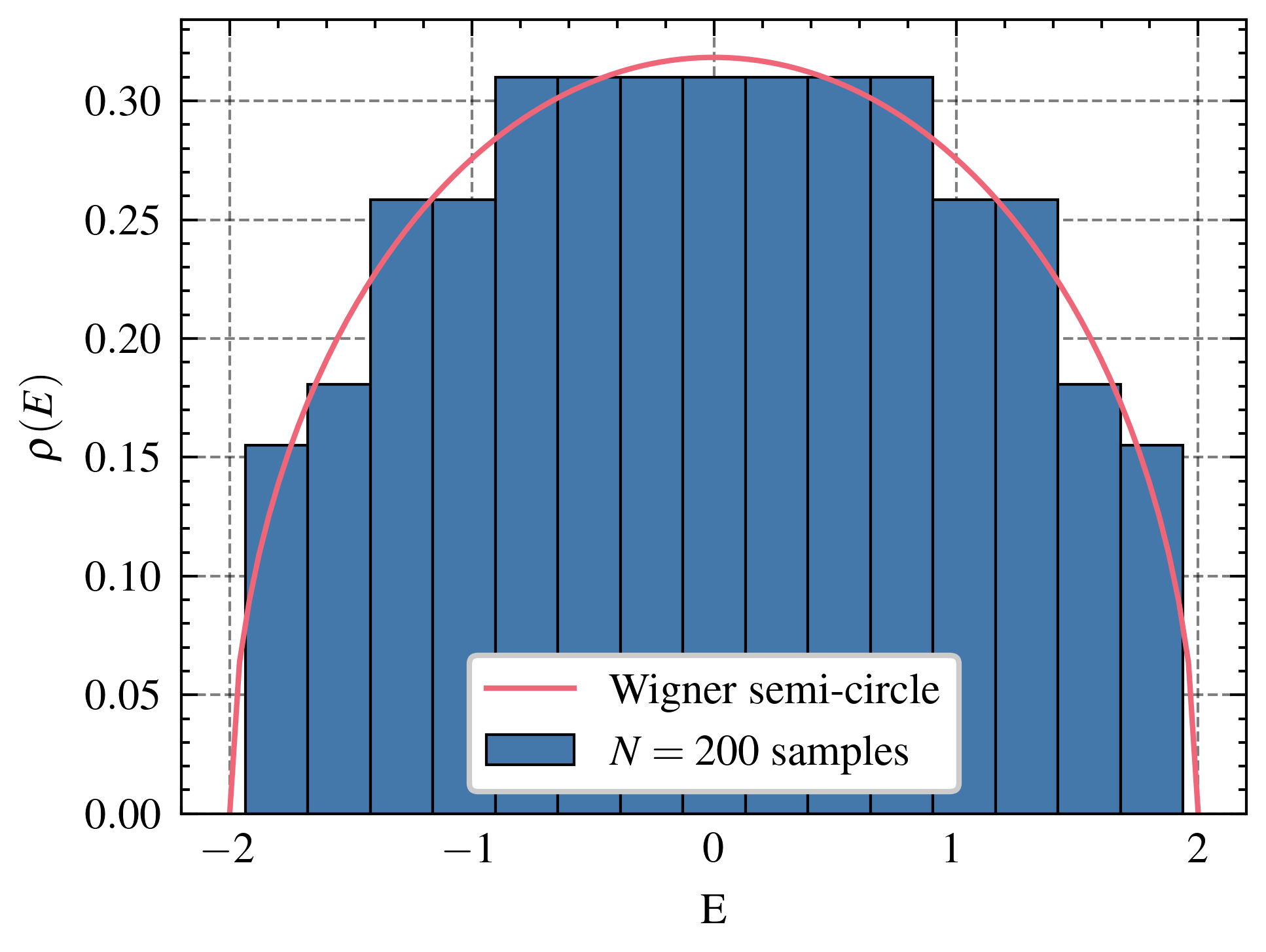}  
    \caption{GUE smoothed density-of-states.}
    \label{fig:wigner_semicircle}
  \end{subfigure}
  \begin{subfigure}{.5\textwidth}
    \centering
    \includegraphics[width=.9\linewidth]{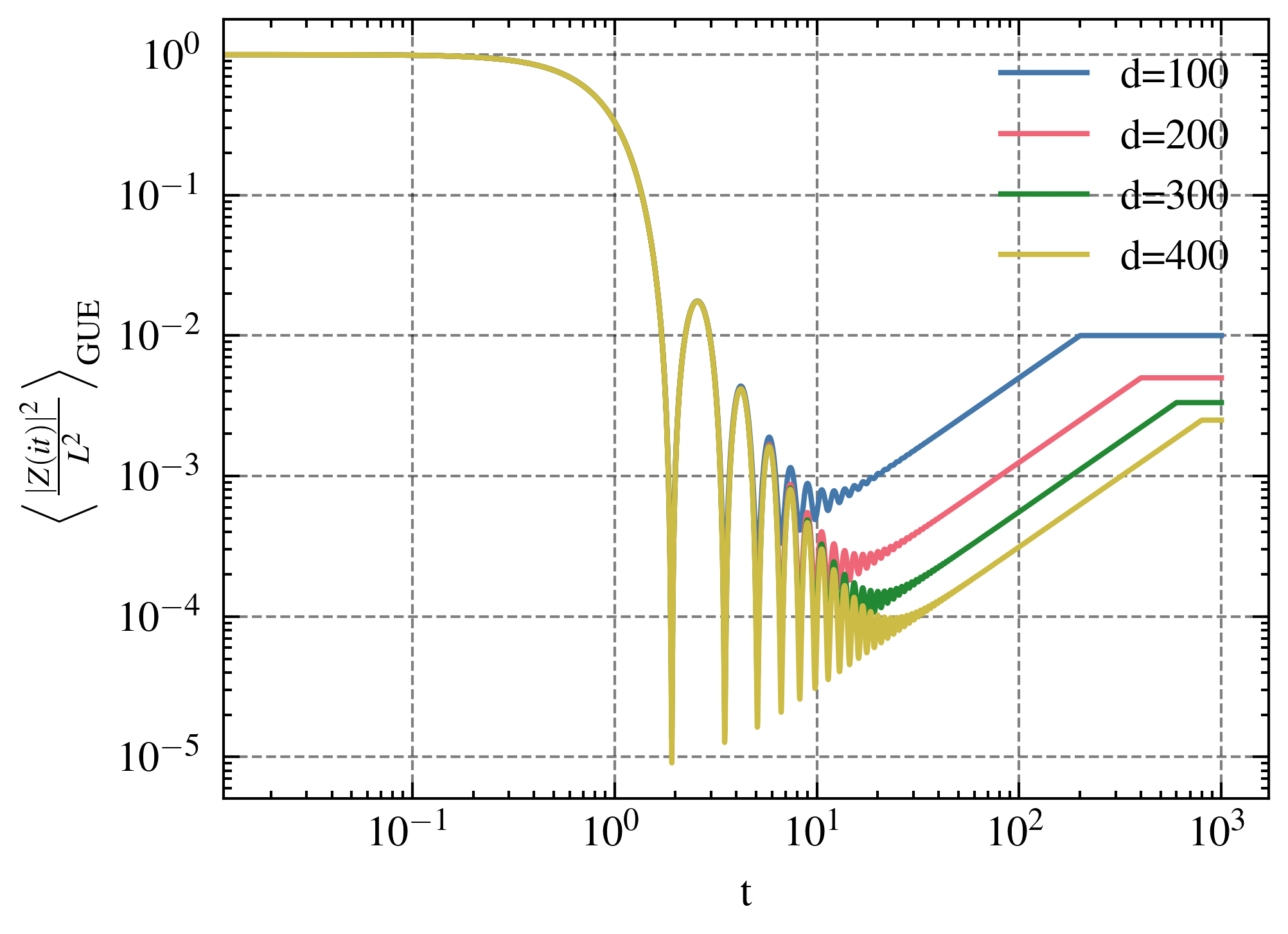}  
    \caption{GUE infinite temperature form factor.}
    \label{fig:GUE_infinite_form_factor}
  \end{subfigure}
  \caption{One-point function and spectral form factor for the Gaussian unitary ensemble. (a) As we choose the variance of the GUE to be $\frac{1}{L}$, the resulting spectrum (in the $N \to \infty$ limit) is supported continuously over $[-2,2]$. (b) The infinite temperature form factor (b) presents the decay-ramp-plateau transition resulting from the universal form of the sine kernel. The integral was performed analytically by using the box approximation \cite{Liu/Spectral_form_factor:lecture_notes/2018}. }
  \label{App_fig:GUE_form_factor}
  \end{figure}
Two-point spectral correlations $\rho(x,y)$ are far more complicated. However, for energy differences near the semicircle center, \emph{i.e.}, $|x-y| \ll 4$, the connected part reduces into the universal sine kernel \cite{Brezin/Universal_correlation/1993,Gaikwad/Spectral_form_factor/2019,Brezin/Spectral_form_factor/1996, Brezin_Hikami/Universal_correlation_extended/1997,Liu/Spectral_form_factor:lecture_notes/2018},

\begin{equation}
 \label{eq:sine_kernel}
 \rho(x,y)=\rho\left(\frac{x+y}{2}\right)\delta(x-y)+\rho(x)\rho(y)-\frac{1}{L^{2}}\frac{\sin^{2}(L(x-y))}{(L\pi(x-y))^{2}}.
\end{equation}
In \eqref{eq:sine_kernel}, we add the $\delta$ term to account for eigenvalue collisions. The spectral form factor is defined as the square norm of the analytically continued partition function $Z(it)=\Tr{e^{it H}}$. For Hamiltonians drawn from the GUE, the spectral form factor presents a distinct decay-ramp-plateau structure \cite{Cotler/Blackholes_SYK/2017} (see Fig. \ref{App_fig:GUE_form_factor}) given by the different energy scales involved in \eqref{eq:sine_kernel},

\begin{equation}
 \Braket{|Z(it)|^{2}}_{\text{GUE}}=\int e^{it(x-y)}\rho(x,y)dx\, dy.
\end{equation}
For late times, the contribution to the form factor is dominated by levels very close to each other, and the collision term in \eqref{eq:sine_kernel} will give rise to the plateau of height $\mathcal{O}(L)$. Alongside, for moderately separated energy levels the main contribution is given by the universal sine-kernel which gives rise to the universal ramp. Notice that for short times, the contributions are given by far apart energy levels, and therefore the disconnected term $\rho(x)\rho(y)$ will dominate. However, this term is not universal, and therefore the initial decay will be strongly system-dependent.

\subsection{Haar unitary ensemble}\label{appendix:unitary}

For the sake of completeness in this appendix we present some of the ``Joys'' of the $U(d)$ Haar measure, focusing on the relevant uses in this work. First things first: The normalized unitary Haar measure $[U^{\dagger}dU]$ is defined as the unique left and right invariant measure over the unitary group. This means in particular that for every (integrable) function $f$ over $U(d)$ and fixed unitary $V$ the following holds,

\begin{equation*}
 \int_{U(d)}f(VU)[U^{\dagger}dU]=\int_{U(d)}f(UV)[U^{\dagger}dU]=\int_{U(d)}f(U)[U^{\dagger}dU] \quad \text{and} \quad \int_{U(d)}[U^{\dagger}dU]=1.
\end{equation*}
Although this notation may seem strange compared with more ``standard'' ones (\emph{e.g.}, \cite{Cotler/spectral_decoupling/2020,Yoshida/Chaos_design/2017}), the notation $[U^{\dagger}dU]$ is adopted \footnote{This notation is also used in \cite{Forrester/Log_gases_rmt/2010}.} here to simply highlight this invariance, \emph{i.e}, $[(VU)^{\dagger}d(VU)]=[U^{\dagger}V^{\dagger}VdU]=[U^{\dagger}dU]$. It is a well-known result due to the seminal work of Collins \cite{Benoit/Weingarten_original/2003} that all moments can be set up in terms of some class functions $\text{Wg}^{U}$ of the symmetric group $S_{d}$,

\begin{equation}
 \label{app:weingarten_formula}
 \int_{U(d)}U_{l_{1}r_{1}}\dots U_{l_{d}r_{d}}U^{\dagger}_{s_{1}q_{1}}\dots U^{\dagger}_{s_{d}q_{d}} [U^{\dagger}dU]=\sum_{\pi, \sigma \in S_{d}}\delta_{\pi}(\vec{l},\vec{q})\,\delta_{\pi}(\vec{r},\vec{s})\, \text{Wg}^{U}(\sigma^{-1}\pi,d),
\end{equation}
where $\delta_{\pi}(\vec{i},\vec{j})=\delta_{i_{1},j_{\pi(1)}} \dots \delta_{i_{d},j_{\pi(d)}}$ represents a contraction of the whole index set and the sum runs over $S_{d}\times S_{d}$. The $\text{Wg}^{U}$ weights are called the unitary Weingarten functions and similar formulas exist for either the orthogonal or symplectic groups \cite{Benoit/Weingarten_symplectic_orthogonal/2006,Matsumoto/Weingarten_coset_spaces/2013}. As the Weingarten functions are class functions of the symmetric group, they only depend upon the cycle structure of the permutation. \emph{i.e.}, for $S_{2}$ we have a swap $(1\,2)$ containing a single length-2 cycle and the identity $(1)(2)$ containing two length one-cycles. For both permutations, the unitary Weingarten functions are $\text{Wg}^{U}((1 \, 2),d)=\text{Wg}^{U}([2],d)=-\frac{1}{d(d^{2}-1)}$ and $\text{Wg}^{U}((1)(2),d)=\text{Wg}^{U}([1,1],d)=\frac{1}{d^{2}-1}$. In Table \ref{table:weingarten_functions} we show a list containing the first of these functions. In particular, the Haar unitary two- and four-moment operators are easily obtained from \eqref{app:weingarten_formula},

\begin{table}[ht]
  \centering
  \begin{tabular}{|l|l|}
   
   $\text{Wg}^{U}([1],d)=\displaystyle{\frac{1}{d}}$   & $\text{Wg}^{U}([4],d)=\displaystyle{\frac{-5d}{d^{2}(d^{2}-1)(d^{2}-4)(d^{2}-9)}}$       \\
   $\text{Wg}^{U}([2],d)=\displaystyle{\frac{-1}{d(d^{2}-1)}}$   & $\text{Wg}^{U}([3,1],d)=\displaystyle{\frac{2d^{2}-3}{d^{2}(d^{2}-1)(d^{2}-4)(d^{2}-9)}}$     \\
   $\text{Wg}^{U}([1,1],d)=\displaystyle{\frac{1}{d^{2}-1}}$ & $\text{Wg}^{U}([2,2],d)=\displaystyle{\frac{d^{2}+6}{d^{2}(d^{2}-1)(d^{2}-4)(d^{2}-9)}}$     \\
   $\text{Wg}^{U}([3],d)=\displaystyle{\frac{2}{d(d^{2}-1)(d^{2}-4)}}$  &  $\text{Wg}^{U}([2,1,1],d)=\displaystyle{\frac{-d^{3}+4d}{d^{2}(d^{2}-1)(d^{2}-4)(d^{2}-9)}}$ \\
   $\text{Wg}^{U}([2,1],d)=\displaystyle{\frac{-1}{(d^{2}-1)(d^{2}-4)}}$  & $\text{Wg}^{U}([1,1,1,1],d)=\displaystyle{\frac{d^{4}-8d^{2}+6}{d^{2}(d^{2}-1)(d^{2}-4)(d^{2}-9)}}$  \\
   $\text{Wg}^{U}([1,1,1],d)=\displaystyle{\frac{d^{2}-2}{d(d^{2}-1)(d^{2}-4)}}$  & \quad \quad \quad $\cdots$
\end{tabular}

\caption{Some unitary Weingarten functions labeled by the cycle structure of the permutation. For a permutation $\pi$ in $S_{n}$, the asymptotic large-$d$ behavior of the functions is  $\mathcal{O}\left(d^{|\pi|-2n} \right)$.}
\label{table:weingarten_functions}
\end{table}

\begin{equation}
 \hat{\Phi}^{U(d)}_{2}=\int_{U(d)}U\otimes U^{\dagger} \,[U^{\dagger}dU]=\frac{1}{d}\,\text{SWAP} 
\end{equation}
for two moments, and

\begin{equation*}
 \begin{tikzpicture}
	\begin{pgfonlayer}{nodelayer}
		\node [style=none] (51) at (-9.25, -0.25) {};
		\node [style=none] (52) at (-9.25, 0.75) {};
		\node [style=none] (53) at (-12.25, -0.25) {};
		\node [style=none] (54) at (-12.25, 0.75) {};
		\node [style=none] (55) at (-9.25, 1.75) {};
		\node [style=none] (56) at (-9.25, 2.75) {};
		\node [style=none] (59) at (-12.25, 1.75) {};
		\node [style=none] (60) at (-12.25, 2.75) {};
		\node [style=none] (68) at (4.5, -0.25) {};
		\node [style=none] (69) at (4.5, 0.75) {};
		\node [style=none] (70) at (1.5, -0.25) {};
		\node [style=none] (71) at (1.5, 0.75) {};
		\node [style=none] (72) at (4.5, 1.75) {};
		\node [style=none] (73) at (4.5, 2.75) {};
		\node [style=none] (76) at (1.5, 1.75) {};
		\node [style=none] (77) at (1.5, 2.75) {};
		\node [style=none] (79) at (-13.75, 1.25) {$\displaystyle{\frac{1}{d^{2}-1}}$};
		\node [style=none] (80) at (-0.5, 1.25) {$\displaystyle{-\frac{1}{d(d^{2}-1)}}$};
		\node [style=none] (84) at (11.5, 2.75) {};
		\node [style=none] (85) at (11.5, 1.75) {};
		\node [style=none] (86) at (8.5, 2.75) {};
		\node [style=none] (87) at (8.5, 1.75) {};
		\node [style=none] (88) at (11.5, 0.75) {};
		\node [style=none] (89) at (11.5, -0.25) {};
		\node [style=none] (92) at (8.5, 0.75) {};
		\node [style=none] (93) at (8.5, -0.25) {};
		\node [style=none] (95) at (-2.75, -0.25) {};
		\node [style=none] (96) at (-2.75, 0.75) {};
		\node [style=none] (97) at (-5.75, -0.25) {};
		\node [style=none] (98) at (-5.75, 0.75) {};
		\node [style=none] (99) at (-2.75, 1.75) {};
		\node [style=none] (100) at (-2.75, 2.75) {};
		\node [style=none] (101) at (-5.75, 1.75) {};
		\node [style=none] (102) at (-5.75, 2.75) {};
		\node [style=none] (103) at (-7.5, 1.25) {$\displaystyle{+\frac{1}{d^{2}-1}}$};
		\node [style=none] (104) at (-16.5, 1.25) {$\displaystyle{\hat{\Phi}_{4}^{U(d)}=}$};
		\node [style=none] (105) at (6.5, 1.25) {$\displaystyle{-\frac{1}{d(d^{2}-1)}}$};
	\end{pgfonlayer}
	\begin{pgfonlayer}{edgelayer}
		\draw (60.center) to (52.center);
		\draw (59.center) to (51.center);
		\draw [style={white_edge}] (54.center) to (56.center);
		\draw [style={white_edge}] (53.center) to (55.center);
		\draw (54.center) to (56.center);
		\draw (53.center) to (55.center);
		\draw [style={white_edge}, in=180, out=0, looseness=0.75] (70.center) to (73.center);
		\draw [in=180, out=0, looseness=0.75] (70.center) to (73.center);
		\draw [style={white_edge}, in=-180, out=0, looseness=0.50] (71.center) to (72.center);
		\draw [in=180, out=0] (71.center) to (72.center);
		\draw [style={white_edge}, in=-180, out=0, looseness=0.75] (86.center) to (89.center);
		\draw [in=-180, out=0, looseness=0.75] (86.center) to (89.center);
		\draw [style={white_edge}, in=180, out=0, looseness=0.50] (87.center) to (88.center);
		\draw [in=-180, out=0] (87.center) to (88.center);
		\draw (101.center) to (96.center);
		\draw [style={white_edge}] (98.center) to (99.center);
		\draw (98.center) to (99.center);
		\draw [style={white_edge}] (102.center) to (95.center);
		\draw (102.center) to (95.center);
		\draw [style={white_edge}] (97.center) to (100.center);
		\draw (97.center) to (100.center);
		\draw [style={white_edge}] (77.center) to (69.center);
		\draw [style={white_edge}] (76.center) to (68.center);
		\draw (77.center) to (69.center);
		\draw (76.center) to (68.center);
		\draw [style={white_edge}] (92.center) to (84.center);
		\draw [style={white_edge}] (93.center) to (85.center);
		\draw (92.center) to (84.center);
		\draw (93.center) to (85.center);
	\end{pgfonlayer}
\end{tikzpicture}
\end{equation*}
for four moments. In \cite{Poland/NFL_quantum_learning/2020} the authors present a very interesting derivation of these results by using only the invariance of the Haar measure and some properties of the $U(d)$ Lie algebra generators. A powerful method to evaluate unitary integrals is given by the Shur-Weyl duality \cite{Fulton_Harris/Representation_theory/1991}, which states that the spans of the images of the symmetric group $S_{d}$ and the general linear group acting on the $k$th fold Hilbert space $\mathcal{H}^{\otimes k}$ are centralizers of each other. Particularly, this means that if for any $A$ in $\text{End}(\mathcal{H}^{\otimes k})$ and every $V$ in $U(d)$ the commutator vanishes, $[A, V^{\otimes k}]=0$, then $A$ must be a linear combination of permutation operators. As an example, consider the unitary twofold channel $\Phi^{U(d)}_{2}(\cdots)$. By the Haar measure invariance, it holds $[\Phi_{2},V^{\otimes 2}]=0$. Therefore, the Shur-Weyl duality states that,

\begin{equation}
 \Phi_{2}^{U(d)}(A)=\alpha(A)\, \mathbb{I}\otimes \mathbb{I} + \beta(A)\, \text{SWAP}
\end{equation}
with both, $\alpha$ and $\beta$ scalar valued functions of $A$, which are easily determined by the trace-preserving properties of the $\Phi_{2}^{U(d)}$ channel, \emph{i.e}, $\Tr \Phi_{2}^{U(d)}(\cdot)=\Tr(\cdot) $ for $A$ and $A\, \text{SWAP}$,

\begin{equation*}
 \Phi_{2}^{U(d)}(A)= \frac{1}{d^{2}-1}\left(\Tr{A}-\frac{\Tr{\text{SWAP}\, A}}{d}\right) \mathbb{I}\otimes \mathbb{I}+\frac{1}{d^{2}-1}\left(\Tr{\text{SWAP}\, A}-\frac{\Tr{A}}{d}\right)\text{SWAP}.
\end{equation*}
As a remark, we use the underlying idea behind this duality (but clearly in a different context) to bootstrap the $C$-ensemble plateau operator in Appendix \ref{Appendix:Scrambling}.

\subsection{Parametrization of Haar}

When dealing with matrix models in some cases it is easier to integrate over Hermitian matrices rather than unitary ones due to their simple flat measure. This can be achieved by representing the unitary matrix in terms of a Hermitian one \cite{Morozov/parametrization_trick/2010},

\begin{align}
\label{eq:A_Haarflat}
U & = \frac{1+iH}{1-iH} \xrightarrow{\quad \quad} [U^{\dagger}dU]\propto \frac{[dH]}{\det^{d}(1+H^{2})} \\
U & = e^{iH} \xrightarrow{\quad \quad} [U^{\dagger}dU] \propto \prod_{i<j}\frac{4 \sin^{2}\frac{1}{2}(\lambda_{i}-\lambda_{j})}{(\lambda_{i}-\lambda_{j})^{2}}[dH],
\end{align}
where $[dH]$ denotes the usual flat measure over Hermitian matrices, and in $e^{iH}$ we assume that $H$ has a spectrum supported over $[-\pi,\pi]$. A highly useful representation is obtained by explicitly imposing the unitary constraint over a generic complex matrix $C$ using the matrix delta distribution,

\begin{equation}
 [U^{\dagger}dU] \propto \delta(C^{\dagger}C-1)[dC], \quad C \in \text{GL}(d,\mathbb{C}),
\end{equation}
where $[dC]$ is the flat measure over the $2d^{2}$ independent matrix elements,

\begin{equation}
 [dC]=\prod_{l,m}^{d}d\text{Re}[C_{lm}]d\text{Im}[C_{lm}].
\end{equation}  
The equivalence between all the aforementioned representations must be interpreted inside the integral symbol, {\em i.e.}
\begin{equation*}
 \int_{U(d)}(\cdots)[U^{\dagger}dU] = \frac{\int_{\text{GL}(d,\mathbb{C})}(\cdots)\delta(C^{\dagger}C-1)\,[dC]}{\int_{\text{GL}(d,\mathbb{C})}\delta(C^{\dagger}C-1)\,[dC]}=\frac{\int_{H=H^{\dagger}}(\cdots)\,\det^{-d}(1+H^{2})[dH]}{\int_{H=H^{\dagger}}\,\det^{-d}(1+H^{2})[dH]},
\end{equation*}
where we divide by the partition function representation to ensure normalization.

\subsection{Proof of the partition function inversion formula}

Here we present the proof of the partition function inversion formula \eqref{theorem:partition_function_inversion_formula} presented in Appendix \ref{appendix:partition_function}. First of all, let us use the delta representation of the ensemble partition function,

\begin{equation}
 \Vol(H)=\frac{1}{Z_{\delta}}\int_{\text{GL}(d,\mathbb{C})} \delta(CC^{\dagger}-1)\delta_{\perp}(CHC^{\dagger})[dC]
\end{equation}
where $Z_{\delta}$ is the normalization constant of the delta representation and the integration is performed over the whole complex general linear group $\text{GL}(d,\mathbb{C})$. The main advantage of this representation is that under arbitrary redefinitions, \emph{i.e.}, $C\rightarrow AC$ with $A$ in $\text{GL}(d,\mathbb{C})$, the measure $[dC]$ simply transforms as $|\det(A)|^{2d}[dC]$. Therefore, for a positive definite and invertible Hamiltonian\footnote{As the $C$-ensemble measure is invariant under shifts, $H\rightarrow H+\mu \mathbb{I}$ we can always furnish this condition.} $H$ the redefinition $C\rightarrow \sqrt{H}^{-1}C$ changes the partition function as,

\begin{equation}
    \label{eq:A_quasi_inverse}
    \Vol(H)=\frac{\det{H}^{-d}}{Z_{\delta}}\int_{\text{GL}(d,\mathbb{C})} \delta(CH^{-1}C^{\dagger}-\mathbb{I})\,\delta_{\perp}(CC^{\dagger})\,[dC].
\end{equation}
However, this redefinition comes with a subtle point: it introduces a dimension of square root of energy to $C$ \footnote{Recall that during this work we use natural units so otherwise stated the Hamiltonian's have energy dimensions denoted by [Energy].}, \emph{e.g.}, $\text{[Energy]}^{1/2}$. To recover a dimensionless $C$-operator, let us make the following trick. First, we rescale each row of the $C$-operator as,

\begin{equation}
    \braket{l|C|k} \rightarrow \frac{1}
    {\sqrt{\beta_{l}}}\braket{l|C|k} \quad \text{where}\;\beta_{l}>0 \;\text{and} \; [\beta_{l}]=[\text{Energy}]^{-1}.
\end{equation}
This rescaling induces a ``false'' $\beta_{k}$-parameter dependence in the partition function as,

\begin{equation}
    \Vol(H)=\frac{\det(\beta_{\parallel})^{d-1}}{Z_{\delta}\det(H)^{d}}\int_{\text{GL}(d,\mathbb{C})} \delta_{\perp}(CH^{-1}C^{\dagger})\delta_{\parallel}(CH^{-1}C^{\dagger}-\beta_{\parallel})\delta_{\perp}(CC^{\dagger}) [dC].
\end{equation}
Here we use the factorization of the full delta as the product of the off-diagonal and diagonal parts, {\em e.g.} $\delta(\cdots)=\delta_{\perp}(\cdots)\delta_{\parallel}(\cdots)$, and $\beta_{\parallel}$ denotes the diagonal matrix of $\beta_{k}$'s. As the partition function is invariant under rotations of the Hamiltonian(see Appendix \ref{appendix:partition_function}), we can replace $H$ with the diagonal matrix of eigenvalues $E$,  

\begin{equation}
  \Vol(H)=\frac{\det(\beta_{\parallel})^{d-2}}{Z_{\delta}\det(E)^{d}}\int_{\text{GL}(d,\mathbb{C})} \delta_{\perp}(CE^{-1}C^{\dagger})\delta_{\parallel}(C\beta_{\parallel}^{-1}E^{-1}C^{\dagger}-\mathbb{I})\delta_{\perp}(CC^{\dagger}) [dC].
\end{equation}
By comparing this expression with the partition function of a $C$-ensemble with the reciprocal Hamiltonian $H^{-1}$,

\begin{equation}
  \Vol(H^{-1})=\frac{1}{Z_{\delta}}\int_{\text{GL}(d,\mathbb{C})} \delta_{\perp}(CE^{-1}C^{\dagger})\delta_{\parallel}(CC^{\dagger}-\mathbb{I})\delta_{\perp}(CC^{\dagger}) [dC],
\end{equation}
we see that apart from pre-factors, both partition functions coincide if we choose the appropriate ``gauge'' where $\beta_{\parallel} = E^{-1}$. As the choice of $\beta_{\parallel}$ is arbitrarily subjected to the condition that it has dimensions of $[\text{Energy}]^{-1}$, we arrive at the desired duality\footnote{The condition $\beta_{\parallel}>0$ is also required. However, as the $C$-ensemble partition function is invariant under energy shifts (see Appendix \ref{appendix:partition_function}), we can always furnish this condition.},

\begin{equation}
 \label{eq:proof_partition_function_duality}
 \Vol(H) = \frac{\Vol{(H^{-1})}}{\det(H)^{2(d-1)}}.
\end{equation}
As a final remark, notice that in the derivation of \eqref{eq:proof_partition_function_duality} we did not make any assumptions on the degeneracy of the spectrum. Therefore, this last result holds for non-degenerate and degenerate Hamiltonians as well.

\section{The  \texorpdfstring{$U(1)^{d} \times S_{d}$}{} ensemble}
\label{Appendix:U1dSd}

In this Appendix we explicitly find the two and four moment operators for the $U(1)^{d}\times S_{d}$ ensemble of unitaries. The results presented here are new as far as we know and have played a central role for the exact computation of the $C$-ensemble moment operators.  

\subsection{Generalities}

\begin{figure}[!ht]
	\centering
	\includegraphics[width=0.6\textwidth]{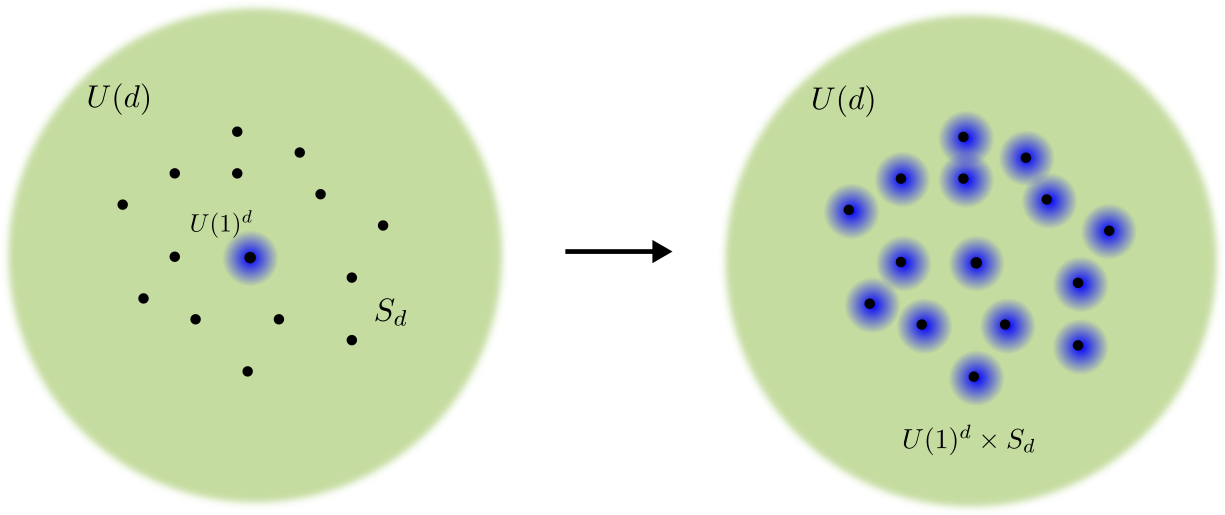}
	\caption{Cartoon of the $U(1)^{d}\times S_{d}$ ensemble as a subset of the unitary group. Separately, the $d!$ permutation matrices represented by black points are not enough to become dense over the whole unitaries, nor is the ``infinite'' number of diagonal unitaries represented by a blurred blue region around the identity, {\em i.e.}, ``the center of the $U(d)$ ball''. In the $U(1)^{d}\times S_{d}$ ensemble, each permutation is blurred by the $U(1)^{d}$ action, and the ensemble becomes sufficiently dense over the unitaries to correctly reproduce two-moments and become a 1-design. However, it is still not dense enough to yield into higher-order designs.}
	\label{fig:u1sd_ensemble_picture}
\end{figure}

Analogously to the Haar distributed unitaries and the $C$- ensemble, the $2k$th-moment operator for the $U(1)^{d}\times S_{d}$ ensemble is defined as,

\begin{equation*}
	\hat{\Phi}_{2k}^{U(1)^{d}\times S_{d}}=\frac{1}{\Vol(U(1)^{d}\times S_{d})}\int_{U(1)^{d}\times S_{d}} U^{\otimes k}\otimes U^{\dagger \otimes k} \, d\mu[U]_{U(1)^{d}\times S_{d}}.
\end{equation*} 
Here, $d\mu[U]_{U(1)^{d}\times S_{d}}$ is the corresponding (not normalized) ensemble Haar measure\footnote{The existence of this measure is already guaranteed by the compactness of the group.} and $\Vol(U(1)^{d}\times S_{d})$ is the normalization factor (a.k.a ensemble partition function). In contrast to the full $U(d)$ distributed case, for diagonal unitaries the Haar measure is basis dependent. However, a simple representation of the measure can be set up by expanding each diagonal unitary in the particularly chosen computational basis as $e^{i\vec{\phi}}=\sum_{l}e^{i\phi_{l}}\ket{l}\bra{l}$. Within this representation the $U(1)^{d}$ phase integral becomes an integral over the $d$-tori, $T^{d}\cong S^{1}\times \dots \times S^{1}$, and the permutations $\sigma$ in $S_{d}$ acts naturally as $\sigma\ket{l}=\ket{\sigma(l)}$,

\begin{equation*}
	\frac{1}{\Vol(U(1)^{d}\times S_{d})}\int_{U(1)^{d}\times S_{d}} (\cdots) \, d\mu[U]_{U(1)^{d}\times S_{d}} \longrightarrow \frac{1}{(2\pi)^{d}\, d!} \sum_{\sigma \in S_{d}} \int_{T^{d}}\,(\cdots) \prod^{d}_{l=1} d\phi_{l}
\end{equation*} 
For two-moments the integral is easily performed and coincides with $\hat{\Phi}_{2}^{U(d)}$,

\begin{equation}
	\label{eq:app_U1_Sd_two_moment}
	\hat{\Phi}_{2}^{U(1)^{d}\times S_{d}}= \frac{1}{d!}\sum_{\sigma \in S_{d}} \ket{\sigma(l)\, l} \bra{l \,\sigma(l)} = \frac{1}{d}\,\text{SWAP}.
\end{equation}
The fact that the $U(1)^{d}\times S_{d}$ ensemble is a unitary 1-design when neither the $S_{d}$ \emph{discrete} ensemble of permutation matrices nor the $U(1)^{d}$ \emph{continuous} ensemble of diagonal unitaries are 1-designs is an example of the requirements to correctly sample random unitaries. {\em i.e.}, the finite
 $d!$ number of separated unitaries in the $S_{d}$ ensemble is not enough to model random correlations, but neither is the close ``infinite'' number of diagonal unitaries. There must be a balance between the two (see Fig.\ref{fig:u1sd_ensemble_picture}).

\subsection{Four moments}

The strategy for the computation of the four moment operator is similar to the one used for the evaluation of the two moment operator. By expanding each diagonal unitary in the computational basis,

\begin{equation*}
	\hat{\Phi}_{4}^{U(1)^{d}\times S_{d}}=\frac{1}{(2\pi)^{d}\, d!}\sum_{\sigma \in S_{d}} \sum_{l,m,k,s} \ket{l\, k\, \sigma(s) \, \sigma(m)} \bra{\sigma(l) \, \sigma(k) \, s \, m} \int_{T^{d}} e^{i(\phi_{l}+\phi_{k}-\phi_{s}-\phi_{m})}\prod_{l=1}^{d}d\phi_{l} 
\end{equation*}
The phase integral only contributes either for $l=k=s=m$, $l=m \, \land \, k=s$ or $l=s \, \land \, k=m$. Therefore, 
	
\begin{align}
	\label{eq:pre_u1sd_fourmoment}
	\hat{\Phi}_{4}^{U(1)^{d}\times S_{d}}=&-\frac{1}{d!}\sum_{\sigma \in S_{d}}\sum_{l} \ket{l \, l \,\sigma(l) \, \sigma(l)}\bra{\sigma(l) \, \sigma(l) \, l \, l} \notag \\ 
	&+ \frac{1}{d!}\sum_{\sigma \in S_{d}}\sum_{l,k}\ket{l \, k \,\sigma(l) \, \sigma(k)}\bra{\sigma(l) \, \sigma(k) \, l \, k}\\ 
	& + \frac{1}{d!}\sum_{\sigma \in S_{d}}\sum_{l,k}\ket{l \, k \,\sigma(k) \, \sigma(l)}\bra{\sigma(l) \, \sigma(k) \, k \, l}. \notag
\end{align}
Notice that while the first sum involves $d (d!)$ terms, the other two have $d^{2} (d!)$. Although those sums seem rather involved one can gain intuition about what they actually represent by looking at their action on an arbitrary computational basis state $\ket{R_{1} \, R_{2} \, R_{3} \, R_{4}}$, {\em e.g.}, for the second term,

\begin{align*}
	&\sum_{\sigma \in S_{d}}\sum_{l,k}\ket{l \, k \,\sigma(l) \, \sigma(k)}\braket{\sigma(l) \, \sigma(k) \, l \, k| R_{1} \, R_{2} \, R_{3} \, R_{4} } \\
	&=\sum_{\sigma \in S_{d}}\delta_{R_{1},\sigma(R_{3})}\delta_{R_{2},\sigma(R_{4})}\ket{R_{3}\, R_{4} \, \sigma(R_{3}) \, \sigma(R_{4})}.
\end{align*}
The number of permutations $\sigma$ that solve $\sigma(R_{3})=R_{1}$ and $\sigma(R_{2})=R_{4}$ simultaneously is easily obtained by counting the cardinality of the restricted invariant sub-groups, {\em i.e.}, $S_{d-2}$ for $R_{1}\neq R_{2}$ and $R_{3} \neq R_{4}$, or $S_{d-1}$ for  $R_{1}=R_{2}$ and $R_{3}=R_{4}$. Hence,

\begin{align}
	\label{eq:copy_example}
	&\sum_{\sigma \in S_{d}}\delta_{R_{1},\sigma(R_{3})}\delta_{R_{2},\sigma(R_{4})}\ket{R_{3}\, R_{4} \, \sigma(R_{3}) \, \sigma(R_{4})} \\ \notag
	&= \left((d-1)! \delta_{R_{1},R_{2}}\delta_{R_{3},R_{4}}+(d-2)!(1-\delta_{R_{1},R_{2}})(1-\delta_{R_{3},R_{4}}) \right)\ket{R_{3} \, R_{4} \, R_{1} \, R_{2}}.
\end{align}
Diagrammatically, the same can be achieved by means of the COPY:\,$\mathcal{H}\rightarrow \mathcal{H}^{\otimes 2}$ gate defined on the computational basis as $\text{COPY}\ket{l}=\ket{ll}$,

\begin{equation*}
	\begin{tikzpicture}
	\begin{pgfonlayer}{nodelayer}
		\node [style=none] (0) at (-0.5, 0) {};
		\node [style=none] (1) at (0.5, 1) {};
		\node [style=none] (2) at (0.5, -1) {};
		\node [style=none] (3) at (-1.75, 0) {};
		\node [style=none] (4) at (-3.75, 0) {$\displaystyle{\text{COPY}=}$};
	\end{pgfonlayer}
	\begin{pgfonlayer}{edgelayer}
		\draw (3.center) to (0.center);
		\draw (0.center) to (1.center);
		\draw (0.center) to (2.center);
	\end{pgfonlayer}
\end{tikzpicture}
\end{equation*}
Therefore, for the five terms in \eqref{eq:copy_example},

\begin{equation*}
	\begin{tikzpicture}
	\begin{pgfonlayer}{nodelayer}
		\node [style=none] (4) at (-10.5, 2) {};
		\node [style=none] (5) at (-10.5, 0) {};
		\node [style=none] (6) at (-8.5, 0) {};
		\node [style=none] (8) at (-8.5, 2) {};
		\node [style=none] (9) at (-8, 2.5) {};
		\node [style=none] (10) at (-8, 1.5) {};
		\node [style=none] (11) at (-10.5, 2) {};
		\node [style=none] (12) at (-11, 2.5) {};
		\node [style=none] (13) at (-11, 1.5) {};
		\node [style=none] (14) at (-8.5, 0) {};
		\node [style=none] (15) at (-8, 0.5) {};
		\node [style=none] (16) at (-8, -0.5) {};
		\node [style=none] (17) at (-10.5, 0) {};
		\node [style=none] (18) at (-10.5, 0) {};
		\node [style=none] (19) at (-11, 0.5) {};
		\node [style=none] (20) at (-11, -0.5) {};
		\node [style=none] (22) at (3, 0) {};
		\node [style=none] (24) at (5, 2) {};
		\node [style=none] (25) at (5.5, 2.5) {};
		\node [style=none] (26) at (5.5, 1.5) {};
		\node [style=none] (28) at (2.5, 2.5) {};
		\node [style=none] (29) at (2.5, 1.5) {};
		\node [style=none] (31) at (5.5, 0.5) {};
		\node [style=none] (32) at (5.5, -0.5) {};
		\node [style=none] (33) at (3, 0) {};
		\node [style=none] (34) at (3, 0) {};
		\node [style=none] (35) at (2.5, 0.5) {};
		\node [style=none] (36) at (2.5, -0.5) {};
		\node [style=none] (37) at (-3.75, 2) {};
		\node [style=none] (38) at (-1.75, 0) {};
		\node [style=none] (39) at (-1.25, -0.5) {};
		\node [style=none] (40) at (-1.25, 0.5) {};
		\node [style=none] (41) at (-4.25, -0.5) {};
		\node [style=none] (42) at (-4.25, 0.5) {};
		\node [style=none] (43) at (-1.25, 1.5) {};
		\node [style=none] (44) at (-1.25, 2.5) {};
		\node [style=none] (45) at (-3.75, 2) {};
		\node [style=none] (46) at (-3.75, 2) {};
		\node [style=none] (47) at (-4.25, 1.5) {};
		\node [style=none] (48) at (-4.25, 2.5) {};
		\node [style=none] (51) at (12.25, -0.5) {};
		\node [style=none] (52) at (12.25, 0.5) {};
		\node [style=none] (53) at (9.25, -0.5) {};
		\node [style=none] (54) at (9.25, 0.5) {};
		\node [style=none] (55) at (12.25, 1.5) {};
		\node [style=none] (56) at (12.25, 2.5) {};
		\node [style=none] (59) at (9.25, 1.5) {};
		\node [style=none] (60) at (9.25, 2.5) {};
		\node [style=none] (62) at (-6, 1) {$\displaystyle{-(d-2)!}$};
		\node [style=none] (63) at (-14.75, 1) {$\displaystyle{(d-1)!\left(1+\frac{1}{d-1}\right)}$};
		\node [style=none] (64) at (0.75, 1) {$\displaystyle{-(d-2)!}$};
		\node [style=none] (65) at (7.5, 1) {$\displaystyle{+(d-2)!}$};
	\end{pgfonlayer}
	\begin{pgfonlayer}{edgelayer}
		\draw [in=-180, out=0] (4.center) to (6.center);
		\draw (8.center) to (9.center);
		\draw (8.center) to (10.center);
		\draw [style={white_edge}, in=180, out=0] (5.center) to (8.center);
		\draw [in=180, out=0, looseness=1.25] (5.center) to (8.center);
		\draw (11.center) to (12.center);
		\draw (11.center) to (13.center);
		\draw (14.center) to (15.center);
		\draw (14.center) to (16.center);
		\draw (18.center) to (19.center);
		\draw (18.center) to (20.center);
		\draw (24.center) to (25.center);
		\draw (24.center) to (26.center);
		\draw [style={white_edge}, in=180, out=0] (22.center) to (24.center);
		\draw [in=180, out=0, looseness=1.25] (22.center) to (24.center);
		\draw (34.center) to (35.center);
		\draw (34.center) to (36.center);
		\draw [style={white_edge}, in=180, out=0, looseness=0.75] (28.center) to (31.center);
		\draw [style={white_edge}, in=180, out=0, looseness=0.75] (29.center) to (32.center);
		\draw [in=-180, out=0, looseness=0.50] (28.center) to (31.center);
		\draw [in=-180, out=0, looseness=0.50] (29.center) to (32.center);
		\draw (38.center) to (39.center);
		\draw (38.center) to (40.center);
		\draw [style={white_edge}, in=-180, out=0] (37.center) to (38.center);
		\draw [in=-180, out=0, looseness=1.25] (37.center) to (38.center);
		\draw (46.center) to (47.center);
		\draw (46.center) to (48.center);
		\draw [style={white_edge}, in=-180, out=0, looseness=0.75] (41.center) to (43.center);
		\draw [style={white_edge}, in=-180, out=0, looseness=0.75] (42.center) to (44.center);
		\draw [in=-180, out=0, looseness=0.50] (41.center) to (43.center);
		\draw [in=-180, out=0, looseness=0.50] (42.center) to (44.center);
		\draw (60.center) to (52.center);
		\draw (59.center) to (51.center);
		\draw [style={white_edge}] (54.center) to (56.center);
		\draw [style={white_edge}] (53.center) to (55.center);
		\draw (54.center) to (56.center);
		\draw (53.center) to (55.center);
	\end{pgfonlayer}
\end{tikzpicture}
\end{equation*}
By repeating the same procedure for all the terms in the right hand side of \eqref{eq:pre_u1sd_fourmoment} the \emph{exact} four-moment operator of the $U(1)^{d}\times S_{d}$ ensemble yields,

\begin{equation}
	\label{eq:Appendix_U1_Sd_fourmoment}
	\begin{tikzpicture}
	\begin{pgfonlayer}{nodelayer}
		\node [style=none] (4) at (-10.5, 2) {};
		\node [style=none] (5) at (-10.5, 0) {};
		\node [style=none] (6) at (-8.5, 0) {};
		\node [style=none] (8) at (-8.5, 2) {};
		\node [style=none] (9) at (-8, 2.5) {};
		\node [style=none] (10) at (-8, 1.5) {};
		\node [style=none] (11) at (-10.5, 2) {};
		\node [style=none] (12) at (-11, 2.5) {};
		\node [style=none] (13) at (-11, 1.5) {};
		\node [style=none] (14) at (-8.5, 0) {};
		\node [style=none] (15) at (-8, 0.5) {};
		\node [style=none] (16) at (-8, -0.5) {};
		\node [style=none] (17) at (-10.5, 0) {};
		\node [style=none] (18) at (-10.5, 0) {};
		\node [style=none] (19) at (-11, 0.5) {};
		\node [style=none] (20) at (-11, -0.5) {};
		\node [style=none] (22) at (3, 0) {};
		\node [style=none] (24) at (5, 2) {};
		\node [style=none] (25) at (5.5, 2.5) {};
		\node [style=none] (26) at (5.5, 1.5) {};
		\node [style=none] (28) at (2.5, 2.5) {};
		\node [style=none] (29) at (2.5, 1.5) {};
		\node [style=none] (31) at (5.5, 0.5) {};
		\node [style=none] (32) at (5.5, -0.5) {};
		\node [style=none] (33) at (3, 0) {};
		\node [style=none] (34) at (3, 0) {};
		\node [style=none] (35) at (2.5, 0.5) {};
		\node [style=none] (36) at (2.5, -0.5) {};
		\node [style=none] (37) at (-3.75, 2) {};
		\node [style=none] (38) at (-1.75, 0) {};
		\node [style=none] (39) at (-1.25, -0.5) {};
		\node [style=none] (40) at (-1.25, 0.5) {};
		\node [style=none] (41) at (-4.25, -0.5) {};
		\node [style=none] (42) at (-4.25, 0.5) {};
		\node [style=none] (43) at (-1.25, 1.5) {};
		\node [style=none] (44) at (-1.25, 2.5) {};
		\node [style=none] (45) at (-3.75, 2) {};
		\node [style=none] (46) at (-3.75, 2) {};
		\node [style=none] (47) at (-4.25, 1.5) {};
		\node [style=none] (48) at (-4.25, 2.5) {};
		\node [style=none] (51) at (12.25, -0.5) {};
		\node [style=none] (52) at (12.25, 0.5) {};
		\node [style=none] (53) at (9.25, -0.5) {};
		\node [style=none] (54) at (9.25, 0.5) {};
		\node [style=none] (55) at (12.25, 1.5) {};
		\node [style=none] (56) at (12.25, 2.5) {};
		\node [style=none] (59) at (9.25, 1.5) {};
		\node [style=none] (60) at (9.25, 2.5) {};
		\node [style=none] (62) at (-6, 1) {$\displaystyle{-\frac{1}{d(d-1)}}$};
		\node [style=none] (63) at (-12.75, 1) {$\displaystyle{\frac{d+1}{d(d-1)}}$};
		\node [style=none] (66) at (-10.5, -2.5) {};
		\node [style=none] (67) at (-8.5, -4.5) {};
		\node [style=none] (68) at (-8, -5) {};
		\node [style=none] (69) at (-8, -4) {};
		\node [style=none] (70) at (-11, -5) {};
		\node [style=none] (71) at (-11, -4) {};
		\node [style=none] (72) at (-8, -3) {};
		\node [style=none] (73) at (-8, -2) {};
		\node [style=none] (74) at (-10.5, -2.5) {};
		\node [style=none] (75) at (-10.5, -2.5) {};
		\node [style=none] (76) at (-11, -3) {};
		\node [style=none] (77) at (-11, -2) {};
		\node [style=none] (78) at (0.75, 1) {$\displaystyle{-\frac{1}{d(d-1)}}$};
		\node [style=none] (79) at (7.5, 1) {$\displaystyle{+\frac{1}{d(d-1)}}$};
		\node [style=none] (80) at (-12.75, -3.5) {$\displaystyle{-\frac{1}{d(d-1)}}$};
		\node [style=none] (82) at (-3.75, -4.5) {};
		\node [style=none] (83) at (-1.75, -2.5) {};
		\node [style=none] (84) at (-1.25, -2) {};
		\node [style=none] (85) at (-1.25, -3) {};
		\node [style=none] (86) at (-4.25, -2) {};
		\node [style=none] (87) at (-4.25, -3) {};
		\node [style=none] (88) at (-1.25, -4) {};
		\node [style=none] (89) at (-1.25, -5) {};
		\node [style=none] (90) at (-3.75, -4.5) {};
		\node [style=none] (91) at (-3.75, -4.5) {};
		\node [style=none] (92) at (-4.25, -4) {};
		\node [style=none] (93) at (-4.25, -5) {};
		\node [style=none] (94) at (-6, -3.5) {$\displaystyle{-\frac{1}{d(d-1)}}$};
		\node [style=none] (95) at (5.5, -5) {};
		\node [style=none] (96) at (5.5, -4) {};
		\node [style=none] (97) at (2.5, -5) {};
		\node [style=none] (98) at (2.5, -4) {};
		\node [style=none] (99) at (5.5, -3) {};
		\node [style=none] (100) at (5.5, -2) {};
		\node [style=none] (101) at (2.5, -3) {};
		\node [style=none] (102) at (2.5, -2) {};
		\node [style=none] (103) at (0.75, -3.5) {$\displaystyle{+\frac{1}{d(d-1)}}$};
		\node [style=none] (104) at (-16.5, 1.25) {$\displaystyle{\hat{\Phi}_{4}^{U(1)^{d}\times S_{d}}=}$};
	\end{pgfonlayer}
	\begin{pgfonlayer}{edgelayer}
		\draw [in=-180, out=0] (4.center) to (6.center);
		\draw (8.center) to (9.center);
		\draw (8.center) to (10.center);
		\draw [style={white_edge}, in=180, out=0] (5.center) to (8.center);
		\draw [in=180, out=0, looseness=1.25] (5.center) to (8.center);
		\draw (11.center) to (12.center);
		\draw (11.center) to (13.center);
		\draw (14.center) to (15.center);
		\draw (14.center) to (16.center);
		\draw (18.center) to (19.center);
		\draw (18.center) to (20.center);
		\draw (24.center) to (25.center);
		\draw (24.center) to (26.center);
		\draw [style={white_edge}, in=180, out=0] (22.center) to (24.center);
		\draw [in=180, out=0, looseness=1.25] (22.center) to (24.center);
		\draw (34.center) to (35.center);
		\draw (34.center) to (36.center);
		\draw [style={white_edge}, in=180, out=0, looseness=0.75] (28.center) to (31.center);
		\draw [style={white_edge}, in=180, out=0, looseness=0.75] (29.center) to (32.center);
		\draw [in=-180, out=0, looseness=0.50] (28.center) to (31.center);
		\draw [in=-180, out=0, looseness=0.50] (29.center) to (32.center);
		\draw (38.center) to (39.center);
		\draw (38.center) to (40.center);
		\draw [style={white_edge}, in=-180, out=0] (37.center) to (38.center);
		\draw [in=-180, out=0, looseness=1.25] (37.center) to (38.center);
		\draw (46.center) to (47.center);
		\draw (46.center) to (48.center);
		\draw [style={white_edge}, in=-180, out=0, looseness=0.75] (41.center) to (43.center);
		\draw [style={white_edge}, in=-180, out=0, looseness=0.75] (42.center) to (44.center);
		\draw [in=-180, out=0, looseness=0.50] (41.center) to (43.center);
		\draw [in=-180, out=0, looseness=0.50] (42.center) to (44.center);
		\draw (60.center) to (52.center);
		\draw (59.center) to (51.center);
		\draw [style={white_edge}] (54.center) to (56.center);
		\draw [style={white_edge}] (53.center) to (55.center);
		\draw (54.center) to (56.center);
		\draw (53.center) to (55.center);
		\draw (67.center) to (68.center);
		\draw (67.center) to (69.center);
		\draw [style={white_edge}, in=-180, out=0] (66.center) to (67.center);
		\draw [in=-180, out=0, looseness=1.25] (66.center) to (67.center);
		\draw (75.center) to (76.center);
		\draw (75.center) to (77.center);
		\draw [style={white_edge}, in=180, out=0, looseness=0.75] (70.center) to (73.center);
		\draw [in=180, out=0, looseness=0.75] (70.center) to (73.center);
		\draw [style={white_edge}, in=-180, out=0, looseness=0.50] (71.center) to (72.center);
		\draw [in=180, out=0] (71.center) to (72.center);
		\draw (83.center) to (84.center);
		\draw (83.center) to (85.center);
		\draw [style={white_edge}, in=180, out=0] (82.center) to (83.center);
		\draw [in=180, out=0, looseness=1.25] (82.center) to (83.center);
		\draw (91.center) to (92.center);
		\draw (91.center) to (93.center);
		\draw [style={white_edge}, in=-180, out=0, looseness=0.75] (86.center) to (89.center);
		\draw [in=-180, out=0, looseness=0.75] (86.center) to (89.center);
		\draw [style={white_edge}, in=180, out=0, looseness=0.50] (87.center) to (88.center);
		\draw [in=-180, out=0] (87.center) to (88.center);
		\draw (101.center) to (96.center);
		\draw [style={white_edge}] (98.center) to (99.center);
		\draw (98.center) to (99.center);
		\draw [style={white_edge}] (102.center) to (95.center);
		\draw (102.center) to (95.center);
		\draw [style={white_edge}] (97.center) to (100.center);
		\draw (97.center) to (100.center);
	\end{pgfonlayer}
\end{tikzpicture}
\end{equation}
Some comments about this result deserve special attention. First, the ensemble yields a unitary 1- design, {\em i.e.}, at the level of two-moments, both the $U(1)^{d}\times S_{d}$ and $U(d)$-Haar ensembles coincide (see \ref{eq:app_U1_Sd_two_moment}). However, it does not yield a higher-order design\footnote{Recall that the Haar theorem implicitly prevents any arbitrary unitary ensemble with a non-uniform measure to become an infinite design (\emph{i.e.}, that all moments coincide), but does not inhibit that both measures share the same moments up to some finite higher order.}. Nevertheless, the tensor structure of four- and two- moment operators $\hat{\Phi}^{U(1)^{d}\times S_{d}}_{2},\,\hat{\Phi}^{U(1)^{d}\times S_{d}}_{4}$ is the one of an spectral decoupled ensemble \eqref{def:spectral_decoupling}, providing a further example of why spectral decoupling does not imply a design.


Second, notice that in the large-$d$ limit the leading  contribution to $\hat{\Phi}_{4}^{U(1)^{d}\times S_{d}}$ in (\ref{eq:Appendix_U1_Sd_fourmoment}) is given by a $\mathcal{O}(d^{-1})$ term, whereas for $\hat{\Phi}_{4}^{U(d)}$ this leading contribution is of order $\mathcal{O}(d^{-2})$ (see Appendix \ref{Appendix:RMT_Haar}). This subtle difference will result in a $\mathcal{O}(d^{-1})$ contribution for the $C$-ensemble four-point operator.  

Finally let us make a short remark about the orthogonal $C$-ensemble. In that case, the conventional time reversal invariance imposes that
the ensemble measure will be invariant under $\mathbb{Z}_{2}^{d}\times S_{d}$ rather than $U(1)^{d}\times S_{d}$ for the unitary case. By similar arguments as the ones presented at the beginning of this Appendix, it is easy to show that the $\mathbb{Z}_{2}^{d}\times S_{d}$ ensemble yields an orthogonal 1-design, \emph{i.e.},

\begin{equation}
	\hat{\Phi}^{\mathbb{Z}_{2}^{d}\times S_{d}}_{2}=\hat{\Phi}^{O(d)}_{2}=\frac{1}{d}\text{SWAP}.
\end{equation}
Therefore, by the factorization \eqref{eq:2kmoment_factorization} the orthogonal $C$-ensemble will also yield an orthogonal 1-design\footnote{This was also expected from the discussion in Section \ref{section:C_ensemble}.}.

\section{\texorpdfstring{$C$}{}-ensemble Partition Function}
\label{appendix:partition_function}

During the main part of the paper we have focused our efforts on finding, and also giving a physical interpretation to the $C$-ensemble moment operators. However, we have not given enough relevance, apart from the suggested entropy/complexity relation, to the normalization factor of the $C$-ensemble measure, \emph{i.e.}, the ensemble partition function,

\begin{equation}
  \label{app_partition_func:eq_vol}
  \Vol(H)=\int_{U(d)}\delta_{\perp}(CHC^{\dagger})[C^{\dagger}dC].
\end{equation}
Although our method for finding the $C$-ensemble first and second moment operators does not require the explicit evaluation of (\ref{app_partition_func:eq_vol}), in this appendix, we will find the $C$-ensemble partition function and show that it is an interesting object by itself.

\subsection{Bootstrapping the partition function}

To set the grounds, first let us discuss the symmetries of the partition function. Due the left and right invariance of the Haar measure, (\ref{app_partition_func:eq_vol}) is invariant under rotations of the Hamiltonian $H\rightarrow UHU^{\dagger}$ for every unitary $U$ and $H$ Hermitian,
 
 \begin{equation}
   \Vol{(UHU^{\dagger})}=\Vol{(H)}.  
 \end{equation}
This implies that $\Vol{(H)}$ must be a symmetric function $f(\{E_{\mu}\})$ of the eigenvalues $\{E_{\mu}\}$. Additionally, for complex-Hermitian matrices the off-diagonal delta $\delta_{\perp}(\cdots)$ of a matrix argument involves $2\binom{d}{2}$ scalar delta functions\footnote{For an orthogonal $C$-ensemble, the appropriate one for a system with conventional time reversal invariance, there will be $\binom{d}{2}$ independent scalar delta functions instead.}, so under a global scaling $H\rightarrow \lambda H$ with $\lambda>0$  , the partition function transforms as,
 
 \begin{equation*}
   \Vol{(\lambda H)}=f(\{\lambda E_{\mu}\})=\lambda^{-2\binom{d}{2}}f(\{E_{\mu}\}).
 \end{equation*}
This scaling dimension $2\binom{d}{2}$ fixes the units of the partition function to be $[\text{Energy}]^{-2\binom{d}{2}}$ where $[\text{Energy}]$ are the energy units of the Hamiltonian (recall that in this work we set $\hbar=1$). A physically relevant property of the ensemble is the invariance under energy shifts $H\rightarrow H+\lambda \mathbb{I}$ with $\lambda$ real, which implies that there is no privileged energy level and that only the relative differences between them matter\footnote{This was also pointed out in the main part of the text, where this eigenstate ``democracy'' is a result of the underlying $S_{d}$ symmetry on the ensemble measure.}. As a consequence the partition function solely depends upon the $\binom{d}{2}$ energy frequencies $\{E_{\mu \nu}=E_{\mu}-E_{\nu}\}$,
 
 \begin{equation*}
   \Vol{(\lambda H)}=f(\{\lambda E_{\nu \mu}\}).
 \end{equation*}
 Before continuing, let us remark that all these three properties, rotation invariance, scaling behavior and shift invariance, hold also for an orthogonal $C$-ensemble with the only difference that the scaling dimension would be $\binom{d}{2}$ instead of $2\binom{d}{2}$ of the unitary case. This observation is relevant and we will comment more on it after the following result.

 \begin{theorem}[Partition function duality]\label{theorem:partition_function_inversion_formula}
 For every $d$-dimensional Hermitian Hamiltonian $H$ with non-zero determinant, $\det(H)\neq 0$, and inverse $H^{-1}$ it holds,
 \begin{equation*}
     \Vol({H})=\det(H)^{-2(d-1)}\Vol{(H^{-1})}.
 \end{equation*}
\end{theorem}
This theorem establishes a duality between the partition functions associated with two different $C$-ensembles, one associated with $H$ and the other with its inverse $H^{-1}$. The theorem is based on a relationship between unitary and Gaussian matrix integrals presented by us in Appendix \ref{Appendix:RMT_Haar}.

This result is surprising for the following reason: Before considering this duality, we already have shown that the partition function is translationally invariant and possesses a well defined scaling behavior. The theorem tell us that the partition function presents some symmetric behavior under inversions $H\rightarrow H^{-1}$ as well. By combining all this information, it is straightforward to see that a possible candidate partition function, which satisfies all those symmetries is given by the inverse square Vandermonde determinant,

\begin{equation}
\label{app_partition_func:eq_dyson_index}
 \frac{1}{\Delta (H)^{2}}=\prod_{l<m} \frac{1}{|E_{l}-E_{m}|^{2
 }}.
\end{equation} 
Here, we highlight the suggestive square over the Vandermonde and conjecture that it must be interpreted as the Dyson index of the Hamiltonian symmetry class, {\em i.e.}, $\beta=1$ for an orthogonal $C$-ensemble and $\beta=4$ for a symplectic one. Returning to the partition function, notice that by multiplying this particular candidate by any invariant function under translations, scalings and inversions we end up with another possible partition function candidate. More formally, we can parametrize any candidate as,

\begin{equation*}
  \Vol{(H)}=\frac{g(\{ E_{l} \})}{\Delta(H)^{2}},
 \end{equation*} 
where $g(\{ E_{l} \})$ is an $\text{SL}(2,\mathbb{R})$-invariant function\footnote{This is the same to say that $g(\{ E_{l} \})$ must be invariant under real M\"obious transformations.}. It is a well known result that $\text{SL}(2,\mathbb{R})$ invariants can be build up from the cross-ratios \cite{Anderson/Cross_ratio/2005}, 

\begin{equation*}
   (x_{1}, x_{2} : x_{3}, x_{4} )=\frac{(x_{3}-x_{1})(x_{4}-x_{2})}{(x_{4}-x_{1})(x_{3}-x_{2})},
\end{equation*}
and their permutations (forming the anharmonic group). Thus, the general form of the partition function, is constrained by these symmetries, must be

\begin{equation}
  \Vol{(H)}=\frac{g(\{ (E_{l}, E_{k} : E_{s}, E_{m} ) \})}{\Delta(H)^{2}}.
  \label{eq:C2_sl2_partition_function}
\end{equation}
Now, let us remark the following: by analytical continue \eqref{eq:C2_sl2_partition_function} to $\mathbb{C}$, {\em i.e.}, by sending each eigenvalue $E_{l}\rightarrow z_{l}$ with $z_{l}$ complex, the partition function (\ref{eq:C2_sl2_partition_function}) looks identical as a typical $d$-point vacuum expectation value of a two-dimensional conformal field theory (CFT) \cite{Blumenhagen/CFT/2009} $\braket{\phi(z_{1})\dots \phi(z_{l})}_{\text{CFT}}$, where $z_{l}$ are the coordinates of the fields. In fact, the global $\text{SL}(2,\mathbb{C})$ symmetry of these theories forces $n$-point functions with $n>3$ to be functions of cross ratios, just like our partition function. Motivated by this observation, we propose that, \emph{the $C$-ensemble partition function $\Vol(H)$ is the same as a $d$-point vacuum expectation value of some primary scalar fields $\phi_{l}$, each with equal scaling dimension, of some two-dimensional conformal field theory}.

It is worth mentioning that Hermitian single matrix models, {\em i.e.}, Hamiltonian RMT ensembles, admit an interpretation as CFT's in some particular limit (see \cite{les_houches_CFT/RMT/2004}). However, what is unexpected, at least for us, is that a unitary matrix model with an ``exotic'' (non-smooth) probability measure, such as the $C$-ensemble, also yields a partition function related to a CFT. 

\subsection{Partition function, RMT evaluation}

In the previous section, we constrained the general form of the partition function in terms of its symmetries. Here, we will use a more direct approach based on RMT machinery. At first sight, this approach may seem to carry less physical content in comparison with the previous one. However, we will show in the next section that both perspectives are, in fact, complementary and only by combining them we can capture a complete picture. For simplicity, in this section, we assume a Hamiltonian with a non-degenerate spectrum.

To start, notice that the off-diagonal delta $\delta_{\perp}(\cdots)$ in the $C$-ensemble measure can be set in a convenient integral representation by inserting a diagonal delta counter-term $\delta_{\parallel}(B)$, 

\begin{equation*}
  \delta_{\perp}(CHC^{\dagger}) \propto \int [dB] \delta_{\parallel}(B) e^{i\Tr(BCHC^{\dagger})},
\end{equation*}
where the Hermitian auxiliary operator $B=B^{\dagger}$ can be interpreted as an associated ``momentum'' operator\footnote{This interpretation arises by noticing that the matrix plane wave $e^{i\Tr CBC^{\dagger}H}$ is an eigenfunction of the differential operator $\sum_{lm}\frac{\partial}{\partial B_{lm}}$ for every unitary $C$.}. The measure $[dB]$ denotes the usual flat-measure, and for simplicity, we have dropped the Fourier transform normalization factor, which will be fixed at the end. In terms of this representation the partition function reads,

\begin{equation*}
  \Vol{(H)} \propto \int [dB]\delta_{\parallel}(B)\int_{U(d)}[C^{\dagger}dC]e^{i\Tr{CBC^{\dagger}H}}.
\end{equation*}
By the Haar measure invariance, both $H$ and $B$ can be replaced in the exponential by their diagonal matrices of eigenvalues $D$ and $E$, respectively. Also, by using the polar decomposition $B\rightarrow UDU^{\dagger}$, the partition function can be cast as an integral over one set of eigenvalues, $D$, and two sets of unitaries $C$ and $U$,

\begin{equation}
  \label{eq:C2_Eigenvalue_rep_partition_function}
  \Vol{(H)}\propto\int_{\mathbb{R}^{d}} dD\: \Delta(D)^2\,\int_{U(d)\times U(d)}[C^{\dagger}dC][U^{\dagger}dU]\delta_{\parallel}(UDU^{\dagger})e^{i\Tr{CDC^{\dagger}E}}.
\end{equation}
%

The diagonal delta $\delta_{\parallel}(UDU^{\dagger})$ can be cast in an integral representation by inserting the diagonal auxiliary operator $K$,

\begin{equation}
  \Vol(H)\propto\int_{\mathbb{R}^{d} \times \mathbb{R}^{d}} dK\, dD\, \Delta(D)^{2}\int_{U(d)\times U(d)}e^{i\Tr K U D U^{\dagger}}e^{i\Tr D C E C^{\dagger}}.
\end{equation}
Both unitary integrals can be exactly evaluated using the famous and elegant $U(d)$ Harish-Chandra-Itzykson-Zuber \cite{Eynard/RMT/2015, PZinn/HCIZ/2003} formula\footnote{For every pair of Hermitian operators $X$,$Y$, the integral $\int_{U(d)} e^{i\Tr X U Y U^{\dagger}}[U^{\dagger}dU] \propto \frac{\det e^{ix_{l}y_{m}}}{\Delta(X)\Delta(Y)}$, with $\{x_{l}\}$, $\{y_{l}\}$ the eigenvalues of $X$ and $Y$, respectively.}. Thus, we can get rid off the angular dependence and express the partition function as an integral involving only the eigenvalues,

\begin{equation}
  \Vol(H) \propto \int_{\mathbb{R}^{d} \times \mathbb{R}^{d}}  \frac{\det e^{i K_{l} D_{m}} \, \det e^{i D_{l} E_{m}}}{\Delta(K) \Delta(E)}\,dK \, dD.
\end{equation}
This integral can be evaluated easily using the following trick: all terms inside the integrand are anti-symmetric functions of their arguments. This implies that after expanding each determinant, the remaining $d!$ integrals per determinant are equal, leading to

\begin{equation}
  \int_{\mathbb{R}^{d} \times \mathbb{R}^{d}}  \frac{\det e^{i K_{l} D_{m}} \, \det e^{i D_{l} E_{m}}}{\Delta(K) \Delta(E)}\,dK \, dD = (d!)^{2} \int_{\mathbb{R}^{d} \times \mathbb{R}^{d}} \frac{e^{i \Tr(K D)}e^{i \Tr(E D)}}{\Delta(K)\Delta(E)} dK \, dD.
\end{equation}
Finally, the $D$ integral yields a delta distribution which sets $K$ to be equal to $E$, and we recover the bootstrap solution presented in the previous section,

\begin{equation}
\label{eq:app_p_function-Normalization}
  \Vol(H) \propto \frac{1}{\Delta^{2}(H)}.
\end{equation}
From here, the proportionality factor can be simply recovered by averaging $\Vol(H)$ over some convenient RMT Hamiltonian ensemble, {\em e.g.}, for a Gaussian one, the off-diagonal delta $\delta_{\perp}(\cdots)$, together with the $U(d)$ invariance of the Gaussian measure, further reduces the averaged volume to $d$ one-dimensional Gaussian integrals over the diagonal elements of $H$, \emph{i.e.},

\begin{equation}
  \int e^{-\Tr H^{2}} \Vol(H)\, [dH]=\int_{U(d)}[C^{\dagger}dC]\int_{\text{GUE}}e^{-\Tr{H^{2}}}\delta_{\perp}(CHC^{\dagger})[dH]= \pi^{\frac{d}{2}}.
\end{equation}
Additionally, inserting \eqref{eq:app_p_function-Normalization}, and explicitly performing the eigenvalue integral \footnote{Notice that unexpectedly the $C$-ensemble partition function coincides exactly with the Jacobian of the Hermitian flat measure, {\em e.g.} $\Vol(H)[dH]=[U^{\dagger}dU]dE$.}, 

\begin{equation}
  \label{app:non_degenerate_partition_function}
  \Vol(H) = \frac{\prod_{l=1}^{d}l!}{\pi^{\binom{d}{2}}} \frac{1}{\Delta^{2}(H)}.
\end{equation}
Apart from the inverse square Vandermonde, we wish to point out that the proportional prefactors in \eqref{app:non_degenerate_partition_function} can be rearranged to explicitly reflect the symmetries of the ensemble. Concretely, by identifying the product $\prod_{l=1}^{d}l!$ as $d! \prod_{l=1}^{d-1}l!$, where $d!$ corresponds to the volume of the $S_{d}$ group and $\prod_{l=1}^{d-1}l!$ is the Barnes $G$-function \cite{Voros/Asymptotic_Barnes/1987}, which arises from the volume of the unitary group \cite{Zhang/Unitary_volume/2017,Zyczkowsky/Unitary_volume/2003}, the partition function reads

\begin{equation}
    \Vol(H) = 2^{\binom{d}{2}}\,\frac{\Vol(U(1)^{d} \times S_{d})}{\Vol(U(d))} \frac{1}{\Delta^{2}(H)}.
\end{equation}
From this result, it is straightforward to confirm our previous conjecture in \ref{app_partition_func:eq_dyson_index} regarding the interpretation of the exponent in the square Vandemonde as the Dyson-index $\beta$ of the symmetry class. Concretely, we propose:

\begin{equation}
    \Vol(H) = (1)^{\binom{d}{2}}\,\frac{\Vol(\mathbb{Z}_{2}^{d} \times S_{d})}{\Vol(O(d))} \frac{1}{|\Delta(H)|^{1}}.
\end{equation}
for the volume of the orthogonal ($\beta=1$) $C$-ensemble, and

\begin{equation}
\Vol(H) = (4)^{\binom{d}{2}}\,\frac{\Vol(U(1)^{d}\times U(1)^{d} \times S_{2d})}{\Vol(\text{SP}(2d))} \frac{1}{|\Delta(H)|^{4}}.
\end{equation}
for the symplectic ($\beta=4$) case.

Before finishing let us make the following remark. From the solution \eqref{app:non_degenerate_partition_function} it
is clear that the partition function $\Vol(H)$ diverges for eigenvalue collisions, {\em i.e.} $E_{l}\to E_{m}$. However, the fact that the integral representation \eqref{app_partition_func:eq_vol} remains valid even for a degenerate Hamiltonian means that this cannot be true at all. The upshot is that, as we mentioned initially, the solution \eqref{app:non_degenerate_partition_function} only holds for a non-degenerate Hamiltonian, and in the degenerate case the correct solution is given in terms of the ansatz presented in equation \eqref{eq:C2_sl2_partition_function}, where the uncertain $g$ function enters by regulating the eigenvalue collision, {\em i.e.}, $g$ factorizes like

\begin{equation}
  g = (E_{l}-E_{m})^{2}\,\Tilde{g}\quad \text{for} \quad E_{l} \to E_{m},
\end{equation}
with $\Tilde{g}$ an eigenvalue function free from singularities in the $E_{l} \to E_{m}$ limit. 

In order to provide a concrete example of how the $\tilde{g}$ function modifies the $C$-ensemble partition function in the presence of degeneracies, let us consider the particular case of two distinct degenerate levels, $\bar{E}_{1}$ and $\bar{E}_{2}$, with degeneracies $h_{1}$ and $h_{2}$, respectively. For this particular scenario, the diagonalized Hamiltonian takes the form 
\[
  H \to \left[\begin{array}{ c | c | c}
    \bar{E}_{1}\,\mathbb{I}_{h_{1}} & 0 & 0\\
    \hline
    0 & \bar{E}_{2}\,\mathbb{I}_{h_{2}} & 0 \\
    \hline
    0 & 0 & E_{\perp}
  \end{array}\right],
\]
where $\mathbb{I}_{h_{l}}$ is the identity in the $l$-subspace ($l=\{1,2\}$) spanned by the degenerate states, and $E_{\perp}$ is the $(d-(h_{1}+h_{2}))$ diagonal matrix of the remaining non-degenerate energy levels. After isolating the contribution from the non-degenerate level, the $C$-ensemble partition function factorizes as

\begin{equation}
\label{eq:p_fuction_splitting_example}
\Vol{(H)}=\Vol{(E_{\perp})}\tilde{g}(\bar{E}_{1},\bar{E}_{2},E_{\perp}),
\end{equation}
where $\Vol{(E_{\perp})}$ denotes the volume of the $C$-ensemble restricted to the non-degenerate levels, \emph{i.e.},

\begin{equation}
    \Vol(E_{\perp}) \propto \prod_{E_{l} < E_{m}\, \in \,E_{\perp}} \frac{1}{|E_{l}-E_{m}|^{2}}.
\end{equation}
To determine $\tilde{g}$, the trick is to employ the inversion formula \eqref{theorem:partition_function_inversion_formula}, which, as previously noted, remains valid even for a degenerate spectrum. After inserting the factorization \eqref{eq:p_fuction_splitting_example} into the inversion formula, $\tilde{g}$ is constrained to satisfy

\begin{equation}
\det(E_{\perp})^{2(h_{1}+h_{2})}\bar{E}_{1}^{2h_{1}(d-1)}\bar{E}_{2}^{2h_{2}(d-1)}\tilde{g}(\bar{E}_{1},\bar{E}_{2},E_{\perp}) = \tilde{g}\left(\frac{1}{\bar{E}_{1}},\frac{1}{\bar{E}_{2}},E_{\perp}^{-1}\right).
\end{equation}
This, combined with the translational and permutational invariance inherited from $\Vol(H)$, constrains $\tilde{g}$
(up to a normalization factor) to take the form

\begin{equation}
    \tilde{g}(\bar{E}_{1},\bar{E}_{2},E_{\perp}) = \delta_{h_{1},h_{2}}\left(\prod_{E \in E_{\perp}}|\bar{E}_{1} - E|\right)^{-\alpha}\left(\prod_{E \in E_{\perp}}|\bar{E}_{2} - E|\right)^{-\beta}|\bar{E}_{1}-\bar{E}_{2}|^{-\gamma},
\end{equation}
with exponents explicitly given by,

\begin{equation}
    \alpha = 2h_{1},\; \beta = 2h_{2},\; \text{and} \; \gamma = (h_{1}+h_{2})(h_{1}+h_{2}-1).
\end{equation}
The previous example can be extended to the case of three degenerate levels with degeneracies $(h_{1},h_{2},h_{3})$, leading to

\begin{equation}
    \tilde{g}(\{\bar{E}_{a}\},E_{\perp}) = \left(\prod_{a=1}^{3}\left(\prod_{E \in E_{\perp}}|\bar{E}_{a} - E|\right)^{-2h_{a}}\right)|\bar{E}_{1}-\bar{E}_{2}|^{x}|\bar{E}_{1}-\bar{E}_{3}|^{y}|\bar{E}_{2}-\bar{E}_{3}|^{z},
\end{equation}
where, again the $x,y,z$ -exponents, can be explicitly given by,

\begin{align*}
    2h_{1}\left(d-\sum_{a=1}^{3}h_{a}\right)+x+y & =2h_{1}(d-1),\\
    2h_{2}\left(d-\sum_{a=1}^{3}h_{a}\right)+y+z & =2h_{2}(d-1),\\
    2h_{3}\left(d-\sum_{a=1}^{3}h_{a}\right)+z+x & =2h_{3}(d-1).
\end{align*}
As one can already infer, the reason why the inversion formula, combined with symmetries, determines the $C$-ensemble partition function—even in the presence of a small number of degenerate energy levels—can be traced back to our conjecture (or identification) of the 
$C$-ensemble volume as a correlation function in some two-dimensional CFT. In the degenerate case, the degree of degeneracy acts as a conformal weight, enabling us to fix the partition function up to three-point correlations. For cases involving additional degeneracies, however, one would need to evaluate the unitary integral \eqref{eq:C2_Eigenvalue_rep_partition_function} for arbitrary configurations of degeneracies, which lies beyond the scope of this work\footnote{The precise relation of the $C$-ensemble as a CFT is work in progress.}.

The key point here is that, although $\tilde{g}$ cannot be explicitly determined solely from symmetries for more than four different degenerate levels—as this would require either: (i) the operator-product-expansion (OPE) coefficients of the (currently unknown) proposed ``$C$-ensemble dual CFT'', or (ii) performing the explicit unitary integral for an arbitrary configuration of degeneracies— all the results presented so far in the main part of the document are independent of $\tilde{g}$. This is because they were obtained by directly bootstrapping the ensemble moment operators $\hat{\Phi}^{\mathcal{E}_{C}}_{2k}$ without the need to divide by the partition function in the first place.

\subsection{Proof of the Vandermonde-Variance statement}

In this section, we present a proof of the equation \eqref{eq:C-ensemble_entropy_leading_term} used in Section.\ref{sub:C_ensemble_complexity} to relate the entropy of the $C$-ensemble with the level-statistics of the arbitrary Many-body Hamiltonian. In terms of the level spacing's, $s_{l}=(E_{l+1}-E_{l})\overline{\Delta E}^{-1}$, the logarithm of the square Vandermonde can be set as

\begin{equation}
\label{app_partition_function:eq_level_spacings}
    \log\left(\Delta^{2}(H)\right) = \sum_{l<m}\log \left( \sum_{k=l}^{m-1} s_{k}\right)^{2}.
\end{equation}
Here the set $\{s_{l}\}_{1}^{d-1}$ can be thought as a sequence of outcomes of a random variable $s$ with distribution $P(s)$. Typical examples include, \emph{e.g.}, $P(s)=e^{-s}$ for a system presenting level clustering, or $P(s)=\frac{32}{\pi^2}s^{2}e^{-\frac{4}{\pi}s^{2}}$ in the case of level repulsion (the chaotic case)\footnote{For our proof, we will leave $P(s)$ to be arbitrary.}. By definition, $s$ is normalized with mean value equal to one, and let $\sigma^{2}$ denote the variance of $s$. Now, due to the central limit theorem, each average 

\begin{equation}
   \overline{s}_{m-1} = \sum_{k=l}^{m-1} \frac{s_{l}+s_{l+1}+\dots +s_{m-1}}{m-l}
\end{equation}
is a random variable following a distribution that is asymptotically close to a normal one of mean $1$ and variance $\frac{\sigma^{2}}{m-l}$. The leading contributions to \eqref{app_partition_function:eq_level_spacings} are given by large energy-differences, \emph{i.e.}, the UV-modes $l-m \gg  1$. As for many-body systems the dimension of the associated Hilbert-space is exponentially large in the number of particles, we expect that the majority of averages inside \eqref{app_partition_function:eq_level_spacings} will be performed over a significantly large number of samples. Therefore,
we can approximate each term inside  \eqref{app_partition_function:eq_level_spacings} by replacing it by their normal average, \emph{i.e.},

\begin{align}
\label{eq:vandermonde_cnetral_limit}
\sum_{l<m}\log \left( \sum_{k=l}^{m-1} s_{k}\right)^{2} & \to \sum_{q=1}^{d-1}(d-q)\braket{\log(q+\sqrt{q}x)^{2}\, )}_{N\left(0,\sigma\right)} \\
&=\sum_{q=1}^{d-1}(d-q)\log(q^{2}) + \sum_{q=1}^{d-1}(d-q)\braket{\log(1+\frac{x}{\sqrt{q}})^{2}}_{N\left(0,\sigma\right)}, \nonumber
\end{align}
where $\braket{\cdots}_{N\left(\mu,\sigma\right)}$ denotes the average over a respective normal-distribution of mean $\mu$ and variance $\sigma^{2}$. The logarithm average can be suitable evaluated by splitting the real integral into the contributions coming from the absolute value:

\begin{align}
    \Bigl \langle{\log\left(1+\frac{x}{\sqrt{q}}\right)^{2} \Bigr \rangle}_{N\left(0,\sigma\right)} &= \frac{2}{\sqrt{2\pi}}\int_{\mathbb{R}}\log \left| 1+\eta x\right|e^{-\frac{x^2}{2}}dx \\
    &= \frac{2}{\sqrt{2\pi}}\int_{\frac{1}{\eta}}^{\infty}\log \left((\eta x)^{2}-1\right)e^{-\frac{x^2}{2}}dx + \, \frac{2}{\sqrt{2\pi}}\int_{-\frac{1}{\eta}}^{\frac{1}{\eta}}\log \left(\eta x +1\right)e^{-\frac{x^2}{2}}dx \nonumber
\end{align}
where $\eta^{2}=\frac{\sigma^{2}}{q}$ denotes the scaled variance. Since the level-spacings are normalized, then $\sigma^{2}\sim \mathcal{O}(1)$, and we are interested in the asymptotics around $\eta \ll 1$, \emph{e.g.},

\begin{equation}
\label{eq:asymptotics_integral_1}
    \int_{\frac{1}{\eta}}^{\infty}\log \left((\eta x)^{2}-1\right)e^{-\frac{x^2}{2}}dx \sim \eta e^{-\frac{1}{2 \eta^2}}\left(\log \left(\eta^2\right) +(\log(4)-2\gamma)\right)
\end{equation}
for the first integral (where $\gamma$ denotes the Euler -gamma constant). For the second integral we have instead,

\begin{equation}
\label{eq:asymptotics_integral_2}
    \int_{-\frac{1}{\eta}}^{\frac{1}{\eta}}\log \left(\eta x+1\right)e^{-\frac{x^2}{2}}dx \sim -\frac{\sqrt{2\pi}}{2}\eta^{2}+e^{-\frac{1}{2\eta^2}}\left(\eta^{3}+\eta\right).
\end{equation}
Therefore, by replacing \eqref{eq:asymptotics_integral_1} and \eqref{eq:asymptotics_integral_2} into \eqref{eq:vandermonde_cnetral_limit}, we obtain the claimed result in \eqref{eq:C-ensemble_entropy_leading_term}, \emph{i.e.},

\begin{align}
\label{eq:QED_vandermonde_claim}
    \log(\Delta^{2}(H)) &\sim \sum_{q=1}^{d-1}(d-q)\log(q^2)-\sigma^{2}\,\sum_{q=1}^{d-1}\frac{(d-q)}{q} \\
    &+\frac{\sqrt{2\pi}}{2}\sum_{q=1}^{d-1}(d-q)e^{-\frac{q}{2\sigma^2}}\left(\left(\frac{\sigma}{\sqrt{q}}\right)^{3}+\frac{\sigma}{\sqrt{q}}\log \left(\frac{\sigma}{\sqrt{q}}\right)^{2}+\frac{\sigma}{\sqrt{q}}(\log(4)-2\gamma +1)\right). \nonumber
\end{align}
Particularly, the first term on the right-hand side of \eqref{eq:QED_vandermonde_claim} acts as a counter-term for the entropy of the unitary-group and the second-term, \emph{e.g.}, 
\begin{equation}
    \sum_{q=1}^{d-1}(d-q)\log(q^2)-\sigma^{2}=\sigma^{2}\left(d(H^{(d-1)}-1)+1\right),
\end{equation}
with $H^{(d-1)}$ the sum of the reciprocals of the first $d-1$ (the Harmonic number), corresponding to the leading contribution from the level-statistics of $H$.

\section{About the plateau operator}
\label{Appendix:Scrambling}

After integrating over the $U(1)^{d}\times S_{d}$ invariant degrees of freedom the C-ensemble dependence on the specific Hamiltonian $H$ gets ``zipped'' in what we call the plateau operator (identified also as the infinite time average of two-point correlations). In this appendix, we introduce two elements: first, we start by presenting some basic identities for the plateau operator, and later, we construct an operator-algebraic equation that, once solved, enables us to find the plateau operator for arbitrary systems.

\subsection{properties of the \texorpdfstring{$s$}{}-tensor}

Before delving directly into the plateau operator, we first present some properties of the $s$-tensor. First, as previously defined in Section \ref{section:C_ensemble}, it is constructed as the product of two COPY$:\mathcal{H} \to \overline{\mathcal{H}}^{\otimes 2}$ operators, which diagrammatically reads: 

\begin{equation}
  \begin{tikzpicture}
	\begin{pgfonlayer}{nodelayer}
		\node [style=none] (0) at (-6.5, 0) {$\displaystyle{s\text{-tensor}=\text{COPY}\,\text{COPY}^{\dagger}=}$};
		\node [style=none] (1) at (3, -1) {};
		\node [style=none] (2) at (3, 1) {};
		\node [style=none] (3) at (1, 0) {};
		\node [style=none] (4) at (2.25, 0) {};
		\node [style=none] (5) at (-1, 1) {};
		\node [style=none] (6) at (-1, -1) {};
		\node [style=none] (7) at (1, 0) {};
		\node [style=none] (8) at (-0.25, 0) {};
	\end{pgfonlayer}
	\begin{pgfonlayer}{edgelayer}
		\draw (1.center) to (4.center);
		\draw (4.center) to (2.center);
		\draw (4.center) to (3.center);
		\draw (5.center) to (8.center);
		\draw (8.center) to (6.center);
		\draw (8.center) to (7.center);
	\end{pgfonlayer}
\end{tikzpicture}
\end{equation}
The motivation behind its name (as mentioned before) is twofold: first, it resembles the $s$-channel from QFT, and second, we are unaware of an specific name in the literature. The $s$-tensor is defined in terms of a basis, and when expressed in the local basis $\{l\}$, it corresponds to the convex sum of diagonal pure states $\ket{l\,l}$ in the twofold Hilbert space, \emph{i.e.},

\begin{equation}
  \label{eq:appendix-s_state_convex}
  s\text{-tensor}=\sum_{l}\ket{l\,l}\bra{l\, l}.
\end{equation}
The $s$-tensor can be related to the $d$-level $\ket{\text{GHZ}}=\frac{1}{\sqrt{d}}\sum_{l}\ket{l\,l\,l}$ state by noticing that the convex combination \eqref{eq:appendix-s_state_convex} is obtained from the partial trace of the $\ket{\text{GHZ}}$ state\footnote{In some sense the $s$-tensor can be identified as ``cousin'' of the maximally mixed state, \emph{i.e.} while the maximally mixed state is obtained from tracing a maximally entangled bipartite state, the $s$-tensor is obtained from tracing a maximally entangled tripartite state \cite{cirac/GHZ/2000}.}
,\emph{i.e.},

\begin{equation}
  \frac{s\text{-tensor}}{d} = \Tr_{1}\ket{\text{GHZ}}\bra{\text{GHZ}} = \Tr_{2}\ket{\text{GHZ}}\bra{\text{GHZ}} = \Tr_{3}\ket{\text{GHZ}}\bra{\text{GHZ}},
\end{equation}
with $\Tr_{i}(\cdots)$ the partial trace over the $i$th Hilbert space. Last but not least, this state can be also used to build the trace-preserving dephasing channel \cite{Tomamichel/quantum_information/2021,Gogolin/Phd_thesis_review/2016} $\mathcal{D}$ over the local basis,

\begin{equation}
  \mathcal{D}(\cdots) = \Tr_{1}(s\text{-tensor}(\mathbb{I}\otimes(\cdots))) = \sum_{l}\ket{l}\bra{l}(\cdots)\ket{l}\bra{l}.
\end{equation}
The relevance behind this identification is twofold. First, the Von-Neumman entropy is non-decreasing under dephasing \cite{Gogolin/Phd_thesis_review/2016}, and therefore every time one sees a dephasing channel, this can be implicitly related with some underlying maximization of entropy. Second, in what directly concerns the plateau operator, we will show in the next section that this operator also defines a dephasing channel, but this time with respect to the energy eigenbasis.

\subsection{Bootstraping the plateau operator}

Although the plateau operator is a fingerprint of the $C$-ensemble, its definition can be naturally extended to other ensembles in a straightforward manner, {\em i.e.}, for a unitary ensemble $\mathcal{E}$, we define the ensemble plateau operator, $G^{\mathcal{E}}:\mathcal{H}^{\otimes 2}\rightarrow \mathcal{H}^{\otimes 2}$, as the right two-fold channel over the $s$-tensor,

\begin{equation}
  \label{eq:general_dephasing_op}
  G^{\mathcal{E}}=\Phi_{2,R}^{\mathcal{E}}(\text{s-tensor}),\quad \text{with} \quad \Phi^{\mathcal{E}}_{2,R}(\cdots)=\int_{\mathcal{E}}U^{\dagger \otimes 2} (\cdots) U^{\otimes 2}d\mu_{\mathcal{E}}[U].
\end{equation}
For Haar distributed unitaries $G^{U(d)}$ is easily obtained using the Shur-Weyl duality\footnote{Alternatively, it can be directly extracted by contractions of the four-moment operator.} (\ref{Appendix:RMT_Haar}), {\em e.g.}

\begin{equation}
  \label{eq:Haar_dephasing_op}
\begin{tikzpicture}
	\begin{pgfonlayer}{nodelayer}
		\node [style=none] (45) at (-2.25, 0.75) {};
		\node [style=none] (46) at (-2.25, -0.75) {};
		\node [style=none] (47) at (-0.75, 0.75) {};
		\node [style=none] (48) at (-0.75, -0.75) {};
		\node [style=none] (49) at (0.75, 0.75) {};
		\node [style=none] (50) at (0.75, -0.75) {};
		\node [style=none] (51) at (2.75, 0.75) {};
		\node [style=none] (52) at (2.75, -0.75) {};
		\node [style=none] (53) at (0, 0) {+};
		\node [style=none] (54) at (-2.75, 1.25) {};
		\node [style=none] (55) at (-2.75, -1.25) {};
		\node [style=none] (56) at (3.5, 1.25) {};
		\node [style=none] (57) at (3.5, -1.25) {};
		\node [style=none] (58) at (-4.25, 0) {$\displaystyle{\frac{1}{d+1}}$};
	\end{pgfonlayer}
	\begin{pgfonlayer}{edgelayer}
		\draw (45.center) to (47.center);
		\draw (46.center) to (48.center);
		\draw [in=-180, out=0] (49.center) to (52.center);
		\draw [style={white_edge}] (50.center) to (51.center);
		\draw [in=180, out=0] (50.center) to (51.center);
		\draw [bend right=15] (54.center) to (55.center);
		\draw [bend left=15] (56.center) to (57.center);
	\end{pgfonlayer}
\end{tikzpicture}  
\end{equation}
From the discussion in section \ref{Chapter:two_points} about the interpretation of the plateau operator as the long time average of two-point correlation functions, \eqref{eq:Haar_dephasing_op} shows the obvious fact that Haar distributed unitaries cannot give an $H$-dependent diagonal ensemble, {\em e.g.}

\begin{equation}
  G^{U(d)}= \left\langle\sum_{l} \ket{E_{l}\, E_{l}} \bra{E_{l}\, E_{l}}\right\rangle_{U(d)}=\frac{1}{d
  +1}\left( \mathbb{I}^{\otimes 2}+\text{SWAP} \right).
\end{equation}
For the $C$-ensemble $H$ plays the role of an external source and $G^{\mathcal{E}_{C}}$ becomes explicitly $H$-dependent, {\em i.e.}, $G^{\mathcal{E}_{C}}=G^{\mathcal{E}_{C}}[H]$. In fact, as a function of the specific Hamiltonian, we will show that we can bootstrap the $C$-ensemble plateau operator without performing the unitary integral \eqref{eq:general_dephasing_op} in first place. To set grounds, notice that for all unitary $U$ and Hermitian $H$ the $C$-ensemble measure imposes the rotation constraint,

\begin{equation}
  \label{eq:dephasing_rotation_covariance}
  (U\otimes U)G^{\mathcal{E}_{C}}[H](U^{\dagger}\otimes U^{\dagger})=G^{\mathcal{E}_{C}}[UHU^{\dagger}],
\end{equation}
which relates the $C$-ensemble for different choices of computational basis. The most general solution of this equation is given as the linear combination\footnote{This can be seen as a consecuence of Shur-Weyl's duality.},

\begin{align*}
  G^{\mathcal{E}_{C}}[H]=&\alpha_{1}\mathbb{I} \otimes \mathbb{I} +\alpha_{2}\text{SWAP} + f_{1} \otimes  \mathbb{I}+ \mathbb{I}\otimes f_{2} + (f_{3} \otimes  \mathbb{I})(\text{SWAP}) + ( \mathbb{I} \otimes f_{4})(\text{SWAP})\\
  & + (f_{5} \otimes f_{6}) + (f_{7} \otimes f_{8})(\text{SWAP}),
\end{align*}
with $\{ \alpha_{l} \}_{l=1}^{2}$ scalar and $\{ f_{l} \}_{l=1}^{8}$ Hermitian operator valued functions of the Hamiltonian\footnote{Strictly speaking, they are scalar and operator valued distributions, respectively.} $H$. Not all these functions are independent, in particular they are mutually constrained by unitarity, \emph{i.e.}, 

\begin{equation}
  \Tr_{1}G^{\mathcal{E}_{C}}=\Tr_{2}G^{\mathcal{E}_{C}}=\Tr_{1}G^{\mathcal{E}_{C}}\text{SWAP}=\Tr_{2}G^{\mathcal{E}_{C}}\text{SWAP},
\end{equation}
with $\Tr_{1(2)}$ the partial trace over the first  (second) Hilbert space. Alongside the algebraic constraints relating the scalar $\{ \alpha_{l} \}_{l=1}^{2}$ with the operator-valued $\{ f_{l} \}_{l=1}^{8}$ functions, there are also symmetries in the $C$-ensemble measure which strongly constraint the $H$ dependence in each one of these functions. Reflection invariance, $H\rightarrow -H$, imposes that  $\{ \alpha_{l} \}_{l=1}^{2}$ and $\{ f_{l} \}_{l=1}^{8}$ must be even, whereas the invariance under energy shifts, $H\rightarrow H+\mu \mathbb{I}$ with $\mu$ real, implies that all those functions do not depend directly upon $H$ but rather on the centering $\Delta H =H-\frac{1}{d}\Tr{H}$. Before introducing a further symmetry let us point out the following: there are two particular limits from where the $C$-ensemble operator yields the Haar unitary one. The first one is obtained by sending $H$ to the identity, {\em i.e.},

\begin{equation}
  \lim_{H \to \mathbb{I}} G^{\mathcal{E}_{C}}[H]=\frac{1}{d+1}\left( \mathbb{I}\otimes \mathbb{I}+\text{SWAP} \right)=G^{U(d)},
\end{equation}
this is clear as $d\mu^{\mathcal{E}_{C}}\rightarrow d\mu^{U(d)}$ and also because every unitary trivially ``diagonalize'' the identity. The second (and perhaps most relevant) limit is obteined by averaging the $C$-ensemble plateau operator over an RMT Hamiltonian ensemble with normalized unitary-invariant measure $\rho(H)[dH]=\rho(UHU^{\dagger})[dH]$,  

\begin{equation}
  \int_{\text{RMT}} \, G^{\mathcal{E}_{C}}[H] \,\rho(H) [dH] = G^{U(d)}\,\int_{\text{RMT}} \frac{\delta_{\perp}(H)}{\Vol(H)}\, \rho(H)[dH],
\end{equation}
with $\Vol(H)$ the $C$-ensemble partition function (see Appendix \ref{appendix:partition_function}). It can be proven that $\int \frac{\delta_{\perp}(H)}{\Vol(H)}\, \rho(H)[dH]=1$ (see Appendix \ref{Appendix:RMT_Haar}) and the interpretation of this second recipe to obtain the Haar unitary scramble operator from the $C$-ensemble one is the following: by averaging over all possible Hamiltonians, as each unitary diagonalizes a single Hamiltonian at least once, we are implicitly averaging over all random unitaries.

Turning back to the bootstrap of $G^{\mathcal{E}_{C}}[H]$, let us remark a relevant property of the s-tensor. As the $s$-tensor is obtained by joining two copies, there are two possible ways to do this: either attaching them vertically or horizontally, {\em i.e.},

\begin{equation}
  \begin{tikzpicture}
	\begin{pgfonlayer}{nodelayer}
		\node [style=none] (0) at (-0.5, 0) {};
		\node [style=none] (1) at (-1, 0.75) {};
		\node [style=none] (2) at (-1, -0.75) {};
		\node [style=none] (3) at (0.25, 0) {};
		\node [style=none] (4) at (1.5, -0.025) {};
		\node [style=none] (5) at (2, 0.725) {};
		\node [style=none] (6) at (2, -0.775) {};
		\node [style=none] (7) at (0.75, -0.025) {};
		\node [style=none] (8) at (4.475, -1) {};
		\node [style=none] (9) at (5.225, -1.5) {};
		\node [style=none] (10) at (3.725, -1.5) {};
		\node [style=none] (11) at (4.475, -0.25) {};
		\node [style=none] (12) at (4.475, 1) {};
		\node [style=none] (13) at (5.225, 1.5) {};
		\node [style=none] (14) at (3.725, 1.5) {};
		\node [style=none] (15) at (4.475, 0.25) {};
		\node [style=none] (16) at (3, 0) {=};
	\end{pgfonlayer}
	\begin{pgfonlayer}{edgelayer}
		\draw (1.center) to (0.center);
		\draw (2.center) to (0.center);
		\draw (0.center) to (3.center);
		\draw (5.center) to (4.center);
		\draw (6.center) to (4.center);
		\draw (4.center) to (7.center);
		\draw (9.center) to (8.center);
		\draw (10.center) to (8.center);
		\draw (8.center) to (11.center);
		\draw (13.center) to (12.center);
		\draw (14.center) to (12.center);
		\draw (12.center) to (15.center);
	\end{pgfonlayer}
\end{tikzpicture}
\end{equation}
For the plateau operator this implies that $G^{\mathcal{E}}$ (for any unitary ensemble $\mathcal{E}$) must be symmetric over the ``cut'' (the way that one chooses to join the COPYs). This kind of ``crossing''\footnote{So named because it resembles the $s-t$ duality in QFT scattering.} symmetry is reflected by the indistinguishability of the Hilbert space replicas $\mathcal{H}^{\otimes 2}$ inside $G^{\mathcal{E}_{C}}$ and is explicitly given by the left and right invariance under swaps,

\begin{equation}
  \text{SWAP}\,G^{\mathcal{E}_{C}}[H] = G^{\mathcal{E}_{C}}[H]\,\text{SWAP}=G^{\mathcal{E}_{C}}[H].
\end{equation}
This, along with the covariance under unitary rotations of the Hamiltonian \eqref{eq:dephasing_rotation_covariance} automatically constraint the tensor structure of the $C$-ensemble plateau operator to be of the form,

\begin{equation}
  \begin{tikzpicture}
	\begin{pgfonlayer}{nodelayer}
		\node [style=none] (0) at (-2.25, 0.75) {};
		\node [style=none] (1) at (-2.25, -0.75) {};
		\node [style=none] (2) at (-0.75, 0.75) {};
		\node [style=none] (3) at (-0.75, -0.75) {};
		\node [style=none] (9) at (-3.25, 0) {$\alpha(H)$};
		\node [style=none] (15) at (0.75, -0.75) {};
		\node [style=none] (17) at (3.25, -0.75) {};
		\node [style=none] (18) at (2, 0.75) {$\psi(H)$};
		\node [style=none] (19) at (0.75, 0.75) {};
		\node [style=none] (20) at (1, 0.75) {};
		\node [style=none] (21) at (3.25, 0.75) {};
		\node [style=none] (22) at (3, 0.75) {};
		\node [style=none] (23) at (4.75, 0.75) {};
		\node [style=none] (24) at (7.25, 0.75) {};
		\node [style=none] (25) at (6, -0.75) {$\psi(H)$};
		\node [style=none] (26) at (4.75, -0.75) {};
		\node [style=none] (27) at (5, -0.75) {};
		\node [style=none] (28) at (7.25, -0.75) {};
		\node [style=none] (29) at (7, -0.75) {};
		\node [style=none] (30) at (0, 0) {+};
		\node [style=none] (31) at (-4.25, 1.25) {};
		\node [style=none] (32) at (-4.25, -1.25) {};
		\node [style=none] (33) at (12.5, 1.25) {};
		\node [style=none] (34) at (12.5, -1.25) {};
		\node [style=none] (35) at (10.25, 0.75) {$\phi(H)$};
		\node [style=none] (36) at (8.75, 0.75) {};
		\node [style=none] (37) at (9.25, 0.75) {};
		\node [style=none] (38) at (11.75, 0.75) {};
		\node [style=none] (39) at (11.25, 0.75) {};
		\node [style=none] (40) at (10.25, -0.75) {$\phi(H)$};
		\node [style=none] (41) at (8.75, -0.75) {};
		\node [style=none] (42) at (9.25, -0.75) {};
		\node [style=none] (43) at (11.75, -0.75) {};
		\node [style=none] (44) at (11.25, -0.75) {};
		\node [style=none] (45) at (13.75, 0.75) {};
		\node [style=none] (46) at (13.75, -0.75) {};
		\node [style=none] (47) at (15.25, 0.75) {};
		\node [style=none] (48) at (15.25, -0.75) {};
		\node [style=none] (49) at (16.75, 0.75) {};
		\node [style=none] (50) at (16.75, -0.75) {};
		\node [style=none] (51) at (18.75, 0.75) {};
		\node [style=none] (52) at (18.75, -0.75) {};
		\node [style=none] (53) at (16, 0) {+};
		\node [style=none] (54) at (13.25, 1.25) {};
		\node [style=none] (55) at (13.25, -1.25) {};
		\node [style=none] (56) at (19.5, 1.25) {};
		\node [style=none] (57) at (19.5, -1.25) {};
		\node [style=none] (58) at (8, 0) {+};
		\node [style=none] (59) at (4, 0) {+};
	\end{pgfonlayer}
	\begin{pgfonlayer}{edgelayer}
		\draw (0.center) to (2.center);
		\draw (1.center) to (3.center);
		\draw (15.center) to (17.center);
		\draw (19.center) to (20.center);
		\draw (22.center) to (21.center);
		\draw (23.center) to (24.center);
		\draw (26.center) to (27.center);
		\draw (29.center) to (28.center);
		\draw [bend right=15] (31.center) to (32.center);
		\draw [bend left=15] (33.center) to (34.center);
		\draw (36.center) to (37.center);
		\draw (39.center) to (38.center);
		\draw (41.center) to (42.center);
		\draw (44.center) to (43.center);
		\draw (45.center) to (47.center);
		\draw (46.center) to (48.center);
		\draw [in=-180, out=0] (49.center) to (52.center);
		\draw [style={white_edge}] (50.center) to (51.center);
		\draw [in=180, out=0] (50.center) to (51.center);
		\draw [bend right=15] (54.center) to (55.center);
		\draw [bend left=15] (56.center) to (57.center);
	\end{pgfonlayer}
\end{tikzpicture}
\end{equation}
where $\alpha(H)$ is a scalar, dimensionless function of the Hamiltonian $H$, and $\psi(H)$, $\phi(H)$ are operator-valued dimensionless functions of $H$. Contractions of $G^{\mathcal{E}_{C}}$ further provide two constraints for $\psi,\phi$ and $\alpha$, {\em e.g.},

\begin{equation}
\label{eq:operator_constraint}
  \alpha(H)(d+1)\, \mathbb{I}+\Tr{\psi(H)}\, \mathbb{I}+(2+d)\psi(H)+\phi(H)\Tr{\phi(H)}+\phi(H)^{2}=\mathbb{I}
\end{equation}
for $\Tr_{1}G^{\mathcal{E}_{C}}$, and

\begin{equation}
\label{eq:scalar_constraint}
  \alpha(H)d(d+1)+2(d+1)\Tr{\psi(H)}+(\Tr{\phi(H)})^{2}+\Tr(\phi(H)^{2})=d
\end{equation}
for $\Tr{G^{\mathcal{E}_{C}}}$. From \eqref{eq:operator_constraint}, \eqref{eq:scalar_constraint} and after some algebra one finds an exact closed form for $G^{\mathcal{E}_{C}}$ solely in terms of $\phi(H)$,

\begin{align}
\label{eq:pre_dephasing_G}
     G^{\mathcal{E}_{C}}[H]=&\left( \frac{1}{d+1}+\frac{\Tr{\phi(H)}+(\Tr{\phi(H)})^{2}}{(d+1)(d+2)} \right)(\mathbb{I}\otimes \mathbb{I}+\text{SWAP})+\phi \otimes \phi(\mathbb{I}\otimes \mathbb{I}+\text{SWAP}) \\ 
     -&\left(\frac{\phi(H)^{2}+\phi(H)\Tr{\phi(H)}}{d+2}\otimes\mathbb{I}+\mathbb{I}\otimes \frac{\phi(H)^{2}+\phi(H)\Tr{\phi(H)}}{d+2}\right)(\mathbb{I}\otimes \mathbb{I}+\text{SWAP}). \notag
\end{align}
As a short comment, it is surprising that the constraints given by the contractions \eqref{eq:operator_constraint} and \eqref{eq:scalar_constraint} cancel the $\alpha$ dependence and further fix the plateau operator  $G^{\mathcal{E}}_{C}$ up to a single operator-valued function $\phi(H)$. 

Returning to the result in \eqref{eq:pre_dephasing_G}, notice that it can be further simplified by going into centering variables\footnote{Another surprising cancellation pops out, that is, the result does not depend upon the trace of $\phi(H)$ at all. This means that we can choose a gauge such that $\Tr{\phi(H)}=0$ in the first place.}, {\em e.g.}, $\Delta \phi(H)=\phi(H)-\frac{1}{d}\Tr{\phi(H)}$,

\begin{equation}
\label{eq:complete_dephasing}
  \begin{tikzpicture}
	\begin{pgfonlayer}{nodelayer}
		\node [style=none] (9) at (-5.5, 0) {$\displaystyle{\frac{ \Tr (\Delta \phi(H)^{2})}{(d+1)(d+2)}}$};
		\node [style=none] (15) at (-8.5, -4.25) {};
		\node [style=none] (17) at (-3.5, -4.25) {};
		\node [style=none] (18) at (-6, -2.75) {$(\Delta \phi(H))^{2}$};
		\node [style=none] (19) at (-8.5, -2.75) {};
		\node [style=none] (20) at (-7.75, -2.75) {};
		\node [style=none] (21) at (-3.5, -2.75) {};
		\node [style=none] (22) at (-4.25, -2.75) {};
		\node [style=none] (31) at (-9.25, -2.25) {};
		\node [style=none] (32) at (-9.25, -4.75) {};
		\node [style=none] (45) at (-2, 0.75) {};
		\node [style=none] (46) at (-2, -0.75) {};
		\node [style=none] (47) at (-0.5, 0.75) {};
		\node [style=none] (48) at (-0.5, -0.75) {};
		\node [style=none] (49) at (1, 0.75) {};
		\node [style=none] (50) at (1, -0.75) {};
		\node [style=none] (51) at (3, 0.75) {};
		\node [style=none] (52) at (3, -0.75) {};
		\node [style=none] (53) at (0.25, 0) {+};
		\node [style=none] (54) at (-2.5, 1.25) {};
		\node [style=none] (55) at (-2.5, -1.25) {};
		\node [style=none] (56) at (3.75, 1.25) {};
		\node [style=none] (57) at (3.75, -1.25) {};
		\node [style=none] (60) at (-11, -3.5) {$\displaystyle{-\frac{1}{d+2}}$};
		\node [style=none] (61) at (3.25, -2.75) {};
		\node [style=none] (62) at (-1.75, -2.75) {};
		\node [style=none] (68) at (-2.5, -3.5) {+};
		\node [style=none] (69) at (4, -2.25) {};
		\node [style=none] (70) at (4, -4.75) {};
		\node [style=none] (71) at (-13.5, 0) {$G^{\mathcal{E}_{C}}[H]$};
		\node [style=none] (72) at (-11.5, 0) {=};
		\node [style=none] (73) at (0.75, -4.25) {$(\Delta \phi(H))^{2}$};
		\node [style=none] (74) at (-1.75, -4.25) {};
		\node [style=none] (75) at (-1, -4.25) {};
		\node [style=none] (76) at (3.25, -4.25) {};
		\node [style=none] (77) at (2.5, -4.25) {};
		\node [style=none] (78) at (-11, -7.25) {+};
		\node [style=none] (79) at (-7.25, -6.5) {$\Delta \phi(H)$};
		\node [style=none] (80) at (-9.5, -6.5) {};
		\node [style=none] (81) at (-8.5, -6.5) {};
		\node [style=none] (82) at (-5, -6.5) {};
		\node [style=none] (83) at (-6, -6.5) {};
		\node [style=none] (84) at (-7.25, -8) {$\Delta \phi(H)$};
		\node [style=none] (85) at (-9.5, -8) {};
		\node [style=none] (86) at (-8.5, -8) {};
		\node [style=none] (87) at (-5, -8) {};
		\node [style=none] (88) at (-6, -8) {};
		\node [style=none] (89) at (-4.5, -6) {};
		\node [style=none] (90) at (-4.5, -8.5) {};
		\node [style=none] (91) at (5.5, -2.75) {};
		\node [style=none] (92) at (5.5, -4.25) {};
		\node [style=none] (93) at (7, -2.75) {};
		\node [style=none] (94) at (7, -4.25) {};
		\node [style=none] (95) at (8.5, -2.75) {};
		\node [style=none] (96) at (8.5, -4.25) {};
		\node [style=none] (97) at (10.5, -2.75) {};
		\node [style=none] (98) at (10.5, -4.25) {};
		\node [style=none] (99) at (7.75, -3.5) {+};
		\node [style=none] (100) at (5, -2.25) {};
		\node [style=none] (101) at (5, -4.75) {};
		\node [style=none] (102) at (11.25, -2.25) {};
		\node [style=none] (103) at (11.25, -4.75) {};
		\node [style=none] (104) at (-10, -6) {};
		\node [style=none] (105) at (-10, -8.5) {};
		\node [style=none] (106) at (-3, -6.5) {};
		\node [style=none] (107) at (-3, -8) {};
		\node [style=none] (108) at (-1.5, -6.5) {};
		\node [style=none] (109) at (-1.5, -8) {};
		\node [style=none] (110) at (0, -6.5) {};
		\node [style=none] (111) at (0, -8) {};
		\node [style=none] (112) at (2, -6.5) {};
		\node [style=none] (113) at (2, -8) {};
		\node [style=none] (114) at (-0.75, -7.25) {+};
		\node [style=none] (115) at (-3.5, -6) {};
		\node [style=none] (116) at (-3.5, -8.5) {};
		\node [style=none] (117) at (2.75, -6) {};
		\node [style=none] (118) at (2.75, -8.5) {};
		\node [style=none] (119) at (-10, 0) {$G^{U(d)}$};
		\node [style=none] (120) at (-8.5, 0) {+};
	\end{pgfonlayer}
	\begin{pgfonlayer}{edgelayer}
		\draw (15.center) to (17.center);
		\draw (19.center) to (20.center);
		\draw (22.center) to (21.center);
		\draw [bend right=15] (31.center) to (32.center);
		\draw (45.center) to (47.center);
		\draw (46.center) to (48.center);
		\draw [in=-180, out=0] (49.center) to (52.center);
		\draw [style={white_edge}] (50.center) to (51.center);
		\draw [in=180, out=0] (50.center) to (51.center);
		\draw [bend right=15] (54.center) to (55.center);
		\draw [bend left=15] (56.center) to (57.center);
		\draw (61.center) to (62.center);
		\draw [bend left=15] (69.center) to (70.center);
		\draw (74.center) to (75.center);
		\draw (77.center) to (76.center);
		\draw (80.center) to (81.center);
		\draw (83.center) to (82.center);
		\draw (85.center) to (86.center);
		\draw (88.center) to (87.center);
		\draw [bend left=15] (89.center) to (90.center);
		\draw (91.center) to (93.center);
		\draw (92.center) to (94.center);
		\draw [in=-180, out=0] (95.center) to (98.center);
		\draw [style={white_edge}] (96.center) to (97.center);
		\draw [in=180, out=0] (96.center) to (97.center);
		\draw [bend right=15] (100.center) to (101.center);
		\draw [bend left=15] (102.center) to (103.center);
		\draw [bend right=15] (104.center) to (105.center);
		\draw (106.center) to (108.center);
		\draw (107.center) to (109.center);
		\draw [in=-180, out=0] (110.center) to (113.center);
		\draw [style={white_edge}] (111.center) to (112.center);
		\draw [in=180, out=0] (111.center) to (112.center);
		\draw [bend right=15] (115.center) to (116.center);
		\draw [bend left=15] (117.center) to (118.center);
	\end{pgfonlayer}
\end{tikzpicture}
\end{equation}
where, in addition, we have explicitly identified the Haar plateau operator $G^{U(d)}$ in \eqref{eq:complete_dephasing}. The separability of the $C$-ensemble plateau operator as a (universal) Haar contribution plus a (non-universal) explicitly Hamiltonian contribution encoded in $\phi(H)$ means in particular that every claim made by the $C$-ensemble coincides with the same claim as it was made by the $U(d)$ Haar, but with additional fine details due to the fixed Hamiltonian\footnote{Stated otherwise, the $C$-ensemble is as good as the unitary Haar one but with additional advantages provided by the specific dependence on the Hamiltonian.} $H$ (at least for two-moments\footnote{We conjecture that this extends for higher moments.}). From the structure of \eqref{eq:complete_dephasing} we can identify the first correction to the unitary Haar ensemble as a scalar perturbation to $G^{U(d)}$, \emph{i.e.},

\begin{equation}
  G^{\mathcal{E}_{C}}[H]=\left(\frac{1}{d+1}+\frac{\Tr{\Delta \phi(H)^{2}}}{(d+1)(d+2)}\right)(\mathbb{I}^{\otimes 2}+\text{SWAP})+\dots,
\end{equation}
where further corrections are given by operator-dependent contributions from the Hamiltonian dependence inside $\phi(H)$.

\subsection{An equation for the plateau operator}

In the previous section we showed that, for the $C$-ensemble, the plateau operator can be surprisingly highly constrained up to a single operator-valued function $\phi(H)$ of the fixed Hamiltonian $H$. In this section, we revisit the interpretation of the $C$-ensemble plateau operator and, in particularly, show that it actually corresponds to the exact $s$-state in the energy eigenbasis rather than its average value. Once this identification is established, we provide an equation for the left unknown operator $\phi(H)$ which (once solved) completely fixes the plateau operator and therefore the $s$-state in the energy eigenbasis for arbitrary systems.            
To set grounds, let us focus on the following integral,

\begin{align}
  & \int_{U(d)}\Tr_{1} \left( C^{\dagger \otimes 2}(\text{s-tensor})C^{\otimes 2}(H\otimes \mathbb{I}) \right)\,  d\mu_{\mathcal{E}_{C}}[C] \\ \notag
  & = \int_{U(d)}\Tr_{1} \left( (\mathbb{I} \otimes C^{\dagger})(\text{s-tensor})(CH C^{\dagger}\otimes C) \right)\,  d\mu_{\mathcal{E}_{C}}[C].
\end{align}
The off-diagonal delta inside the $C$-ensemble measure imposes that the $CHC^{\dagger}$ term on the integrand must be diagonal, {\em i.e.}, $CHC^{\dagger}=D_{C}$, with $D_{C}$ the matrix of eigenvalues piked at some particular ordering. By tracing over the first Hilbert space, the $s$-tensor imposes that only the integral of $C^{\dagger}D_{C}C$ survives and therefore,

\begin{equation}
  \label{eq:pre_H_equation}
  \Tr_{1}G^{\mathcal{E}_{C}}(H\otimes \mathbb{I})=H.
\end{equation}
The interpretation of this result is the following: From the discussion in chapter \ref{Chapter:two_points}, the plateau operator of a unitary eigenvector ensemble $\mathcal{E}$ can be interpreted as the ensemble average of the $s$-state in the energy eigenbasis,

\begin{equation}
  \label{eq:def_dephasing_diagonal}
  G^{\mathcal{E}}=\left\langle\sum_{l}\ket{E_{l}\, E_{l}}\bra{E_{l}\, E_{l}} \right\rangle_{\mathcal{E}}.
\end{equation}
Additionally, from chapter \ref{section:C_ensemble}, the $C$-ensemble can also be identified with the equivalence class $\{ C \sim U(1)^{d} \times S_{d}\,C, \; \text{with} \; CHC^{\dagger}=E \}$ of all  $d\times d$ unitary matrices that diagonalize a given Hamiltonian $H$ without a particular ordering of the eigenvalues $E$. Therefore, in the specific case of the $C$-ensemble, averaging over all eigenvectors in the diagonal ensemble is the same as averaging over the ``$U(1)^{d} \times S_{d}\, C$'' equivalence class, {\em e.g.},

\begin{equation}
  \label{eq:diagonal_to_C}
  G^{\mathcal{E}_{C}}=\left\langle\sum_{l}\ket{E_{l}\, E_{l}}\bra{E_{l}\, E_{l}} \right \rangle_{\mathcal{E}_{C}}=\sum_{l}\ket{E_{l}\, E_{l}}\bra{E_{l}\, E_{l}},
\end{equation}
and from this identification \eqref{eq:pre_H_equation} follows straightforwardly. The difference between the equality in \eqref{eq:diagonal_to_C} and the one in \eqref{eq:def_dephasing_diagonal} is that the latter is exact, {\em i.e.}, for the Haar unitary ensemble we can only assert about the average of the $s$-eigenvector state \eqref{eq:Haar_dephasing_op}, whereas in the $C$-ensemble case we are dealing with the exact state for a given fixed Hamiltonian $H$. In that direction, it is remarkable that we can compute an exact observable for a single system (particular Hamiltonian) from an eigenvector ensemble perspective ($C$-ensemble). 

We want to remark that this identification between the averaged and the exact diagonal ensembles provided by \eqref{eq:diagonal_to_C} was also expected by looking the symmetries of $\sum_{l}\ket{E_{l} \, E_{l}}\bra{E_{l} \, E_{l}}$ in first place. Stated otherwise, if $\{\ket{E_{l}}\}$ are the eigenstates of $H$, then $\{U\ket{E_{l}}\}$ will be the eigenstates of $UHU^{\dagger}$ for every unitary $U$. Therefore, for $UHU^{\dagger}$,

\begin{equation}
  U\otimes U\left( \sum_{l}\ket{E_{l} \, E_{l}}\bra{E_{l} \, E_{l}} \right)U^{\dagger}\otimes U^{\dagger}=\sum_{l}(U\ket{E_{l}}\otimes U\ket{E_{l}})(\bra{E_{l}}U^{\dagger} \otimes \bra{E_{l}}U^{\dagger}),
\end{equation}
which is the same as the rotation covariance of $G^{\mathcal{E}_{C}}$ \eqref{eq:dephasing_rotation_covariance}. Additionally, $\sum_{l}\ket{E_{l} \, E_{l}}\bra{E_{l} \, E_{l}}$ fulfills the crossing symmetry and also the same contraction constraints as $G^{\mathcal{E}_{C}}$, therefore by bootstrapping the $s$-state in the energy eigenbasis directly, we would get the same result as the one in \eqref{eq:pre_dephasing_G}\footnote{More generally, for every (either scalar or operator-valued) Hamiltonian function $f(H)$  invariant under the $U(1)^{d}\times S_{d}$ group action we will conclude that  $f(H)=\braket{f(H)}_{\mathcal{E}_{C}}$. }. From this perspective, the physical meaning of \eqref{eq:pre_H_equation} becomes clear: It is just the energy dephasing map $\mathcal{D}$ acting on the Hamiltonian, \emph{i.e.},

\begin{equation}
  \mathcal{D}(H)=\Tr_{1}\left(\sum_{l}\ket{E_{l}\,E_{l}}\bra{E_{l}\,E_{l}} (H\otimes \mathbb{I})\right)=\Tr_{1}\left(G^{\mathcal{E}_{C}}(H\otimes \mathbb{I})\right)=H.
\end{equation}
Finally, notice that once the constraint \eqref{eq:pre_H_equation}, the identification \eqref{eq:diagonal_to_C}, and the bootstrapping solution \eqref{eq:complete_dephasing} are provided, there is a unique operator-valued equation for $\Delta \phi(H)$,

\begin{equation}
\label{eq:dephasing_equation_full}
  \Delta H\left(\frac{d}{d+2}\Delta \phi^{2}+\frac{\Tr{(\Delta \phi)^{2}}}{(d+2)(d+1)} -\frac{d}{d+1}\right)+\left( \Delta \phi \Tr{\Delta \phi \Delta H}-\frac{(\Tr{(\Delta \phi)^{2}\Delta H})}{d+2}\right)=0.
\end{equation}
We will call this equation the \emph{plateau equation}, and solving it amounts to finding the $s$-state in the energy eigenbasis for an arbitrary given Hamiltonian $H$. In particular, this equation ``completes'' the $C$-ensemble in the following sense: Before the plateau equation \eqref{eq:dephasing_equation_full}, the plateau operator was a black box that we used to identify differences between the predictions made by the $C$-ensemble with respect to those made by the Haar one. Even with the identification of the plateau operator as the $s$-state in the energy eigenbasis, we will need to diagonalize the Hamiltonian to actually be able to compute ensemble averaged two- and four-point correlation functions. However, if we were able to diagonalize the Hamiltonian in the first place, what would be the purpose of defining the $C$-ensemble? With the plateau equation, the $C$-ensemble is now complete because it provides closed-form expressions for correlation functions without the need of diagonalizing the Hamiltonian explicitly\footnote{Although the dependence on the energy levels still appears in the spectral form factors, the RMT universality allows to estimate those spectral observables from the GUE. }.

\subsubsection{Qubit Solution for the plateau Equation}

In the previous section, we presented an equation for the plateau operator $G^{\mathcal{E}_{C}}[H]$. We know that this equation has a solution because we have already identified the plateau operator as the $s$-state in the energy eigenbasis, \emph{i.e.},

\begin{equation}
  G^{\mathcal{E}_{C}}[H]=\sum_{l}\ket{E_{l} E_{l}}\bra{E_{l} E_{l}}.
\end{equation}
However, we have not yet derived a solution. Here, we will solve the simplest case, namely, a two-dimensional Hilbert space. As both $\Delta \Phi$ and $\Delta H$ are traceless, they can be set as a linear combination of single-qubit Paulis $\{X,Y,Z\}$, \emph{i.e.},

\begin{align}
  \Delta H = h_{x}X+h_{y}Y+h_{z}Z  \quad \text{and} \quad \Delta \phi =\lambda_{x}H+\lambda_{y}Y+\lambda_{z}Z.
\end{align}
with $\vec{\lambda}=\{\lambda_{x},\lambda_{y},\lambda_{z}\}$ and $\vec{h}=\{h_{x},h_{y},h_{z}\}$ vectors in $\mathbb{R}^{3}$. Within this parametrization, the single-qubit plateau-equation reduces to $\vec{\lambda}$ and $\vec{h}$ being parallel, \emph{i.e.}

\begin{equation}
\label{eq:single_qubit_solution}
  \vec{\lambda}=\pm\frac{\vec{h}}{2 \| \vec{h} \|}, \quad \text{ equivalently to } \quad \Delta \phi = \frac{\Delta H}{\sqrt{2\Tr (\Delta H)^{2} }}. 
\end{equation}
This last strategy to solve the plateau equation can be further generalized to larger Hilbert spaces. Concretely, consider a system of $N$ qubits interacting through some arbitrary $k$-local Hamiltonian \cite{Adan_Susskind/second_law_complexity/2018}. The Hilbert space dimension is $2^{N}$, and a suitable basis for traceless functions is given by the $4^{N}-1$ traceless Pauli strings\footnote{Of course, if the Hamiltonian is strictly $k$-local, only $3^{k}\binom{N}{k}$ strings are actually needed. We will not bother with these fine details and we will consider instead the full basis. Another possible basis can be given by the standard generators of the $\text{SU}(2^{N})$ algebra.}. On this basis, the parametrization of the Hamiltonian and the unknown function $\Delta \phi$ read,

\begin{eqnarray}
\label{eq:potential_paulis_solution}
  \Delta H = \sum_{\mathcal{P} \in \{1,X,Y,Z\}^{\otimes N}} h_{\mathcal{P}} \mathcal{P},  \quad \text{and}\quad \Delta \phi = \sum_{\mathcal{P} \in \{1,X,Y,Z\}^{N} } \lambda_{\mathcal{P}} \mathcal{P},
\end{eqnarray}
where we implicitly set $\lambda_{11\dots1}=h_{11\dots1}=0$ in the sum. Additionally, as the dimension grows exponentially with the number of qubits, we can further neglect the subleading terms in \eqref{eq:complete_dephasing} and work directly at first order in $d=2^{N}$. After being able to determine all the $4^{N}-1$ coefficients in \eqref{eq:potential_paulis_solution}, the terms must reorganize as a linear combination of $d-1$ traceless powers\footnote{This follows from a corollary of the Cayley-Hamilton theorem.}, \emph{i.e.},

\begin{equation}
\label{eq:potential_Hamilton_calye_solution}
    \Delta \phi= \sum_{l=1}^{d-1} \alpha_{l} \left(\Delta H^{l} -\Tr{\left( \Delta H ^{l}\right)} \right).
\end{equation}
For the single qubit case, and as we showed in \eqref{eq:single_qubit_solution}, the expansion of $\Delta \phi$ can only go at first order in $\Delta H$. However, as also shown in \eqref{eq:single_qubit_solution}, we expect that the coefficients \eqref{eq:potential_Hamilton_calye_solution} will be highly non-linear functions of the Hamiltonian. Finally, although a closed form for such coefficients is intended to be a perspective for future work, we recall that the plateau operator (under a proper normalization) is a convex combination of the pure states $ \ket{E_{l}\, E_{l}} \bra{E_{l} \, E_{l}}$ in the twofold Hilbert space $\mathcal{H}^{\otimes 2}$. This implies, in particular, that $G^{\mathcal{E}_{C}}[H]$ must be positive definite, which sets additional constraints for the coefficients of the linear combination in \eqref{eq:potential_Hamilton_calye_solution}.

\pagebreak

\acknowledgments
W.S. devotes special thanks to Hernan Ocampo, and Carlos Viviescas for stimulating discussions and comments. The authors also thank Silvia Pappalardi for the initial collaboration and useful discussions.

\bibliographystyle{utphys}
\bibliography{C_ensemble}

\end{document}